\documentclass[]{aa}
\usepackage{graphicx}
\usepackage{txfonts}
\usepackage{natbib}

\newcommand{\bfmath}[1]
{\mbox{\boldmath${\rm #1}$\unboldmath}}

\newcommand{\bea}{\begin{eqnarray}}
\newcommand{\eea}{\end{eqnarray}}
\newcommand{\be}{\begin{equation}}
\newcommand{\ee}{\end{equation}}
\newcommand{\ave}[1]{\langle #1 \rangle}
\newcommand{\map}{M_{\rm ap}}
\newcommand{\dif}{{\rm d}}
\newcommand{\re}{\mathcal{R}e}
\newcommand{\im}{\mathcal{I}m}
\def\e {\mathrm{e}}
\def\i {\mathrm{i}}
\def\Real{{\rm I\mathchoice{\kern-0.70mm}{\kern-0.70mm}{\kern-0.65mm}%
  {\kern-0.50mm}R}}
\def\arcsecf {\hbox{$.\!\!^{\prime\prime}$}}

\bibpunct{(}{)}{;}{a}{}{,}

\begin{document}
%
%   \title{Weak lensing analysis of the dark clump near Abell 1942 \\
%   using HST/WFPC2 observations}

\title{The dark clump near Abell 1942:\\
 dark matter halo or statistical fluke?
\thanks{Based on observations made with the NASA/ESA Hubble Space Telescope,
  obtained at the Space Telescope Science Institute, which is operated by
  the Association of Universities for Research in Astronomy, Inc., under
  NASA contract NAS 5-26555; on observations made with the Chandra X-ray
  Observatory, operated  by the Smithsonian Astrophysical Observatory for
  and on behalf of the National Aeronautics Space Administration under
  contract NAS8-03060; and on observations made with the
  Canada-France-Hawaii Telescope (CFHT) operated by the National Research
  Council of Canada, the Institut des Sciences de l'Univers of the Centre
  National de la Recherche Scientifique and the University of Hawaii.
}
%\thanks{based on observations obtained with the Hubble Space Telescope, the
%  Chandra X-ray Observatory, and the Canada-France-Hawaii Telescope}
}

%   \subtitle{}

   \author{A. von der Linden
          \inst{1,2}
          \and
          T. Erben
          \inst{1}
          \and
          P. Schneider
          \inst{1}
          \and
          F.J. Castander
          \inst{3}
          }

   \offprints{A. von der Linden\\
             \email{anja@mpa-garching.mpg.de}
             }

   \institute{Institut f\"ur Astrophysik und Extraterrestrische
              Forschung, Universit\"at Bonn, 
              Auf dem H\"ugel 71, 53121 Bonn, Germany
              \and
              Max Planck Institut f\"ur Astrophysik, 
              Karl-Schwarzschild-Str. 1, Postfach 1317,
              85741 Garching, Germany
              \and
              Institut d'Estudis Espacials de Catalunya/CSIC,
              Gran Capita 2-4, 08034 Barcelona, Spain
}

   \date{Received 20 January 2005; accepted ??}

   \abstract{
Weak lensing surveys
%, which are based on ellipticity measurements of
%faint galaxies, 
provide the possibility of identifying dark
matter halos based on their total matter content rather than just the
luminous matter content. On the basis of two sets of observations
carried out with the CFHT, Erben et al. (2000) presented the first
candidate \textit{dark clump}, i.e. a dark matter concentration
identified by its significant weak lensing signal
without a corresponding galaxy overdensity or X-ray emission.

We present a set of HST mosaic observations which confirms the presence of an
alignment signal at the dark clump position. The signal strength, however,
is weaker than in the ground-based data. It is therefore still unclear
whether the signal is caused by a lensing mass or is just a chance alignment.
We also present Chandra observations of the dark clump, which fail to reveal
any significant extended emission.

A comparison of the ellipticity measurements from the space-based HST data
and the ground-based CFHT data shows a remarkable agreement on average,
demonstrating that weak lensing studies from high-quality ground-based
observations yield reliable results.
   }

   \maketitle
%   \clearpage
%   \tableofcontents
%   \clearpage
%
%________________________________________________________________
\section{Introduction}

In the currently favored cosmological model,
  structure formation in the universe is dominated by collisionless
  Cold Dark Matter (CDM). The model of structure formation by
  gravitational collapse in a pressure-less fluid is able to
  successfully reproduce the filamentary large scale structure
  observed in the universe \citep[e.g.][]{pea99}. However, for the formation
  of galaxies and galaxy clusters, gas dynamics play an important
  role. It seems obvious that galaxy formation is triggered when gas
  falls into the potential wells of dark matter concentrations. We
  therefore expect to find galaxies at the high-density peaks of the
  dark matter distribution. In the CDM scenario, small halos collapse
  earlier and merge to larger halos later. For galaxy formation, this
  implies that the large galaxies we see today formed from mergers of
  protogalaxies. Observations support this theory: we see more
  irregular, small galaxies at higher redshifts and many merger
  systems and galaxies showing evidence for recent mergers.  This
  \textit{bottom-up} scenario also calls for galaxy clusters to build
  up through the merger of smaller halos.

While it may be possible to (temporarily) drive gas from galaxy-sized
halos, when such halos merge to form cluster-sized objects, the
majority of them should contain galaxies and/or hot gas, so that the
resulting massive halo is expected to contain a substantial amount of
luminous matter. A cluster-sized halo very poor of luminous matter
(\textit{dark clump}) would
require a mechanism to drive the gas out of all the smaller halos from which
it assembled or from the massive halo itself. Both cases are highly unlikely:
the first is very improbable, the second very difficult due to the high mass of
the object. At the moment, there are no well-motivated physical processes to
explain either scenario.

The discovery of a dark clump would therefore call for a critical
reevaluation of our current understanding of structure formation in the
universe. Currently, the only tool available to search for such
objects is gravitational lensing, as it probes matter concentrations
independent of their nature.

In the course of a weak lensing study, \cite{ewm00} announced
the possible discovery of a dark clump, about $7 \arcmin$ south of the
galaxy cluster Abell 1942. This assertion is based on
significant alignment signals seen in two
independent high-quality images, taken with the MOCAM and UH8K cameras
at the CFHT. There is no associated apparent galaxy overdensity
visible in these images nor in deep H-band images analyzed by
\cite{gel01}. There is faint X-ray emission about $1\arcmin$ from 
the lensing centroid detected by the ROSAT survey, but it is unclear
whether this could be associated with a lensing object. If the alignment
signal is due to a lensing mass at a similar redshift as the cluster,
$z=0.223$, this halo would have a mass of the order of  
$10^{14}h^{-1} M_{\odot}$. At a higher redshift (0.8 - 1.0), it would
require a mass of the order of $10^{15}h^{-1} M_{\odot}$.

There are currently three more such dark clump candidates in the literature: 
\begin{itemize}
\renewcommand{\labelitemi}{-}
\item \citet{umf00} find a candidate in their weak lensing analysis of the
  galaxy cluster CL1604+4304 using data from the WFPC2 camera of the
  HST. In two separate datasets, they 
  find a peak $1\farcm7$ southwest from the cluster center, which
  corresponds to about 830 $h^{-1}$ kpc at the redshift of the cluster
  ($z=0.897$). They estimate the mass of the object to be about $4.8 \times
  10^{14}h^{-1} M_{\odot}$, assuming it is located at a similar redshift as
  the cluster.
\item \citet{meh02} found a conspicuous tangential alignment of
  galaxies in an image taken by the STIS camera aboard the HST as a
  parallel observation. However, follow-up wide-field observations
  with the VLT failed to detect a weak lensing signal \citep{emc03},
  so that a chance alignment of 52 galaxies in the original STIS
  analysis is at this point considered the most plausible explanation
  for this candidate.  
\item \citet{dpl03} identify a dark clump candidate about 6$\arcmin$
  southwest of the galaxy cluster Abell 959 ($z=0.286$) in images
  taken with the UH8K camera at the CFHT with evidence that this is a 
  dark sub-clump of the cluster. If this is indeed an object at the redshift
  of the cluster, they deduce a mass of $(1.1 \pm 0.3) \times 10^{14}h^{-1}
  M_{\odot}$. 
\end{itemize}

\citet{wek02} argue that about one out of five Dark Matter halos
identified by weak lensing should be a non-virialized halo, i.e. a halo
which is in the process of collapsing and has not yet reached dynamical
equilibrium. Such objects should have only very little X-ray emission
and about half the projected galaxy density of virialized
halos. The luminosity of such objects would therefore be very
difficult to determine. Accordingly, distinguishing between pure Dark
Matter halos and normal, non-virialized halos may be almost impossible
in these cases.

However, the noise in weak lensing analyses due to intrinsic
ellipticities of galaxies can have a profound effect on the statistics
of the number of halos detected per area. Intrinsic ellipticities may
mimic tangential alignment, thereby causing false peaks or boosting
the significance of lensing signals \citep[e.g.][]{hty03}.

To determine the nature of the dark clump near Abell 1942, we obtained
a set of HST observations of the field (General Observer Program, Proposal
ID 9132, PI Erben). The HST probes fainter and
thus more distant galaxies, for which the distortions of a foreground
lensing mass are larger; additionally, due to the lack of seeing, its
shape measurements should be more reliable. If the alignment signal
seen in the ground-based data is due to a lensing mass, it should thus
be even more significant in the HST data.

The structure of this paper is as follows. In Sect. \ref{sc:data-method} we
present the optical data available to us, namely the CFHT images of the
original detection and the HST data, along with our data reduction methods to
extract object catalogs suitable for lensing studies. Sect. \ref{sc:lensing}
gives a 
brief overview of the weak lensing methods employed in this paper.  
In Sect. \ref{sc:ground} we present a re-analysis of the $I$-band image of
the CFHT data. Our weak 
lensing analysis of the HST data, which confirms the alignment
signal, but not its strength, is described in
Sect. \ref{sc:hst}. In Sect. \ref{sc:chandra} we use a deep Chandra image to
show that the ROSAT source is likely to be a spurious detection. The
appendices illustrate various tests for systematics
(App. \ref{sc:hst-systematics}) and a comparison of
the individual shape measurements of objects common to the CFHT and
HST datasets (App. \ref{sc:compare}).

%%%%%%%%%%%%%%%%%%%%%%%%%%%%%%%%%%%%%%%%%%%%%%%%%%%%

\section{Optical data}
\label{sc:data-method}

The goal of this work is to understand the origin of the lensing signal seen by
\cite{ewm00}. We therefore consider both the ground-based dataset used in
the original discovery as well as the HST data. Such a treatment
also allows for a direct comparison of the ellipticity
measurements of objects detected in both datasets.

\subsection{Ground-based data}
\label{sc:desc-ground}

Our analysis concentrates on the same $I$-band image as used in
\citet{ewm00}, as 
this covers most of the area imaged by the HST. We use Chip 3 of a mosaic
observation of Abell 1942 taken with the UH8K camera, with a pixel scale of
$0\farcs206$.  9 exposures of 1200~s went into the final
image, which has a seeing of $0\farcs74$. Unfortunately, a
photometric calibration is missing.

\subsection{Space-based data}
\label{sc:desc-space}

\begin{figure}[htbp]
\begin{center}
%\setlength{\fboxsep}{-\fboxrule}
%\fbox{
%\includegraphics*[bb=3.7cm 2cm 16.8cm 25.5cm,width=0.6\hsize]
\includegraphics[width=0.9\hsize]
{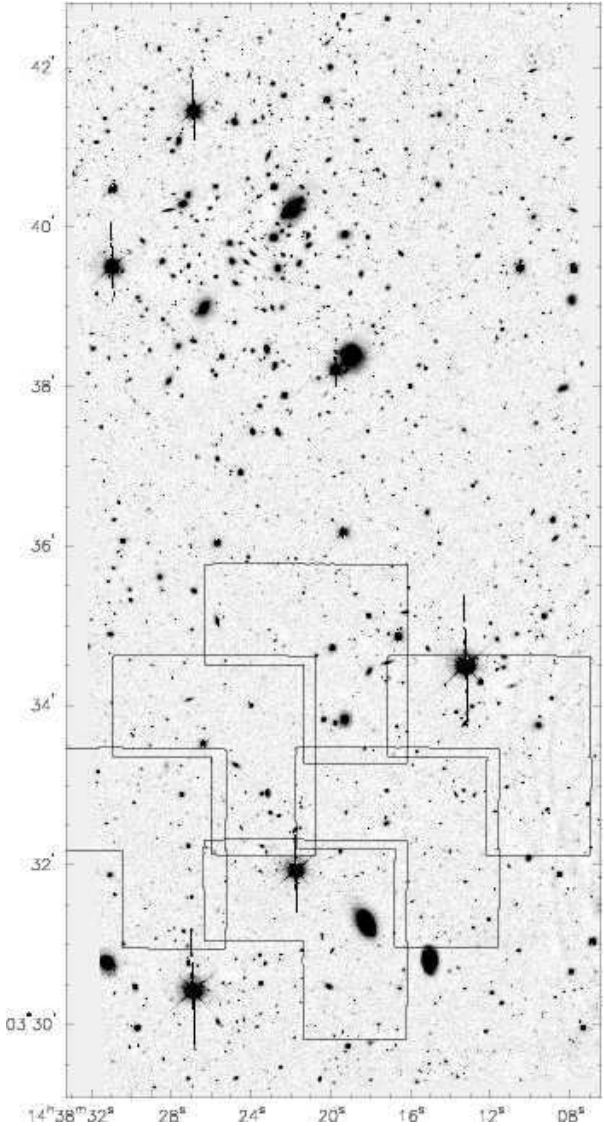}
%}
\caption{The outlines of the six HST pointings superposed on the CFHT
$I$-band image. The center of Abell 1942 lies in the top half of the CFHT
image, the HST mosaic is centered on the dark clump position.}
\label{fig:hst_mosaic}
\end{center}
\end{figure}

Our HST data is a WFPC2 mosaic (approximately 5$\arcmin \times
4\arcmin$) of six pointings, each
consisting of 12 dithered exposures with an exposure time of
400s each, taken between May 20th and June 1st, 2001. The position of the
mosaic with respect to the 
$I$-band image from the CFHT is shown in Fig. \ref{fig:hst_mosaic} The filter
employed was F702W.

%%%%%%%%%%%%%%%%%%%%%%%%%%%%%%

\subsubsection{Data reduction}
\label{sc:hst-data-redu}

Our reduction of the HST data is based largely on the {\tt dither} 
package \citep{frh02}
for {\tt IRAF}. Each of the six pointings was reduced
separately. Simultaneous processing of the four individual chips is done
automatically by the {\ttfamily dither} 
routines. Due to their better signal-to-noise behavior, only
the chips of the Wide Field Camera, namely Chips 2, 3,
and 4 of WFPC2, were used for the later analysis. The dither pattern of the
images allows us to achieve a higher resolution in the coadded image via the
{\tt drizzle} algorithm \citep{gon98}.

\begin{figure*}[htbp]
%%\vspace{0.4cm}
\begin{center}
%%\setlength{\fboxsep}{-\fboxrule}
%%\fbox{
\includegraphics[width=1\hsize]{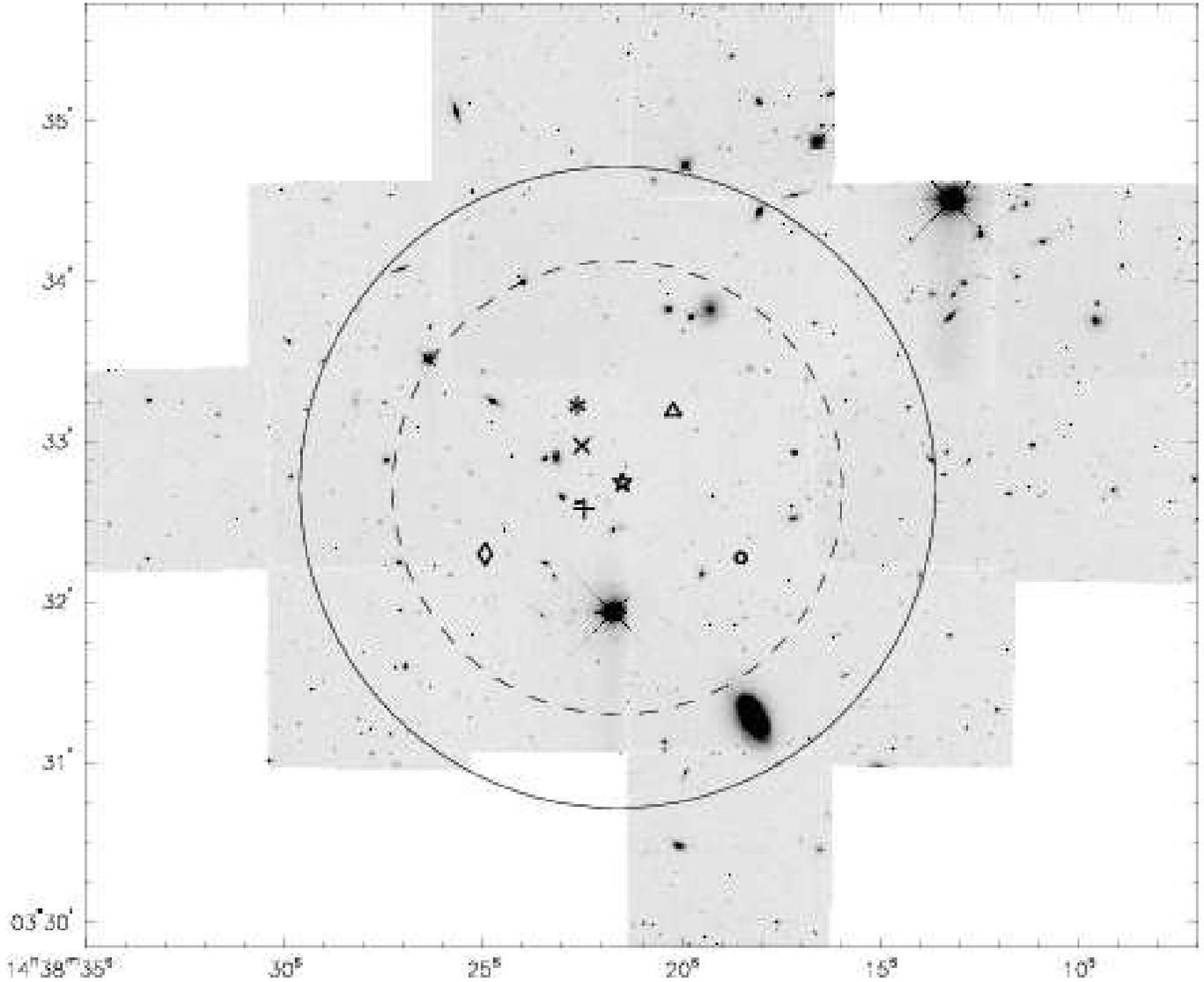}
%%}
\caption{The HST mosaic with an illustration of the various dark clump
  centroids cited in the 
text: the centroid we found in the ground-based data ($\:\star\:$);  the
original position given by \citet{ewm00} ($\:+\:$); the peak position in
the HST data ($\:\triangle\:$); the center of the X-ray emission as found by
ROSAT ($\:*\:$) and by Chandra ($\:\times\:$); the position of the peak
found in the 200$\arcsec$ filter
scale for the faint ground-based galaxies ($\:\circ\:$); and the galaxy
number overdensity of medium-bright HST galaxies
($\:\Diamond\:$). Additionally, 
we plot a circle of $120 \arcsec$ radius around the first position. For our
$\map$ filter function of this radius, the maximal weight is then
assigned along the dashed circle. Assuming a typical uncertainty of
$1\arcmin$ for centroids found via weak lensing, all cited position are
compatible with the dark clump.}
\label{fig:mosaic}
\end{center}
\end{figure*}

The steps of the reduction are outlined in the following:
\begin{description}
\item[Rough cosmic ray removal:] To find the offsets between the
  images, they first have to be cleaned of cosmic rays, which would
  otherwise falsify a cross-correlation.
  Each frame is cleaned using the
  {\tt precor} task. This leaves only objects of a minimal size in the
  image, which should be stars and galaxies, with little
  contamination by cosmic rays.
\item[Offset estimation:] As the individual frames are dithered with
  respect to each other, it is necessary to find their relative 
  offsets. This is done by performing a cross
  correlation of the cosmic ray cleaned images produced in the
  previous step.
\item[Median coaddition:] Using the previously
  determined offsets, the original images are median combined. They
  are mapped via the {\ttfamily drizzle} algorithm onto an output grid
  with pixels of half the original size.
\item[Mask creation:] The median image is mapped back onto the
  original resolution and offset of the original frames using the
  {\ttfamily blot} algorithm, the inverse of {\ttfamily drizzle}. The
  original frame is then compared to the median image to identify
  cosmic rays via the {\ttfamily deriv} and {\ttfamily driz\_cr}
  tasks. Thus, for each frame a mask is created identifying cosmic
  rays. This is combined with a mask identifying defect or possibly
  problematic pixels, which is supplied with each raw frame.
\item[Offset determination:] For each frame, those pixels that are
  flagged in the mask are substituted by their value in the blotted
  image (the transformed median image). These images are then
  cross-correlated to determine the offsets more precisely. Possible
  small rotation angles have to be found manually.
\item[Coaddition:] The images are drizzled onto an output grid of half
  the pixel size ({\tt pix-frac} = 0.5), i.e. twice the original
  resolution. The value of each output pixel is obtained via averaging,
  where pixels which are flagged in the mask image are omitted.
  The {\tt drop-size} (i.e. a scaling applie to the input pixels before
  being mapped) used is 0.6 .
\end{description}
Performing this routine on all of the six pointings, we obtain 18
single-chip images. A mosaic of these is shown in Figure \ref{fig:mosaic}.

%\begin{figure*}[htbp]
%\begin{center}
%%\vspace{18cm}
%%\setlength{\fboxsep}{-\fboxrule}
%%\fbox{
%\includegraphics[width=\hsize]{figures/hst-ddp.ps}
%%}
%\caption{(DDP-filtered) mosaic of the HST images.}
%\label{fig:mosaic}
%\end{center}
%\end{figure*}

%%%%%%%%%%%%%%%%%%%%%%%%%%%%%%%%%%%%%%%%%%%%%%%%%%%%

\subsection{Catalog extraction}
\label{sc:catalog}

For both datasets, we used a similar process to build the object
catalogs. Differences in the procedure arise mainly from the small
field-of-view and mosaic nature of the HST data.

\subsubsection{Preparatory steps}
\label{sc:cat-0}

We manually updated the astrometric information of the ground-based image
until its bright objects coincide with their positions as given in the USNO-A2
catalog \citep{mcd98}. This step is not necessary for the lensing analysis of
the 
ground-based image, but provides us with a reference catalog for the
astrometric calibration of the HST mosaic.

The HST images are aligned roughly with sky coordinates, but since each chip
is read out along a different chip border, the individual images are rotated
with respect to each other.  To avoid
confusion, we first rotate the images of Chips 2 and 4 by $\mp 90^{\circ}$,
respectively, such that north is up and east to the left (approximately).
This is performed solely as a rearrangement of pixel values to avoid
any additional resampling process.

\subsubsection{Masking}
\label{sc:cat-1}

In order to avoid noise signals and distorted ellipticity measurements, we
mask out bright stars and the artefacts they cause 
(diffraction spikes, blooming, CTE trails, a ripple-like structure at the
eastern edge of the $I$-band image, and brightening of the background along
columns of the HST chips in two cases). In the HST image, we also mask the
bright galaxy in the field.

\subsubsection{Source extraction}
\label{sc:cat-2}

We used {\tt SExtractor} \citep{bea96} to identify objects in the
image. {\tt SExtractor} considers $N$ connected pixels that are at a level
$k\cdot\sigma_{\rm sky}$ above the sky background as an object, with
$\sigma_{\rm sky}$ being the standard deviation of the background noise.
For the CFHT image, we used $N = 3$ and $k=1.0$. These are very low
thresholds, but since we later want to correlate objects present in both
datasets, we strive to obtain a high number density of objects. For the HST
images we used $N = 3$ and $k=1.5$.\\

%\paragraph{HST: astrometric and photometric calibration}
\noindent{\bf HST: astrometric and photometric calibration}
%\label{sc:hst-astro}

The astrometric calibration plays an important role for a mosaic dataset
such as our HST images,
as it gives the position of the images with respect to each
other. Indeed, the positions on the sky will be used later on in the lensing
analysis rather than $x$ and $y$ position on the chip.

For a reference catalog, we had extracted a catalog of bright objects from the
astrometrically calibrated ground-based $I$-band image (see
Sect. \ref{sc:cat-0}). We match the entries of the reference
catalog to objects found in the HST images. This allows the 
determination of the pointing and the distortion of the image. Objects that
are detected both in the $I$-band and the HST image can then be identified
by sky coordinates.

The photometric calibration is done based on the relevant keywords of the
HST image headers.

\subsubsection{Ellipticity measurement}
\label{sc:cat-3}

For the objects in the SExtracted catalogs, we measure the ellipticities
using a modified version of the {\tt imcat} software, following the method
of \citet{ksb95}. We used the half-light radius as
measured by {\tt SExtractor} as the radius of the
weight function with which the brightness profile is weighted.

The measured image ellipticity $\bfmath{\chi}$ is related to the source
ellipticity $\bfmath{\chi}^{(s)}$ by 
\be
\bfmath{\chi} = \bfmath{\chi}^{(s)} + \tens{P}^{\rm g} \bf{g} + 
\tens{P}^{\rm sm}\bfmath{q}^{\star}\;,
\label{ksb}
\ee
where $\bf{g}$ is the reduced shear induced by a lensing mass which we
ultimately want to determine.
$\bfmath{q}^{\star}$ is the stellar anisotropy kernel, i.e. the anisotropy of
the PSF for point-like objects.
$\tens{P}^{\rm g}$
and $\tens{P}^{\rm sm}$ (smear polarizibility tensor) are tensors describing
a galaxy's susceptibility to 
the two distorting effects. They are
also measured from a galaxy's light distribution \citep[see][for a
derivation]{bas01}. 

\subsubsection{Anisotropy correction}
\label{sc:cat-4}

To correct for the anisotropy induced by the telescope--detector system, we
measure the ellipticities of the stars present in the field; for these
eq. (\ref{ksb}) simplifies to
\be
\bfmath{\chi}^{\star} = \tens{P}^{\star \rm sm} \bfmath{q}^{\star}\;.
\ee
Stars are selected from a magnitude vs. radius plot. We fit a third-order
polynomial in chip position $(x,y)$ to the quantity
$\left(\frac{\bfmath{\chi}^{\star}}{0.5 \,{\rm tr}(\tens{P}^{\star \rm sm})}
\right) (x,y)$ to estimate $\bfmath{q}^{\star} (x,y)$. 

Thus, we obtain anisotropy-corrected ellipticity measurements:
\be
\bfmath{\chi}^{\rm aniso} = \bfmath{\chi} - \tens{P}^{\rm
  sm}\bfmath{q}^{\star}\;.
\label{eq:aniso}
\ee

\subsubsection{Modification for the HST images}
\label{sc:hst-aniso}

Since the PSF cannot be described analytically across chip-borders,
the anisotropy correction for a mosaic has to be applied to the single
chips. With the small
field of view of the WF chips, we have the added difficulty that for
each image, there are only about five stars that could be used for the
polynomial fitting, obviously not enough. Since the images were taken
consecutively, we can assume that the PSF does not change considerably
between the six pointings. We therefore apply an anisotropy correction
for each chip based upon all the stars that were 
found in the six images taken by that chip. 

The 18 single-chip catalogs are combined to three catalogs, one for each
chip. Because there are still few stars even in these catalogs, the
stellar sequence is selected manually. For
each catalog, a third-order polynomial is fitted.

The anisotropy correction of the HST images is further discussed in Appendix
\ref{app-aniso}.

\subsubsection{Shear estimation}
\label{sc:cat-5}
After the anisotropy correction, the second step in retrieving an estimate
of the local shear from 
ellipticity measurements is the correction for the $\tens{P}^{\rm g}$ tensor.
It is a combination of $\tens{P}^{\rm sm}$ and the shear polarizibility
tensor $\tens{P}^{\rm sh}$:
\be
\tens{P}^{\rm g} = \tens{P}^{\rm sh} - \tens{P}^{\rm sm}(\tens{P}^{\star \rm sm})^{-1} \tens{P}^{\star \rm sh}\;,
\label{pg}
\ee
where the starred quantities are the corresponding tensors as measured on
stellar-sized objects. Note that the weight
function with which $\tens{P}^{\star \rm sh}$ and $\tens{P}^{\star \rm sm}$
are measured 
should be the same as that used for the respective object.

The basic assumption of weak lensing studies is that the average source
ellipticities vanish, i.e. $\ave{\bfmath{\chi}^{(s)}} = 0$. We also assume
that $\ave{(\tens{P}^{\rm g})^{-1} \bfmath{\chi}^{s}} = 0$, so by averaging 
 eq. (\ref{ksb}) we obtain:
\be
{\bf g} = \ave{(\tens{P}^{\rm g})^{-1} \bfmath{\chi}^{\rm aniso}}\;.
\label{ksb-g}
\ee
Thus the expectation value of the quantity $(\tens{P}^{\rm g})^{-1}
\bfmath{\chi}^{\rm aniso}$ is the reduced shear ${\bf g}$ at the respective
point. 

$\tens{P}^{\rm g}$ is an almost diagonal tensor with
similar elements on the diagonal. In fact, in the absence of a weight
function and a PSF its elements would be: $P^{\rm g}_{11} = P^{\rm g}_{22} =
2$ , 
$P^{\rm g}_{12} = P^{\rm g}_{21} = 0$. We can therefore approximate the
tensor $\tens{P}^{\rm g}$ by a scalar quantity:
\be
P^{\rm g}_s \;=\; \frac{1}{2}\, {\rm tr}(\tens{P}^{\rm g})\;.
\label{eq:pgs}
\ee
It has been shown that while the full $\tens{P}^{\rm g}$ tensor can
overestimate the 
shear, this approximation is more conservative and will only
underestimate the true shear \citep{ewb01}.

We therefore use a variant of eq. (\ref{ksb-g}) as our estimate of the local
shear at each galaxy's position:
\be
%{\bf g}
\bfmath{\varepsilon}
 = \frac{\bfmath{\chi}^{\rm aniso}}{P^{\rm g}_s}\;.
\label{eq:ksb-shear}
\ee
At this point, we reject those objects from the catalog that have a
radius equal to or smaller than the stellar locus, are saturated
stars, or have a final ellipticity of 
$|\bfmath{\varepsilon}| > 1$.
%$|{\bf g}| > 1$.

\subsubsection{Rotation of ellipticities}
\label{sc:cat-6}

The ellipticities were measured with respect to the $x$-axis of each
image, which for all images runs approximately along the east axis. However,
the lensing analyses are done in sky coordinates. We therefore need to
transform the ellipticity measurements so that their position angle is
measured relative to the right ascension axis. This is done with the
transformation
$$
\bfmath{\varepsilon} \;\rightarrow\; \bfmath{\varepsilon}\, \e ^{2 \i \varphi}
%{\bf g} \;\rightarrow\; {\bf g}\, \e ^{2 \i \varphi}
$$
where $\varphi$ is the angle of
rotation of the image.
%The ellipticities were measured with respect to the
%$x$-axis of the image, to which the rectascension runs in the opposite
%direction - hence the positive sign of the exponent.

After this step, the 18 single-image catalogs of the HST data are merged
into one catalog.

\subsubsection{Weighting}
\label{sc:cat-7}

To describe the reliability of its shape measurement, we want to assign
a weight to each galaxy, based on its noise properties. Since our shear
estimates are gained from averages over ellipticities, a good weight
estimate is
\be
w_i \;\propto\; \frac{1}{\sigma_i^2}\;.
\label{eq:weight-elli-variance}
\ee
where $\sigma_i$ is the (two-dimensional) dispersion of image ellipticities
in the ensemble
over which is averaged. For intrinsic ellipticities, $\sigma_{\varepsilon}
\approx 0.3 - 0.4$ (measured from data). Measurement errors cause $\sigma_i$
to be higher than 
this. 
We follow the weighting scheme of \citet{ewb01}, and select as the
ensemble to average over the 20 closest neighbors of the respective
galaxy in the parameter space spanned by half-light radius $r_g$ and
magnitude $m$.

\subsubsection{Final cuts}

After the weighting, we remove objects with $|\varepsilon|>0.8$. Such objects
would dominate the shear signal, but 
these are also the objects that are most afflicted by noise in the
$P^{g}$ tensor. Additionally, we use only objects for which $P^{g}>0.3$.
This leaves about 2000 objects in both catalogs, which
corresponds to 20 galaxies/arcmin$^2$ for the $I$-band image and 65
galaxies/arcmin$^2$ for the HST image.

%%%%%%%%%%%%%%%%%%%%%%%%%%%%%%%%%%%%%%%%%%%%%%%%%%%%
%                    LENSING                       %
%%%%%%%%%%%%%%%%%%%%%%%%%%%%%%%%%%%%%%%%%%%%%%%%%%%%

\section{Weak lensing methods}
\label{sc:lensing}

Weak lensing analyses are based on using estimates of the local shear
$\bfmath{\gamma}$ to reconstruct information on the convergence 
$\kappa$, which is a dimensionless measure of the surface mass density. 
In the weak lensing regime, $\kappa \ll 1$, so that $\ave{\bfmath{\varepsilon}}
= {\bf g} = 
\bfmath{\gamma}/(1-\kappa) \approx \bfmath{\gamma}$.

\subsection{Mass reconstruction}
\label{sc:massreco}

Both the shear $\bfmath{\gamma}$ and the convergence $\kappa$ are linear
combinations of second derivatives of the lensing potential, so that it is
possible to express $\kappa$ as an integral over $\bfmath{\gamma}$ via the
Kaiser--Squires Inversion \citep{kas93}.
This method is usually not applied directly, as the shot noise introduced by
summing over individual galaxies (shear measurements) produces infinite
noise. This can be avoided by first smoothing the shear measurements;
however, such a smoothing scale introduces correlated errors. Another
problem arises from the limited field-of-view of any
observations. \citet{ses01} express this as a von Neumann boundary problem, leading to
the so-called \textit{finite--field inversion}. We rely on this method for
mass reconstructions throughout the paper.

\subsection{Mass-aperture statistics}
\label{sc:map}

The \textit{mass-aperture}, or $\map$, {\it Statistics}, developed by
\citet{sch96_2}, provides a method with defined noise properties to identify
mass concentrations. It is based upon the relation
\be
\map(\vec{\theta}_0) =
\!\int_{\Real^2}\! \dif^2 \theta\;
\kappa(\vec{\theta}_0+\vec{\theta}) \: w(|\vec{\theta}|)
= \!\int_{\Real^2}\! \dif^2 \vartheta 
\: \gamma_{\rm t}(\vec{\theta}_0;\vec{\vartheta})
\,Q(|\vec{\vartheta}|)\;.
\label{eq:map}
\ee
The aperture mass $\map(\vec{\theta}_0)$ presents a measure of the average
convergence, mulitplied with a filter function $w$, around a position
$\vec{\theta}_0$ in the lens plane. If $w(|\vec{\theta}|)$ is a compensated
filter, $\map$ avoids 
the mass sheet degeneracy. The right side of eq. (\ref{eq:map}) expresses
$\map$ in terms of the tangential shear $\gamma_{\rm t}$ at the position
$\vec{\theta}_0+\vec{\vartheta}$ with respect to $\vec{\theta}_0$:
\be
\gamma_{\rm t} (\vec{\theta}_0;\vec{\vartheta}) = 
- \re \left(\bfmath{\gamma}(\vec{\theta}_0+\vec{\vartheta})
  \e^{-2\i\phi}\right) \;,
\ee
where $\phi$ is the polar angle of the vector $\vec{\vartheta}$.
The weight function $Q$ is determined in terms of $w$.

Eq. (\ref{eq:map}) is intuitively clear: a lens
most often deforms images so they align tangentially to the center of
mass. An average over the
tangential components of galaxy ellipticities must therefore be a measure of
the surface mass. With this interpretation, $\map$ is a useful quantity in
its own right even if the weak lensing assumption, $\bfmath{\gamma} =
\ave{\bfmath{\varepsilon}}$, breaks down or if part of the weight function lies
outside the field.

The imaginary shear component is the cross component:
\be
\gamma_{\times}(\vec{\theta}_0;\vec{\vartheta}) = 
- \im \left(\bfmath{\gamma}(\vec{\theta}_0+\vec{\vartheta}) 
\e^{-2\i\phi}\right) \;.
\ee
Substituting it for the tangential shear in eq. (\ref{eq:map})
yields $\map^{\times}$ whose expectation value vanishes. Evaluating it
analogous to $\map$ can thus be used as a method to check the quality of the
dataset.

\subsubsection{Application to real data}

\begin{itemize}
\item
In order to apply the weight function $Q$ to 
finite data fields, a cut-off radius $\theta_{\rm
  out}$ should be used, beyond which the filter function
vanishes. Otherwise, the area of 
integration is not well sampled by galaxy images. A compensated filter 
$w(|\vec{\vartheta}|)$, for which $w(|\vec{\vartheta}|) \equiv 0$ for
$|\vec{\vartheta}| > \theta_{\rm out}$ yields a weight function
$Q(|\vec{\vartheta}|)$ which vanishes beyond the same cut-off radius.
We use filter and weight function as introduced in \citet{svj98}, with $l=1$.
\begin{figure}[hbtp]
%\vspace{0.4cm}
\begin{center}
%\setlength{\fboxsep}{-\fboxrule}
%\fbox{
%\includegraphics[bb=1.2cm 5.8cm 20.2cm 24.7cm,width=7cm]
\includegraphics[trim=0 0 0 0.2cm,width=9cm]
{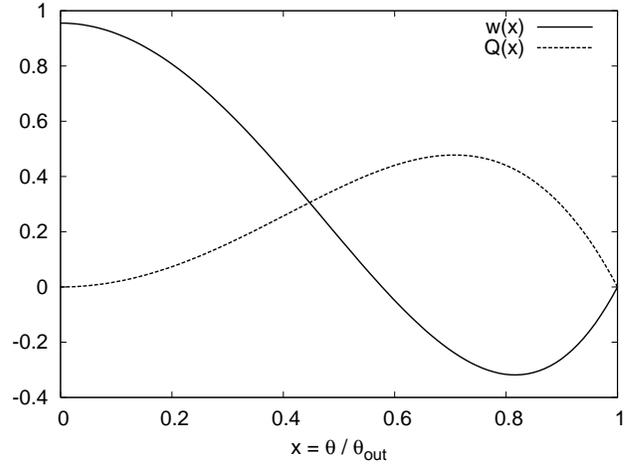}
%}
\caption{The filter function $w$ (solid line) we used and its
  corresponding weight function $Q$ (dashed line) shown in units of
  $\theta_{\rm out}$.
}
\label{wQ}
\end{center}
\end{figure}
%\bea
%w(\theta) &\;=\;& \frac{9}{\pi \theta_{\rm out}^2}
%\left( 1-\left(\frac{\theta}{\theta_{\rm out}}\right)^2 \right)
%\left( \frac{1}{3} - \left(\frac{\theta}{\theta_{\rm out}}\right)^2 \right)\\
%Q(\theta) &\;=\;& \frac{6}{\pi \theta_{\rm out}^2}
%\left(\frac{\theta}{\theta_{\rm out}}\right)^2
%\left( 1-\left(\frac{\theta}{\theta_{\rm out}}\right)^2 \right)\;.
%\eea
\item
We can sample the shear field only at discrete points, namely those where
there is a measured galaxy image. In the weak lensing regime, the image
ellipticity is on average a direct measure of the shear, so that we can use
the tangential ellipticity $\varepsilon_{\rm t}$ as an estimate for the
tangential shear $\gamma_{\rm t}$ 
($\varepsilon_{\rm t}$ defined analogously to $\gamma_{\rm t}$). 
\end{itemize}
We thus estimate $\map$ by using
\be
\map (\vec{\theta}_0;\theta_{\rm out}) \;=\; 
\frac{\pi \theta_{\rm out}^2}{N} \sum\limits_i 
\varepsilon_{\rm t} (\vec{\theta}_i) \, Q(|\vec{\theta}_i-\vec{\theta}_0|)\;,
\label{map}
\ee
where $i$ runs over all $N$ galaxies within a radius $\theta_{\rm out}$ from
the point $\vec{\theta}_0$.

With the weighting scheme introduced in eq. (\ref{eq:weight-elli-variance}),
this becomes: 
\be
\map (\vec{\theta}_0;\theta_{\rm out}) \;=\; 
\pi \theta_{\rm out}^2 \,
\frac{
\sum_i 
%\varepsilon_t(\vec{\theta}_i) \, Q(|\vec{\theta}_i|) / \sigma_i^2}
\varepsilon_{\rm t} (\vec{\theta}_i)\, \sigma_i^{-2} \, 
Q(|\vec{\theta}_i-\vec{\theta}_0|)}
%{\sum\limits 1/\sigma_i^2 }
{\sum_i \sigma_i^{-2} }\;.
\label{mapw}
\ee

\subsubsection{Significances}

Any $\map$ value is incomplete without an estimate of its significance,
i.e. how it compares to the typical noise level of the $\map$ estimator.
The \textit{signal-to-noise ratio} is given by
\be
\frac{S}{N} = \frac{\map}{\sigma_{\map}} = 
\frac{\map}{\sqrt{\langle \map^2 \rangle - \langle \map \rangle^2}}\;,
\ee
where the denominator should represent the case of no lensing.
%which is
%approximated by letting the ensemble average run over the whole data field
The expectation
value $\langle \map \rangle$ of $\map$ then vanishes since 
the galaxies are oriented completely randomly and thus
$\ave{\varepsilon_{\rm t}}$, the expectation value of tangential ellipticities,
vanishes. In the case of a non-weighted $\map$ estimator, we obtain: 
\be
\ave{\map^2} \;=\; \frac{\left(\pi \theta_{\rm out}^2\right)^2}{N^2}\,
\sum\limits_i \,\ave{\varepsilon_{\rm t}^2(\vec{\theta_i})} \,
Q^2(|\vec{\theta}_i-\vec{\theta}_0|)\;.
\ee
In the case of no lensing, $\ave{\varepsilon_{{\rm t},i}^2}$ is the
one-dimensional variance of the intrinsic ellipticities:
\be
\ave{\varepsilon_{{\rm t},i}^2} 
= \frac{1}{2}\,\sigma_{\varepsilon}^2
\ee
and the signal-to-noise ratio becomes:
\be
\frac{S}{N} \;=\; \frac{\sqrt{2}}{\sigma_{\varepsilon}} \:
\frac{\sum_i \varepsilon_{\rm t} (\vec{\theta}_i) \, 
Q(|\vec{\theta}_i-\vec{\theta}_0|)}
{\sqrt{\sum_i Q^2(|\vec{\theta}_i-\vec{\theta}_0|)}}\;.
\ee

For the weighted estimator, one faces the problem that the weights
$1/\sigma_i^2$  
are in general not completely independent of the tangential ellipticity
$\varepsilon_{t,i}$. We assign weights to galaxy images by considering the
variance of ellipticities of an ensemble of galaxy images with similar noise
properties. But in general, large ellipticities are often noise-afflicted,
so that they are assigned a lower weight.
The expression for $\ave{\map^2}$ then cannot be simplified ad hoc:
\be
\ave{\map^2} \;=\; 
(\pi \theta_{\rm out}^2)^2 \,
\left\langle
\frac{
\sum_{i,j} 
%\varepsilon_t(\vec{\theta}_i) \, Q(|\vec{\theta}_i|) / \sigma_i^2}
\varepsilon_{{\rm t},i} \varepsilon_{{\rm t},j}\,\sigma_i^{-2} \sigma_j^{-2}
\,  Q_i Q_j}
%{\sum\limits 1/\sigma_i^2 }
{\sum_{i,j} \sigma_i^{-2} \sigma_j^{-2} }
\right \rangle\;.
\label{map2}
\ee

The significance of a detection is related to the probability that the
observed alignment of tangential ellipticities can be mimicked by a
random distribution of galaxy ellipticities. A commonly used possibility to
determine the significance is therefore to randomize the
position angles of the galaxy images and calculate $\map$ of these. This is
repeated $N_{\rm rand}$ times.

With this in mind, we reconsider eq. (\ref{map2}). Such randomizations
represent a possible ensemble average, which we denote by
$\ave{\ldots}_{\phi}$.  For each realization, the
ellipticity modulus remains the same, only the orientation changes. In this
case, the weights also remain the same and we can simplify the expression:
$$
\ave{\map^2}_{\phi} = \left(\pi \theta_{\rm out}^2 \right)^2 \frac
{\sum_{i,j} \frac{\ave{\varepsilon_{{\rm t},i} \varepsilon_{{\rm
          t},j}}_{\phi}}{\sigma_i^2 
    \sigma_j^2} \; Q_i Q_j} 
{\left(\sum_i \frac{1}{\sigma_i^2}\right)^2} =
\frac{(\pi \theta_{\rm out}^2)^2}{2} \; \frac
{\sum_i \frac{|\bfmath{\varepsilon}_i|^2}{\sigma_i^2}\: Q_i^2}
{\left(\sum_i \frac{1}{\sigma_i^2}\right)^2}
$$
since, as the ellipticity modulus is fixed,
$$
\ave{\varepsilon_{{\rm t},i} \varepsilon_{{\rm t},j}}_{\phi} =
\frac{|\bfmath{\varepsilon}_i|^2}{2} \: \delta_{ij}
\;.
$$
The signal-to-noise ratio for the weighted estimator is therefore:
\be
\frac{S}{N} \;=\; \sqrt{2} \:
\sum_i \frac{\varepsilon_{\rm t} (\vec{\theta}_i) \: \sigma_i^{-2} \, Q(|\vec{\theta}_i-\vec{\theta}_0|)}
{\sqrt{\sum_i |\bfmath{\varepsilon}_i|^2 \, \sigma_i^{-2}
   \: Q^2(|\vec{\theta}_i-\vec{\theta}_0|)}}\;. 
\ee
To check the validity of the assumptions we made, we compared results from
randomizations and this analytic formula and found them to be equivalent.

%%%%%%%%%%%%%%%%%%%%%%%%%%%%%%%%%%%%%%%%%%%%%%%%%%%%
%                 GROUND-BASED                     %
%%%%%%%%%%%%%%%%%%%%%%%%%%%%%%%%%%%%%%%%%%%%%%%%%%%%

\section{Re-analysis of the CFHT data}
\label{sc:ground}

In Table \ref{tab:ewm00}, we summarize the differences between the analyses of
\citet{ewm00} and this work.
\begin{table}[tbp]
\begin{center}
\caption{Summary of the differences of the object catalogs used for the
  lensing analyses of \citet{ewm00} and this work. The {\it brightness} key
  refers to the magnitude bins used to split the sample (note that the
  magnitudes are not calibrated).}
%\vspace{2mm}
\renewcommand{\arraystretch}{1.2}
\begin{tabular}{l|c|c}
 & \citet{ewm00} & this work\\
\hline
masking & $-$ & $\surd$\\
source extraction & $6\, {\rm px}\ge 1.0\sigma$ & $3\, {\rm px}\ge 1.0\sigma$\\
weighting & $-$ & $\surd$\\
$|\bfmath{\varepsilon}|$ & $\le 1.0$ & $\le 0.8$ \\
brightness & $m \gtrsim 23$ & $all$ \\
 & & $m \ge 23.67$ \\
 & & $22.54 < m < 23.67$ \\
 & & $m \le 22.54$
\end{tabular}
\label{tab:ewm00}
\end{center}
\end{table}

\subsection{Mass reconstruction}

\begin{figure}[hbtp]
\begin{center}
%\setlength{\fboxsep}{-\fboxrule}
%\fbox{
\includegraphics[width=0.6\hsize]
{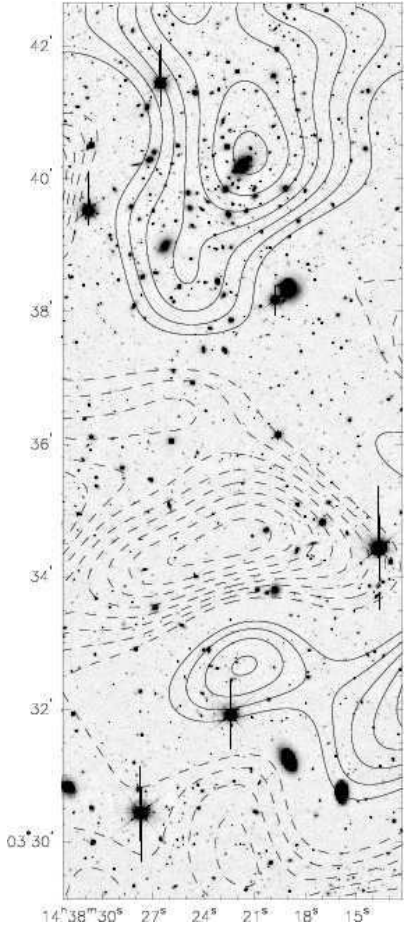}
%}
\caption{Mass reconstruction of the CFHT $I$-band image according to
  \citet{ses01}. Solid (dashed) lines give positive (negative) $\kappa$
  contours, starting at $\pm 0.02$ and in-(de-)creasing in 0.01
  intervals. The smoothing scale is 60\arcsec.}
\label{dclump-I-massreco}
\end{center}
\end{figure}

In order to apply the finite-field inversion (Sect. \ref{sc:massreco}), we
further cut the $I$-band image to avoid the ripple-like reflection artefact 
at the eastern edge altogether. This narrows the available field, but avoids
problems at the boundaries. A resulting mass reconstruction is shown in
Fig. \ref{dclump-I-massreco}.

Abell 1942 shows up prominently in the top half of the field, with the peak
of the mass map centered approximately on the cD galaxy. In the lower half
of the image, there is a second, albeit lower peak at the same position as
detected in \citet{ewm00}. Relative to the peak $\kappa$ of A1942, our dark
clump signal is slightly larger than given by \citet{ewm00}. However, north
of the dark clump, there is a ``hole''  - a region of significantly negative
$\kappa$ values. 
Although this is a somewhat disconcerting result, it must be stressed that
$\kappa$ is underestimated in the whole field due to the mass sheet
degeneracy and the cluster in the field (two of the three field boundaries
close to the cluster and well within its extent display nearly vanishing
$\kappa$ values).
Unfortunately, the original analysis of \citet{ewm00}
only investigates regions of positive $\kappa$, so that this result cannot
be compared.

\subsection{$\map$ analysis}

\subsubsection{Complete sample}

\begin{figure*}[hbtp]
\begin{center}
%\setlength{\fboxsep}{-\fboxrule}
%\fbox{
\includegraphics[width=1.0\hsize]
{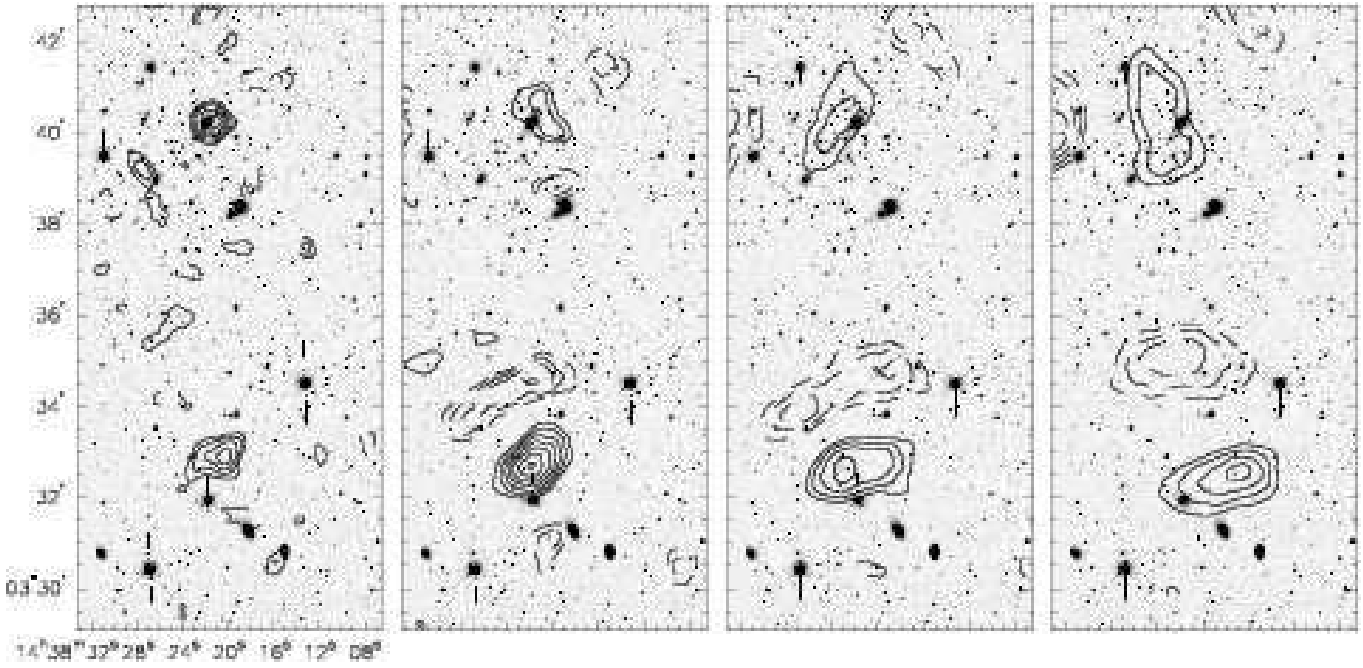}
%}
\caption{Results of a weighted $\map$ analysis of the $I$-band data. Shown
  are positive (negative) S/N contours as solid (dashed) lines starting 
at $\pm$ 2.0$\sigma$ in 0.5$\sigma$ intervals. The filter scales from left
to right are 80$\arcsec$, 120$\arcsec$, 160$\arcsec$, and 200$\arcsec$.
The dark clump signal shows up strongest in the 120$\arcsec$ filter, with a
peak significance of 4.9$\sigma$.}
\label{dclump-I-maps}
\end{center}
\end{figure*}

We perform the $\map$ analysis as described in
Sect. \ref{sc:map} and evaluate the $\map$ statistic on a grid with
grid spacings of 2$\arcsec$. The result is shown in Fig. \ref{dclump-I-maps}.
The dark clump signal is seen significantly at all filter scales, but it is
particularly strong for the 120$\arcsec$ filter, where it reaches a
peak significance of 5$\sigma$. In the other filters, the
significance is at the 3.5$\sigma$ level, as in \citet{ewm00}.

Abell 1942 is detected only weakly for large filter scales, a result that is
consistent with \citet{ewm00}. The mass reconstruction from the previous
paragraph demonstrates why the dark clump reaches a much higher $\map$
significance than the cluster: at the dark clump position, the negative part
of the filter function (Fig. \ref{wQ}) is evaluated largely at the position
of the hole, thereby boosting the signal. The same in reverse is true for
the hole itself: its significance is boosted by its proximity to the Dark
Clump. Yet its significance remains lower than that of the dark clump.

%It is a bit worrying that Abell 1942 is detected only weakly for large
%filter scales. But also this low signal is consistent with
%\citet{ewm00}. Using a filter function which is motivated by the
%suspected density profile of the cluster (NFW profile) yields a
%stronger signal for the galaxy cluster, but similar results for the
%dark clump.

\subsubsection{Rough redshift dependence}

\begin{figure*}[bthp]
\begin{center}
%\setlength{\fboxsep}{-\fboxrule}
%\fbox{
%\includegraphics[bb=1cm 10cm 20cm 18cm,width=1.0\hsize]
\includegraphics[width=1.0\hsize]
{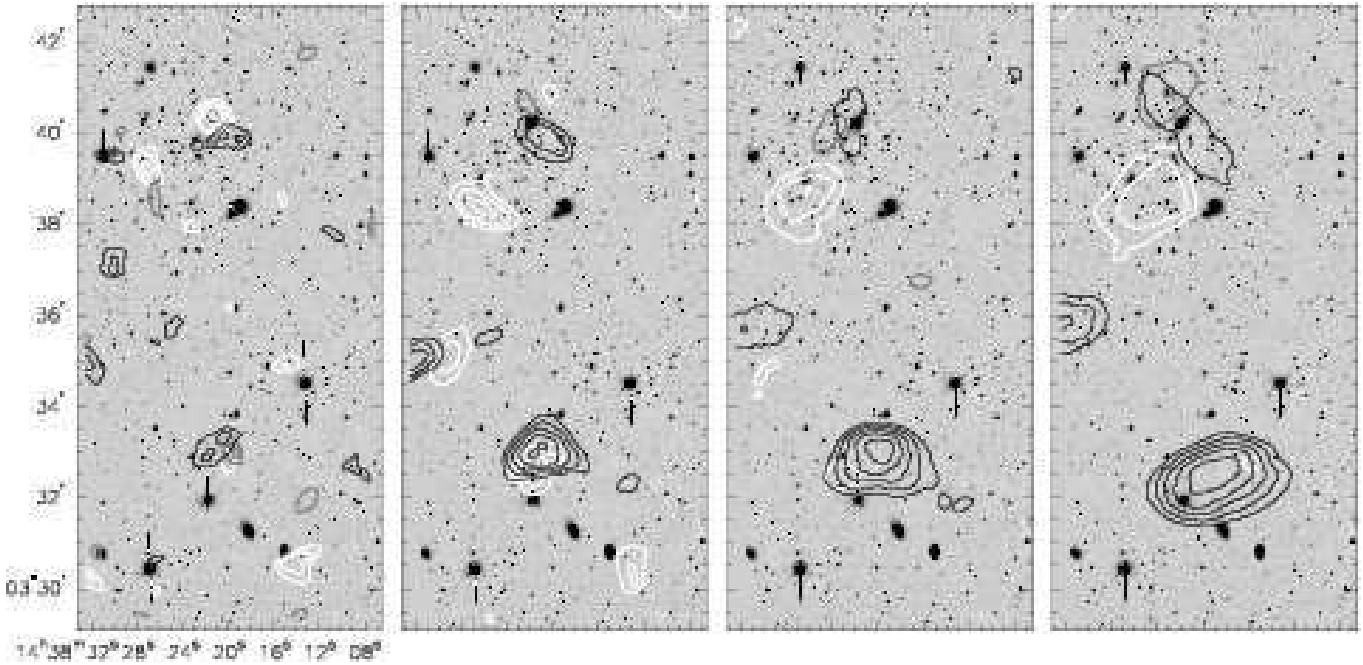}
%}
\caption{The same as the previous figure (Fig. \ref{dclump-I-maps}), but
  with the galaxies divided into three magnitude bins and only positive
  contours. White contours
  correspond to the 
  brightest galaxies, gray contours to those of medium brightness, and black
  contours to the faintest.}
\label{dclump-I-redshift-maps}
\end{center}
\end{figure*}

On average, the more distant a galaxy, the fainter it is. By introducing
magnitude cuts we split the galaxy sample
into three parts of about 660 galaxies each with different mean redshift.
This is a very crude redshift distinction, but should reveal any trend of
lens strength with redshift.

The results of this analysis are shown in Fig. \ref{dclump-I-redshift-maps}. 
We see that the dark clump signal stems mostly from faint galaxies, which
supports the notion that this is a high-redshift object. However, these are
also those objects that are most subject to noise effects.

At the 120$\arcsec$ filter scale, there is also a 3$\sigma$ contribution
from bright galaxies. Assuming that the lensing mass is indeed a high
redshift object, these bright galaxies are unlikely to be at higher
redshifts. Thus, this is probably not a lensing signal.
%The signal might stem from the possible galaxy group close
%by: if these galaxies are physically interacting, their ellipticities are
%no longer independent of each other, so that their mean ellipticity might not
%vanish. 
Yet, this ``contamination'' can explain the high signal-to-noise ratio we
see at this filter scale.
\\

For Abell 1942, there is a strong signal at the smallest filter scale,
centered on the cD galaxy which exhibits a strong lensing arc. It may well
be that at these radii we are not in the weak lensing regime any more and
the tangential alignment is already rather distinct. In
the other filter radii, there is no particularly strong signal. This might
be due to the generic weight function which is 
not adapted to the NFW profile.

Our plot does not show the negative $\map$ contours to avoid
overcrowding. Unlike to the dark clump, all three magnitude bins contribute
to the ``hole''.

\subsubsection{$\map$ cross component}

In Sect. \ref{sc:map}, we argued that $\map^{\times}$,
i.e. $\map$ calculated with the cross component of the shear instead
of the tangential one, must vanish. By
checking the validity of this assumption in the dataset, we can
identify possible problems.

\begin{figure*}[htbp]
%\vspace{0.4cm}
\begin{center}
%\sidecaption
%\setlength{\fboxsep}{-\fboxrule}
%\fbox{
%\includegraphics[width=\hsize, trim=0 0.2cm 0 0]
\includegraphics[trim=0 0.2cm 0 0,width=12cm]
{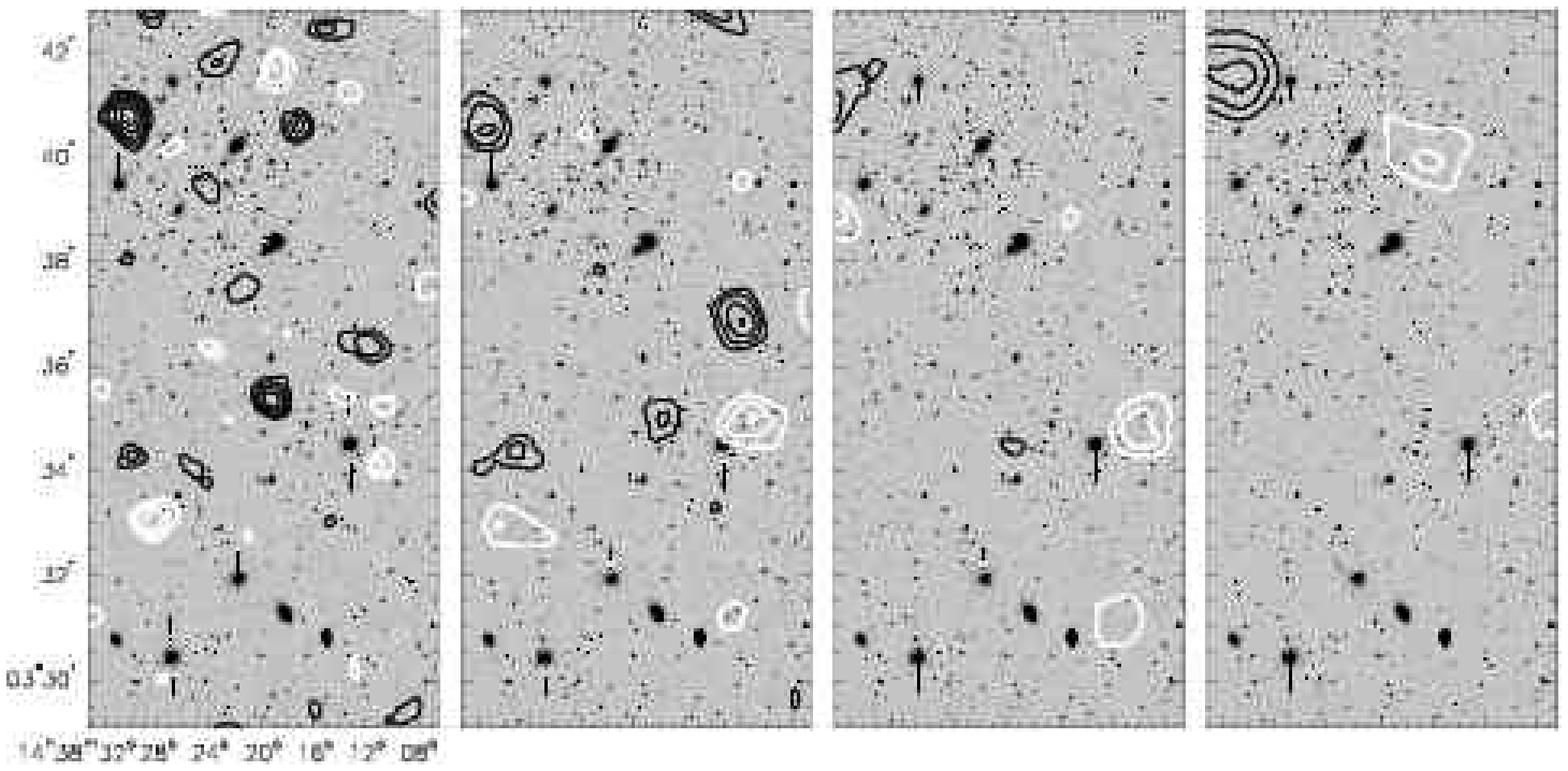}
%}
\caption{Results of a $\map^{\times}$ analysis of the $I$-band image. Shown
  are  
both positive (white) and negative (black) contours, starting at
$\pm2\sigma$ in 0.5$\sigma$ increments, for filter scales of
80$\arcsec$, 100$\arcsec$, 160$\arcsec$, and 200$\arcsec$ (left to right).}
\label{a1942I-cross}
\end{center}
\end{figure*}

The results are shown in Fig. \ref{a1942I-cross}. Particularly at
small filter scales, there are some positive and negative peaks 
with significances ${\scriptstyle \gtrsim} |3\sigma|$
(negative values are indicated by negative
signal-to-noise ratios). However, most of the peaks are at the edge of
the field, where a part of the weight function lies outside the
field. While (the real part of) $\map$ retains its justification
at these places as being simply a measure of the tangential alignment,
we can no longer assume that $\map^{\times}$ vanishes.

There is no peak with a significance larger than 2$\sigma$ in the
vicinity of the dark clump, so the detection passes this test well.

%%%%%%%%%%%%%%%%%%%%%%%%%%%%%%%%%%%%%%%%%%%
%                                         %
%              Radial Profile             %
%                                         %
%%%%%%%%%%%%%%%%%%%%%%%%%%%%%%%%%%%%%%%%%%%

\subsection{Radial profile}
\label{sect:ground-rad-profile}

\begin{figure}[bthp]
%\vspace{0.4cm}
\begin{center}
%\setlength{\fboxsep}{-\fboxrule}
%\fbox{
\includegraphics[bb=1cm 5.8cm 20cm 24.4cm,width=1\hsize]
{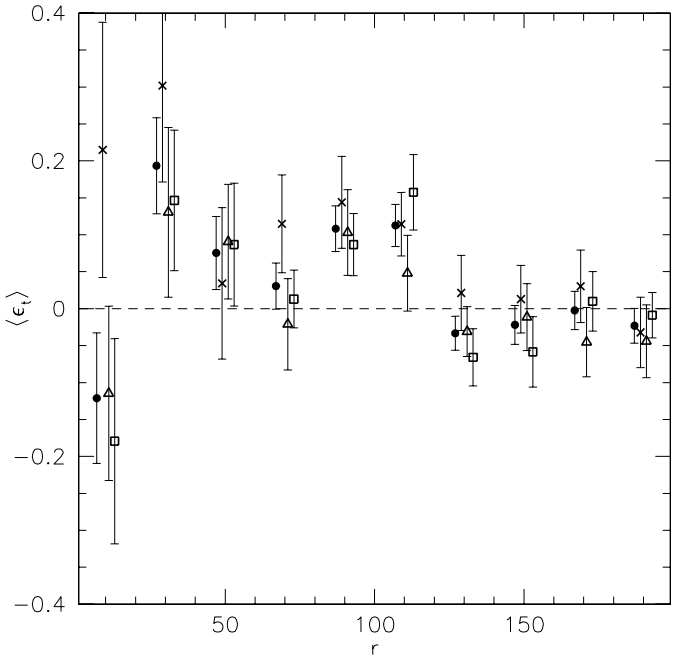}
%}
\caption{The mean tangential ellipticity $\ave{\bfmath{\varepsilon}_t}$
  relative to the  position of the Dark 
Clump as a function of distance from it (from the ground-based data). Each
point corresponds to the 
mean of a 20$\arcsec$ wide bin, where filled circles denote the complete
sample, 
crosses the faint galaxies, triangles the medium bright ones, and squares the
bright galaxies. The error bars represent the standard
deviation of $\ave{\bfmath{\varepsilon}_t}$ estimated from randomization of the
position angles of the galaxies. Note that the first bin is very sensitive
to the choice of centroid - its deviation from the shear profile in the
latter bins is thus not problematic.}
\label{a1942I-gamt}
\end{center}
\end{figure}

Despite several differences to the lensing analysis of \citet{ewm00}, the
$\map$ results agree at least qualitatively. We are curious whether we
can also 
reproduce the radial dependence of the mean tangential ellipticity shown in
Fig. 9 of \citet{ewm00}.

We determine the position of the dark clump from the $\map$ peak at the
120$\arcsec$ filter scale, where the signal is strongest. We find it to be 
$\alpha = 14^{\rm h}\,38^{\rm m}\, 21.6^{\rm s},\: \delta = 3^{\circ}\,
32\arcmin\, 43\farcs$ , which is at a distance of $18\farcs6$ from the
position given by \citet{ewm00}, and thus just at the 1$\sigma$ level they
give for the uncertainty of the centroid's position.

The tangential ellipticity relative to this position is calculated for each
galaxy within 200$\arcsec$. They are then binned according to their distance
from the dark clump and the weighted mean is calculated for each bin. To
estimate the standard deviation, we randomize the position angles of
these galaxies 1000 times and calculate the mean tangential
ellipticity each time, thus gaining an estimate for the standard deviation.
This analysis is done for the complete
galaxy sample as well as the three samples split according to brightness. 
The results of it are shown in Fig. \ref{a1942I-gamt}.

Particularly for the faintest galaxies, we find positive values out to
120$\arcsec$. This agrees well with the strong shear signal seen for these
galaxies. We can also identify the cause of the signal seen for bright
galaxies at the 120$\arcsec$ filter scale: the two significantly positive
bins at 90$\arcsec$ and 110$\arcsec$ (the filter function employed
assigns the highest weight around a radius $\theta_{\rm out}/\sqrt{2}$). 
For the medium bright galaxies $\ave{\bfmath{\varepsilon}_t}$ is largely consistent
with zero.\\

Compared to \citet{ewm00}, who measure $\ave{\bfmath{\varepsilon}_t} \approx
0.06$ at 100$\arcsec$, we find a higher value ($\approx 0.1$). On the other
hand, we find positive values only out to 
120$\arcsec$ rather than 150$\arcsec$. And since the centroid positions do
not coincide, the inner two bins are not comparable. Yet, we can also be
confident that the signal is not caused just by a few galaxies.

\begin{table*}[bthp]
\begin{center}
\caption{Overview of the $\map$ values and signal-to-noise ratios of
  the peaks found in the analyses of the CFHT data. We also give
  the number of galaxies located 
  in the aperture ($N$), and the (weighted) average galaxy ellipticity dispersion
  $\sigma_{\varepsilon}$, as well as the
  offset to the assumed centroid
  position, measured in the
  sky coordinate system. In two cases, there was no $\map$ peak in the
  vicinity of the dark clump - we then quote the values at the reference
  position (in italics).}
\vspace{2mm}
\renewcommand{\arraystretch}{1.2}
\begin{tabular}{|l|r||c|c||c|c||c|c|}
\hline
& $\theta_{\rm out}$ & N & $\sigma_{\varepsilon}$ & $\Delta \alpha [\arcsec]$ & $\Delta \delta
[\arcsec]$ & $\map$ & SNR \\
\hline
\hline
$|\bfmath{\varepsilon}| < 0.8$, & 80$\arcsec$ & 113 & 0.34 & 28 & 16 & 0.096 & 3.7 \\ 
weighted,          & 120$\arcsec$ & 268 & 0.34 & 0   & 0  & 0.078 & 5.1 \\
all galaxies       & 160$\arcsec$ & 468 & 0.35 & $-16$ & $-2$  & 0.046 & 3.7 \\
                   & 200$\arcsec$ & 648 & 0.34 & 58  & $-12$ & 0.034 & 3.5 \\
\hline
$|\bfmath{\varepsilon}| < 0.8$, & 80$\arcsec$ & 54  & 0.41 & 22 & 26  & 0.120 & 2.7 \\ 
weighted,          & 120$\arcsec$ & 110 & 0.39 & 4  & 4   & 0.109 & 4.1 \\
$m > 23.67$        & 160$\arcsec$ & 167 & 0.38 & 36 & 18  & 0.092 & 4.3 \\
                   & 200$\arcsec$ & 241 & 0.39 & 40 & $-22$ & 0.088 & 4.5 \\
\hline
$|\bfmath{\varepsilon}| < 0.8$,  & 80$\arcsec$ & 33 & 0.34 & 40 & 8 & 0.118 & 2.6 \\ 
weighted,           & 120$\arcsec$ & 86 & 0.35 & 6 & $-18$ & 0.058 & 2.0 \\
$22.54 < m < 23.67$ & 160$\arcsec$ & 159 & 0.36 & $-26$ & 2 & 0.031 & 1.5 \\
                    & 200$\arcsec$ & \textit{222} & \textit{0.34}& $-$ & $-$ & \textit{$-$0.003} & \textit{$-$0.1} \\
\hline
$|\bfmath{\varepsilon}| < 0.8$, & 80$\arcsec$ & 35 & 0.29 & 38 & 4 & 0.084 & 2.2 \\ 
weighted,          & 120$\arcsec$ & 74 & 0.29 & 16 & 0 & 0.075 & 3.5 \\
$m < 22.54$        & 160$\arcsec$ & 126 & 0.30 & $-14$ & $-20$ & 0.046 & 2.3 \\
                   & 200$\arcsec$ & 170 & 0.30 & $-48$ & 34 & 0.026 & 1.6 \\
\hline
%\hline
%$|\bfmath{\varepsilon}|<0.8$, & 80$\arcsec$ & 110 & 0.35 &  24 &   8 & 0.100 & 3.7 \\ 
%weighted,        & 120$\arcsec$ & 268 & 0.34 &   0 &   2 & 0.078 & 5.1 \\
%$P_g^s>0.2$      & 160$\arcsec$ & 466 & 0.35 & -14 &  -2 & 0.046 & 3.7 \\
%                 & 200$\arcsec$ & 648 & 0.34 &  58 & -12 & 0.034 & 3.5 \\
%\hline
%\hline
%$|\bfmath{\varepsilon}|<0.8$,&  80$\arcsec$ &  82 & 0.38 &  24 &   8 & 0.106 & 2.9 \\ 
%weighted,        & 120$\arcsec$ & 199 & 0.37 &   0 &   0 & 0.078 & 3.8 \\
%$P_g^s>0.2$      & 160$\arcsec$ & 344 & 0.37 & -18 &   2 & 0.048 & 3.1 \\
%$m>19.5$         & 200$\arcsec$ & 468 & 0.36 &  32 & -26 & 0.049 & 3.7 \\
%\hline
%%\hline
%%$|\bfmath{\varepsilon}| < 1$, & 80$\arcsec$ & && 34 & 16 & 0.116 & 3.2 \\ 
%%non-weighted, & 120$\arcsec$ & && 0 & 0 & 0.092 & 4.3 \\
%%$m > 20$ & 160$\arcsec$ & && -14 & 8 & 0.068 & 4.3 \\
%% & 200$\arcsec$ & && 26 & -28 & 0.060 & 4.3 \\
%%\hline
%\hline
%$|\bfmath{\varepsilon}| < 1$, & 80$\arcsec$  &  82 & 0.47 & 46 & 22 & 0.123 & 2.9 \\ 
%non-weighted, & 120$\arcsec$     & 190 & 0.42 & 4  & -2 & 0.080 & 3.4 \\
%$m > 19.5$ & 160$\arcsec$        & 318 & 0.41 & -14 & -22 & 0.054 & 3.0 \\
%$\nu_{\rm max}>7$ & 200$\arcsec$ & 438 & 0.41 & 36 & -26 & 0.057 & 3.7 \\
%\hline
%\hline
%\hline
%$V$-band & 80$\arcsec$           & 121 & 0.23 & -46 & 22 & 0.054 & 3.5 \\ 
%$|\bfmath{\varepsilon}|<0.8$, & 120$\arcsec$ & 256 & 0.22 & 8   & 32 & 0.038 & 3.6 \\
%weighted & 160$\arcsec$          & 455 & 0.22 & -14 & 38 & 0.034 & 4.4 \\
%$m > 19.5$ & 200$\arcsec$        & 588 & 0.22 & 22  & 14 & 0.028 & 4.1 \\
%\hline
\end{tabular}
\label{tab:map-results}
\end{center}
\end{table*}

\subsection{Summary}
We have successfully confirmed the weak lensing signal seen in two sets of
CFHT observations (our re-analysis of the $V$-band data are not shown here
but agree well with \citet{ewm00}). We show that the alignment signal comes
from faint galaxies, 
which supports the hypothesis that it is caused by a lensing mass at high
redshifts. One must keep in mind, however, that these are also those objects
most affected by noise.

With several variations of the catalog that enters the $\map$ analysis,
we tested that the detection of the dark clump is resistant against these and
consistently recovered at all filter scales. It
reaches a peak significance of about 5$\sigma$, although this signal is
contaminated by a tangential alignment of bright objects, which is unlikely
to be a lensing effect.

%%%%%%%%%%%%%%%%%%%%%%%%%%%%%%%%%%%%%%%%%%%%%%%%%%%%

\section{Analysis of the HST data}
\label{sc:hst}

The HST catalog extends to fainter and thus more distant galaxies than the
ground-based catalog. If the alignment
signal found in the ground-based data does indeed stem from a mass
concentration, its lens strength should increase with source
redshift. Additionally, since the HST is space-based and 
thus not afflicted by atmospheric seeing, its ellipticity measurements are
more reliable than those from ground-based telescopes. If the ground-based
signal is not just a noise peak, the signal should therefore be even
stronger in the HST images.

\subsection{Mass reconstruction}

\begin{figure}[tbp]
%\vspace{0.4cm}
\begin{center}
%\setlength{\fboxsep}{-\fboxrule}
%\fbox{
\includegraphics[width=1\hsize, trim=0 0.2cm 0 0]
{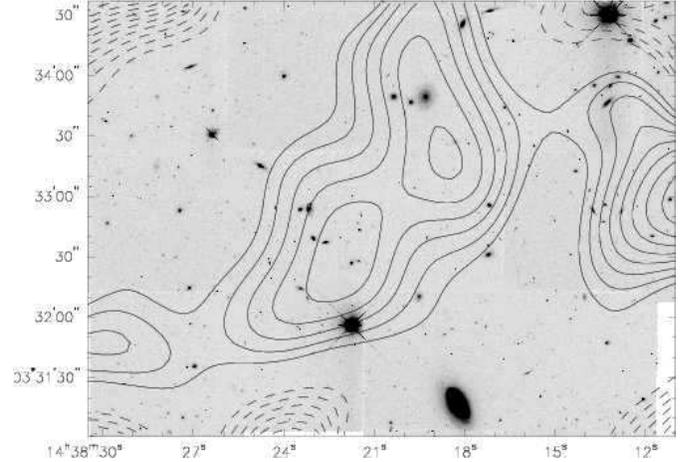}
%}
\caption{The mass density map of the inner region of the HST image
  reconstructed according to the method given in 
\citet{ses01}. Shown are $\kappa$ contours starting at $\pm0.02$ and
in-/decreasing in 0.01 decrements. The smoothing length is $30\arcsec$.}
\label{fig:hst-massreco}
\end{center}
\end{figure}

Fig. \ref{fig:hst-massreco} shows the results of a mass reconstruction of
the inner rectangle of the HST data field. The dark clump shows up
prominently, slightly westward of the position found in the ground-based
analyisis.

\subsection{$\map$ analysis}
\label{sc:hst-map}

\begin{figure*}[tbph]
%\vspace{0.4cm}
\begin{center}
%\sidecaption
%\setlength{\fboxsep}{-\fboxrule}
%\fbox{
\includegraphics[width=12cm]
{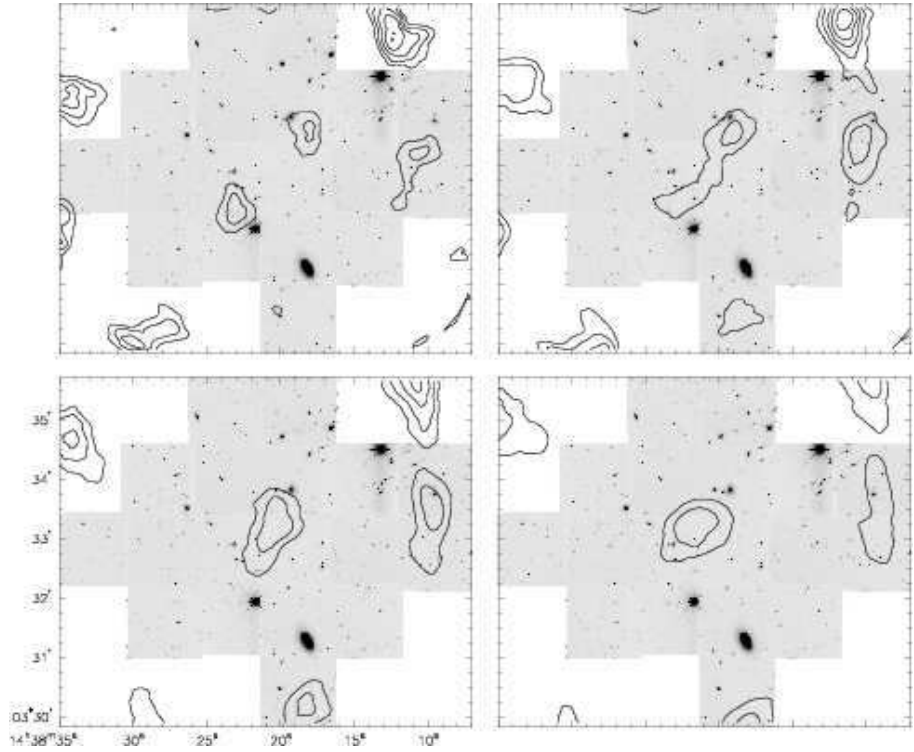}
%}
\caption{Results of the $\map$ analysis of the complete catalog from the HST
images. The filter scales are 80$\arcsec$, 100$\arcsec$, 120$\arcsec$,
and 140$\arcsec$ (from top left to lower right). 
The contours start at 1.5$\sigma$ and increase in 0.5$\sigma$ increments.
}
\label{hst-map}
\end{center}
\end{figure*}

The results of the weighted $\map$ analysis of the complete catalog is shown
in Fig. \ref{hst-map} and summarized in Table
\ref{tab:hst-map-results}. Indeed, we find a peak at approximately the 
same position as in the ground-based images in the 120$\arcsec$
filter, but with only 2.9$\sigma$ significance it is considerably weaker.
At the smallest filter scale used ($80\arcsec$), this peak diminishes and
dissolves into two peaks.

\begin{figure*}[tbph]
%\vspace{0.4cm}
\begin{center}
%\sidecaption
%\setlength{\fboxsep}{-\fboxrule}
%\fbox{
\includegraphics[width=12cm]
{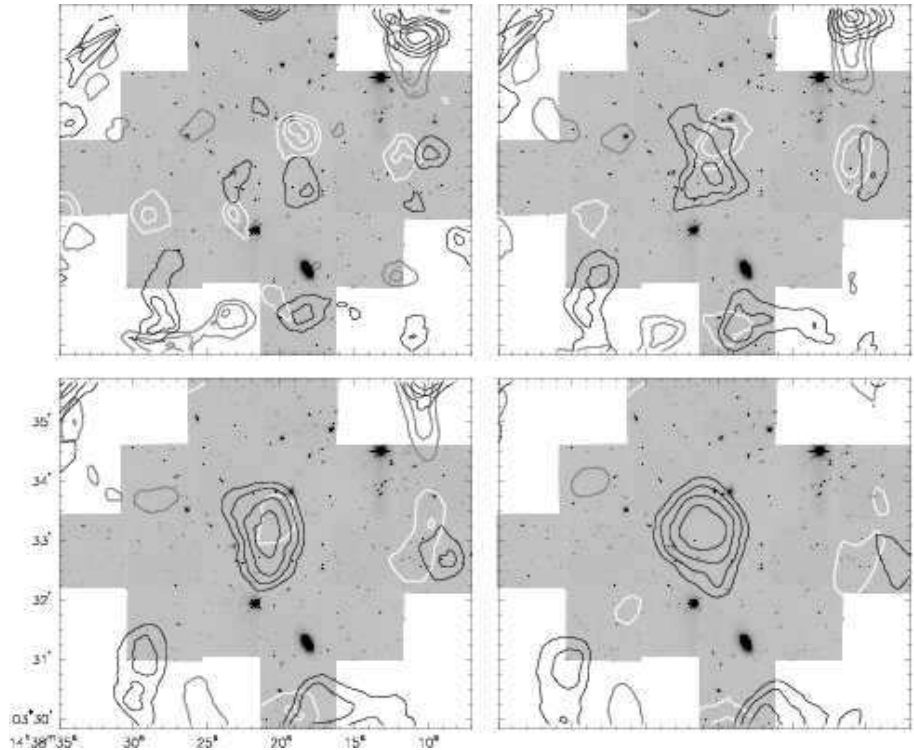}
%}
\caption{The same as the previous figure (Fig. \ref{hst-map}), but
  with the galaxies divided into three bins. White contours correspond to the
  brightest galaxies, gray contours to those of medium brightness, and black
  contours to the faintest. The contours start at 1.5$\sigma$ and increase in 0.5$\sigma$ increments.}
\label{hst-tomo}
\end{center}
\end{figure*}

Just as for the ground-based data, we split the catalog into three bins
according to brightness to probe the redshift dependence of the lens
strength. With cuts at $m=25.46$ and $m=26.43$, the three samples contain
approximately equal numbers of galaxies. The respective $\map$ analyses are
shown in Fig. \ref{hst-tomo}.
The alignment signal is carried by the galaxies in the bright and in the
faint bin, but there is a lack of signal in the
medium-bright bin. The galaxies that were detected in the ground-based
image should be mostly contained in the HST's bright bin. Although the
$\map$ values differ by a factor of about 2, the measurement in the
bright bin therefore confirms that there is tangential
alignment around the dark clump candidate. However, the
lack of alignment in the medium bin is difficult to explain with the lensing
hypothesis. 

\subsection{$\map$ cross component}

\begin{figure*}[tbph]
%\vspace{0.4cm}
\begin{center}
%\sidecaption
%\setlength{\fboxsep}{-\fboxrule}
%\fbox{
\includegraphics[width=12cm]
{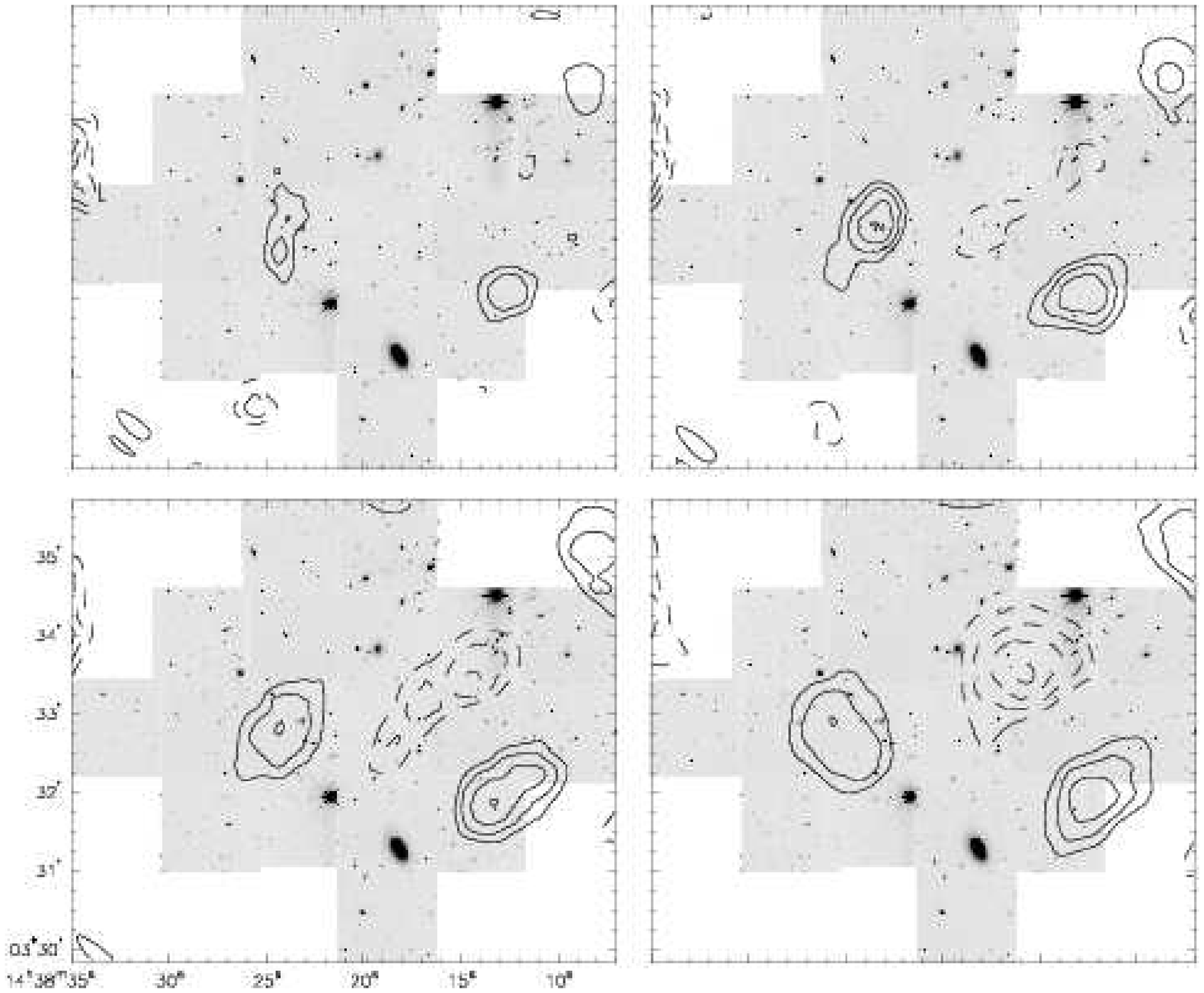}
%}
\caption{The $\map^{\times}$ analysis of the HST
  data. Shown are  
both positive (solid) and negative (dashed) contours, starting at
$\pm1.5\sigma$ in 0.5$\sigma$ increments, for filter scales of
80$\arcsec$, 100$\arcsec$, 120$\arcsec$, and 140$\arcsec$ (upper left to
lower right).}
\label{hst-cross}
\end{center}
\end{figure*}
\begin{figure*}[p]
%\vspace{0.4cm}
\begin{center}
%\sidecaption
%\setlength{\fboxsep}{-\fboxrule}
%\fbox{
\includegraphics[width=12cm]
{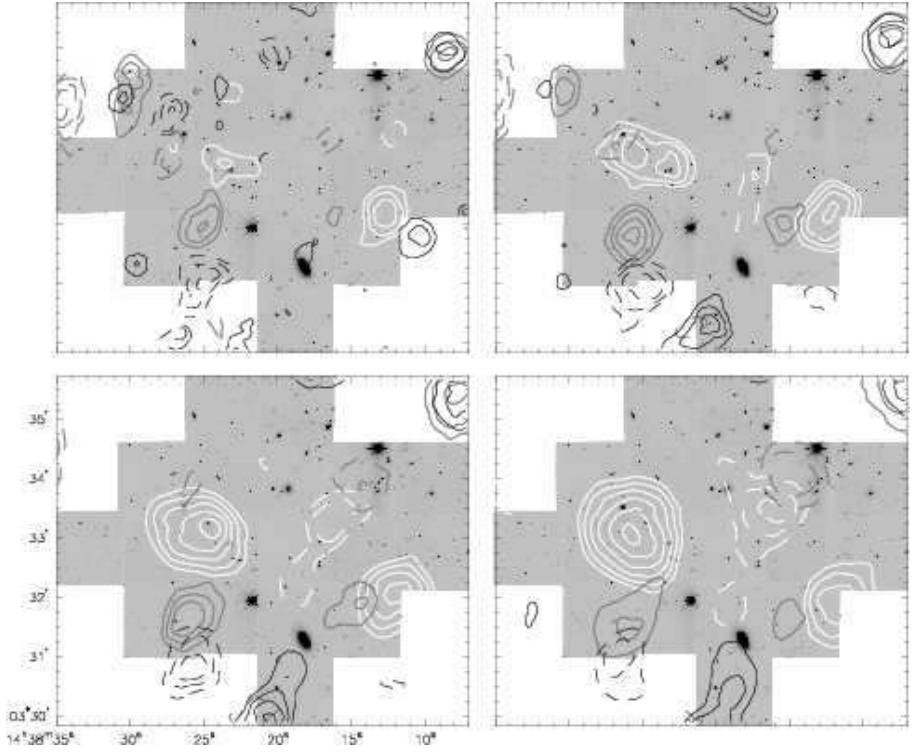}
%}
\caption{The same as Fig. \ref{hst-cross}, but with the galaxies split into
  three samples according to brightness. The color coding is identical to
  that of Fig. \ref{hst-tomo}. The contours start at $\pm1.5\sigma$
  and in-/decrease in 0.5$\sigma$ increments.
%{\it what is cross-tomo.ps? why the difference?}
}
\label{hst-tomo-cross}
\end{center}
\end{figure*}

\begin{table*}[btph]
\begin{center}
\caption{Overview of the $\map$ values and signal-to-noise ratios of
  the peaks found in the analyses of the HST data.
  We also give the number of galaxies located
  in the aperture ($N$), and the (weighted) average galaxy ellipticity
  $\sigma_{\varepsilon}$, as well as the
  offset to the assumed centroid
  position, measured in the
  sky coordinate system. For the small filter scales ($60 \arcsec$ and $80
  \arcsec$), there were often more than one
  peak in the vicinity of the dark clump position. For the medium bright
  bin, there is no peak close to the dark clump in the largest two filter
  scales  - we then quote the values at the reference
  position (in italics).}
\vspace{2mm}
\renewcommand{\arraystretch}{1.2}
\begin{tabular}{|l|r||c|c||c|c||c|c|}
\hline
& $\theta_{\rm out}$ & N & $\sigma_{\varepsilon}$ & $\Delta \alpha [\arcsec]$ & $\Delta \delta
[\arcsec]$ & $\map$ & SNR \\
\hline
\hline
$|\bfmath{\varepsilon}| < 0.8$, &  80$\arcsec$ & 352 & 0.39 & 34 & 32 & 0.037 & 2.2 \\
$P^{\rm g}_s > 0.3$,&              & 410 & 0.38 &  $-42$ &  $-52$ & 0.033 & 2.3 \\
weighted,           & 100$\arcsec$ & 560 & 0.39 &  20 &  26 & 0.030 & 2.2 \\
all galaxies        & 120$\arcsec$ & 907 & 0.39 &  0 &  0 & 0.025 & 2.4 \\
                    & 140$\arcsec$ & 1188 & 0.38 & $-22$ &   6 & 0.022 & 2.5 \\
\hline
$|\bfmath{\varepsilon}| < 0.8$, &  80$\arcsec$ & 138 & 0.40 &  32 & $-32$ & 0.060 & 2.1 \\
$P^{\rm g}_s > 0.3$,&   100$\arcsec$ & 212 & 0.41 &  0 &  $-18$ & 0.060 & 2.7 \\ 
weighted,           & 120$\arcsec$ & 314 & 0.42 &  $-12$ &  $-10$ & 0.063 & 3.3 \\
$m > 26.5$         & 140$\arcsec$ & 392 & 0.42 &  $-12$ &  12 & 0.060 & 3.5 \\
\hline
$|\bfmath{\varepsilon}| < 0.8$, & 80$\arcsec$ & \textit{126} & \textit{0.40} & $-$ & $-$ & \textit{$-$0.034} & \textit{$-$1.3} \\
$P^{\rm g}_s > 0.3$,& 100$\arcsec$ & \textit{204} & \textit{0.40} & $-$ & $-$ & \textit{$-$0.032} & \textit{$-$1.5} \\
weighted,           & 120$\arcsec$ & \textit{294} & \textit{0.39} & $-$ & $-$ & \textit{$-$0.014} & \textit{$-$0.7} \\
$25.5 < m < 26.5$ & 140$\arcsec$ & \textit{374} & \textit{0.40} & $-$ & $-$ & \textit{$-$0.008} & \textit{$-$0.5} \\
\hline

$|\bfmath{\varepsilon}| < 0.8$, & 80$\arcsec$ & 134 & 0.36 &  18 &  34 & 0.068 & 2.8 \\
$P^{\rm g}_s > 0.3$,             & 100$\arcsec$ & 220 & 0.36 &   4 &  16 & 0.045 & 2.4 \\
weighted,                      &  120$\arcsec$ & 309 & 0.35 &   0 &   0 & 0.031 & 1.9 \\
$m < 25.5$ & 140$\arcsec$ & \textit{402} & \textit{0.35} & $-$ & $-$ & \textit{0.009} & \textit{0.7} \\
\hline

%$|\bfmath{\varepsilon}| < 0.8$, &  60$\arcsec$ &  &  &  &  &  &  \\
%                    &  80$\arcsec$ &  &  &  &  &  &  \\
%                    & 100$\arcsec$ &  &  &  &  &  &  \\
%                    & 120$\arcsec$ &  &  &  &  &  &  \\
%\hline
\end{tabular}
\label{tab:hst-map-results}
\end{center}
\end{table*}

As described in Appendix \ref{app-aniso}, it is difficult to judge the quality
of the anisotropy 
correction of the HST data. A faulty anisotropy correction is likely to
cause a non-vanishing $\map^{\times}$ component. Also the CTE problem of the
camera might do 
so. We therefore calculate $\map$ with the shear cross component
instead of the tangential component; the result of this analysis is
shown in Fig. \ref{hst-cross}. 
 
Indeed, there is a 3$\sigma$ peak roughly coincident with the group of galaxies
close to the dark clump, and one close to the border of the image. For
the latter one, a large part of the aperture is evaluated outside of
the field, so that $\map^{\times}$ is not necessarily expected to vanish in
this case. This argument applies only weakly to the first one, which
is affected only by masks of stars.

The stellar anisotropy increases with distance from the center of the chips,
cf. Fig. \ref{anisocorr-chip3-glob}. At the edges of the chips we also
expect the largest deviations of the fit from the actual values. Therefore,
the quality of the anisotropy correction should decrease with increasing
distance from the chip center. To test this effect, we reject those objects
which are more than 
700 pixels from the center of their chip (343 objects) and redo the
$\map^{\times}$ analysis. The result differs only very little from the
previous one.

We also calculate the $\map$ cross component for the three brightness bins,
Fig. \ref{hst-tomo-cross}. We find that the $\map^{\times}$ peaks
are caused solely by the bright galaxies. 
If they were caused by an
insufficient anisotropy correction, we would expect these peaks to show up
for all three bins. Additionally, as the brightest objects are generally
also the largest, the effect of the anisotropy correction is smallest for
these. 

These results indicate that the $\map^{\times}$ peaks are unlikely to be
caused by a poor anisotropy correction. Rather, they seem to be caused by an
intrinsic alignment of some of the bright galaxies.

\subsection{Radial profile}

\begin{figure}[tbhp]
%\vspace{0.4cm}
\begin{center}
%\setlength{\fboxsep}{-\fboxrule}
%\fbox{
\includegraphics[bb=1cm 7cm 20cm 24.4cm,width=1\hsize]
{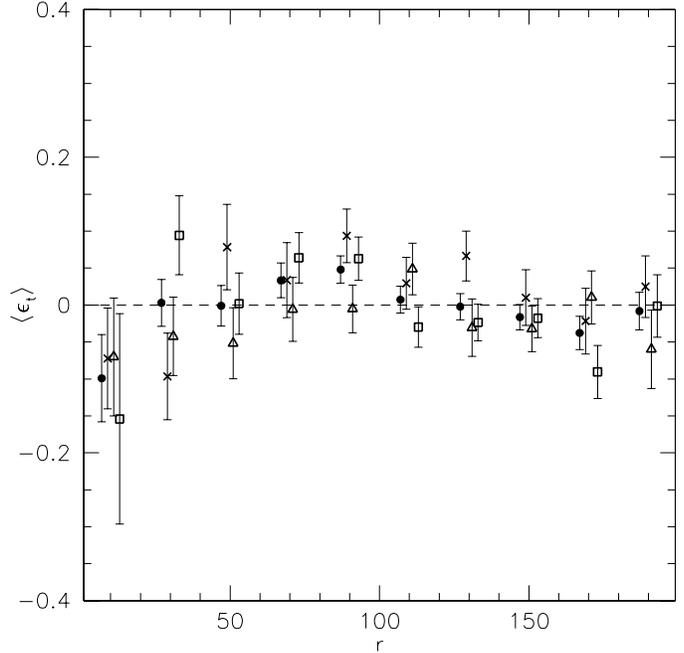}
%}
\end{center}
\caption{The mean tangential ellipticity $\ave{\varepsilon_{\rm t}}$ to the position
of the dark clump (determined from the $\map$ analysis of the HST
data) as a function of distance 
from it. Each point corresponds 
to the mean of a 20$\arcsec$ bin, where filled circles denote the complete sample,
crosses the faint galaxies, triangles the medium bright ones, and squares the
bright galaxies. The error bars represent the standard
deviation of $\ave{\varepsilon_{\rm t}}$ estimated from randomization of the
position angles of the galaxies.}
\label{hst-gamt}
\end{figure}

\begin{figure}[htbp]
%\vspace{0.4cm}
\begin{center}
%\setlength{\fboxsep}{-\fboxrule}
%\fbox{
\includegraphics[bb=1cm 6cm 20cm 24.4cm,width=1\hsize]
{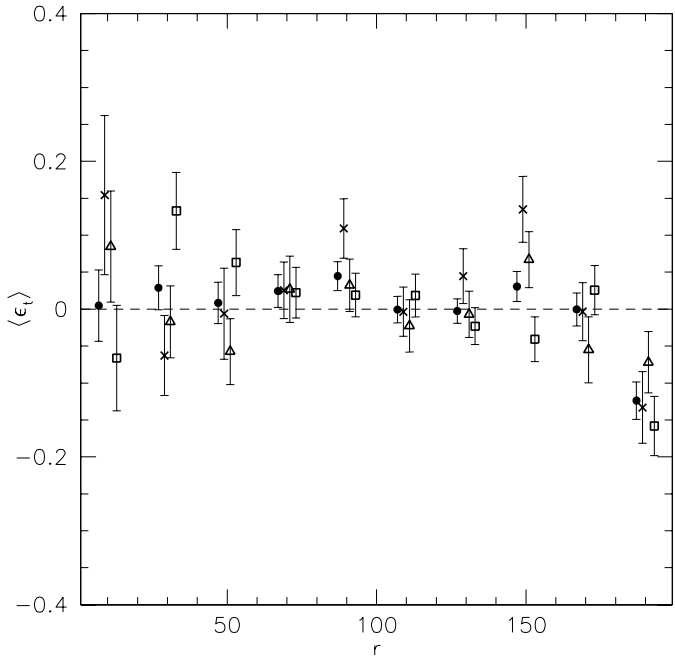}
%}
\caption{Same as Fig. \ref{hst-gamt}, but with the distance measured
  relative to the position of the dark clump (as measured in the ground-based
  data).}
\label{hst-I-gamt}
\end{center}
\end{figure}

Just as described in Sect \ref{sect:ground-rad-profile}, we calculate the
average tangential ellipticity $\ave{\varepsilon_t}$ as a function of the distance from the Dark
Clump position. The position of the strongest lensing signal is $\alpha =
14^{\rm h}\,38^{\rm m}\, 20.3^{\rm s},\: \delta = 3^{\circ}\,33\arcmin\,
10\arcsec$, as measured from the peak in the 120$\arcsec$ filter of the
complete sample.
This is $33\arcsecf3$ from the position we measured in the ground-based data
and $51\arcsecf0$ from the position originally cited by \citet{ewm00}.

The radial profile is shown in Fig. \ref{hst-gamt}.  For the complete
sample of galaxies, $\ave{\varepsilon_{\rm t}}$ is positive or very close to
zero between 20$\arcsec$ and 140$\arcsec$ radii. For this range, it is
inconsistent with zero only for
radii between 60$\arcsec$ and 100$\arcsec$.

%For bright galaxies, $\ave{g_{\rm t}} \approx 0.05$ between 20$\arcsec$ and
%100$\arcsec$. This is actually similar to the value quoted by \citet{ewm00},
%even though the center positions are so different. 
For bright galaxies, there is some excess $\ave{\varepsilon_{\rm t}}$
between 20$\arcsec$ and 100$\arcsec$, which is clearly the cause of
the peak we find.
For the medium bright
galaxies, $\ave{\varepsilon_{\rm t}}$ is largely consistent with zero within in the
$1\sigma$ error bars. However, the value is positive only in two bins. For
the faintest galaxies, $\ave{\varepsilon_{\rm t}}$ is positive over a fairly
large range of radii, but with varying significance and no clear
resemblance of a shear profile.\\

We also calculate the radial profile around the dark clump center we found
in the ground-based images; it is shown in Fig. \ref{hst-I-gamt}. For the
complete galaxy sample, it is largely consistent with zero but with a
trend to positive values. But for the bright galaxies, 
$\ave{\varepsilon_{\rm t}}$ follows a 
typical shear profile between 20$\arcsec$ and 100$\arcsec$, with
$\ave{\varepsilon_{\rm t}} \approx 0.025$ at $r \approx 100\arcsec$
\citet{ewm00} use the radial profile, namely $\ave{\varepsilon_{\rm t}} \approx
0.06$ at $r = 100\arcsec$, to deduce a mass estimate of the Dark
Clump. This further illustrates the difference of a factor of 2 between the
signal strengths in the ground-based and space-based image, which relates
directly to the mass estimate.

However, for the medium bright and faint galaxies, there is no obvious
NFW-like trend in the profile.

\subsection{Projected galaxy density}
\label{sc:gal-dens}

\begin{figure*}[htbp]
%\vspace{0.4cm}
\begin{center}
%\sidecaption
%\setlength{\fboxsep}{-\fboxrule}
%\fbox{
\includegraphics[width=12cm]
{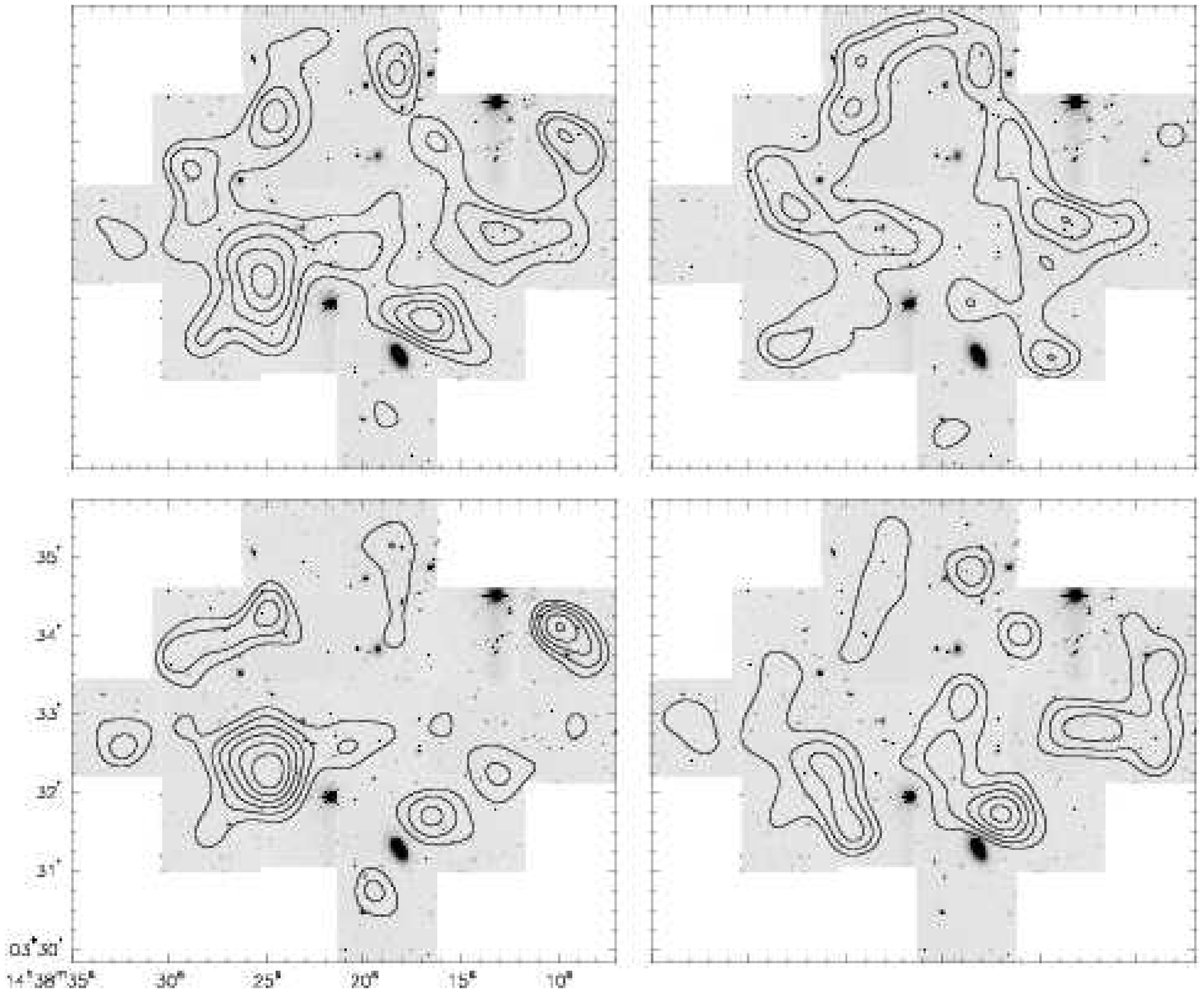}
%}
\caption{ Projected number density of galaxies in the HST image. The upper
  left image shows contours of surface densities, smoothed with a Gaussian
  of width $\sigma = 15\arcsec$, for all galaxies in the
  catalog, starting at 70~arcmin$^{-2}$, and increasing in 10~arcmin$^{-2}$
  intervals. For the other three plots, the galaxy catalog was split as
  before into magnitude bins. Bright galaxies are shown in the upper right
  panel, medium bright ones in the lower left, faint ones in the lower
  right. The contours for these panels start at 25~arcmin$^{-2}$, and
  increase in 5~arcmin$^{-2}$ steps.}
\label{fig:gal-dens}
\end{center}
\end{figure*}

Neither \citet[$I$-band]{ewm00} nor \citet[$H$-band]{gel01} find a
significant overdensity of galaxies which could be associated with the Dark
Clump. With the substantially deeper HST data, we can probe the 
galaxy number density to fainter magnitudes. In Fig. \ref{fig:gal-dens}, we present
surface number densities for the complete galaxy sample, as well as split up
according to brightness, as done before for the lensing analysis.

The bin of brightest HST galaxies corresponds roughly to those objects
detected also in the ground-based image. It is quite puzzling that rather
than an overdensity of galaxies at the dark clump, the galaxies seem to form
a ring around that position. However, this is hardly significant.

The most significant feature we find is a $5\sigma$ galaxy overdensity in
the bin of medium bright galaxies, located about $1\arcmin$ from the Dark
Clump. Given that these objects are fainter than the ones that carry the
lensing signal in the ground-based data, this is rather unlikely to be
associated with a lensing mass. However, it could explain the alignment
signal seen in the faintest HST bin. It would be highly interesting to
investigate whether this overdensity is present also in color space;
unfortunately, the currently available data sets are not deep enough.

\subsection{Summary}

We have shown that we also detect tangential alignment around the Dark
Clump in the HST data. However, it is considerably less significant
than in the ground-based data. It is particularly intriguing that we
do not detect alignment in the medium bright HST galaxies.

We refrain from trying to deduce a mass for the dark clump from these
measurements, as it is obvious that our results are not
unambiguous. As an upper limit, the HST data suggest that the mass estimate
of the dark clump has to be corrected at least by a factor of 1/2 compared
to the original value.

%One problematic issue of the HST images is the anisotropy correction, which
%had to be carried out with only a few stars. Yet, with the $\map^{\times}$
%analysis we have presented evidence that the correction is satisfactory.

A comparison on an object-to-object basis may be able to shed some light on
the question whether the ground-based lensing signal is only a noise peak or
whether there are other, uncorrected problems with the HST mosaic. Such a
comparison is carried out in App. \ref{sc:compare}.

\section{X-ray analysis}
\label{sc:chandra}

\begin{figure*}[bthp]
%\vspace{0.4cm}
\begin{center}
%\sidecaption
%\setlength{\fboxsep}{-\fboxrule}
%\fbox{
\includegraphics[width=12cm]
{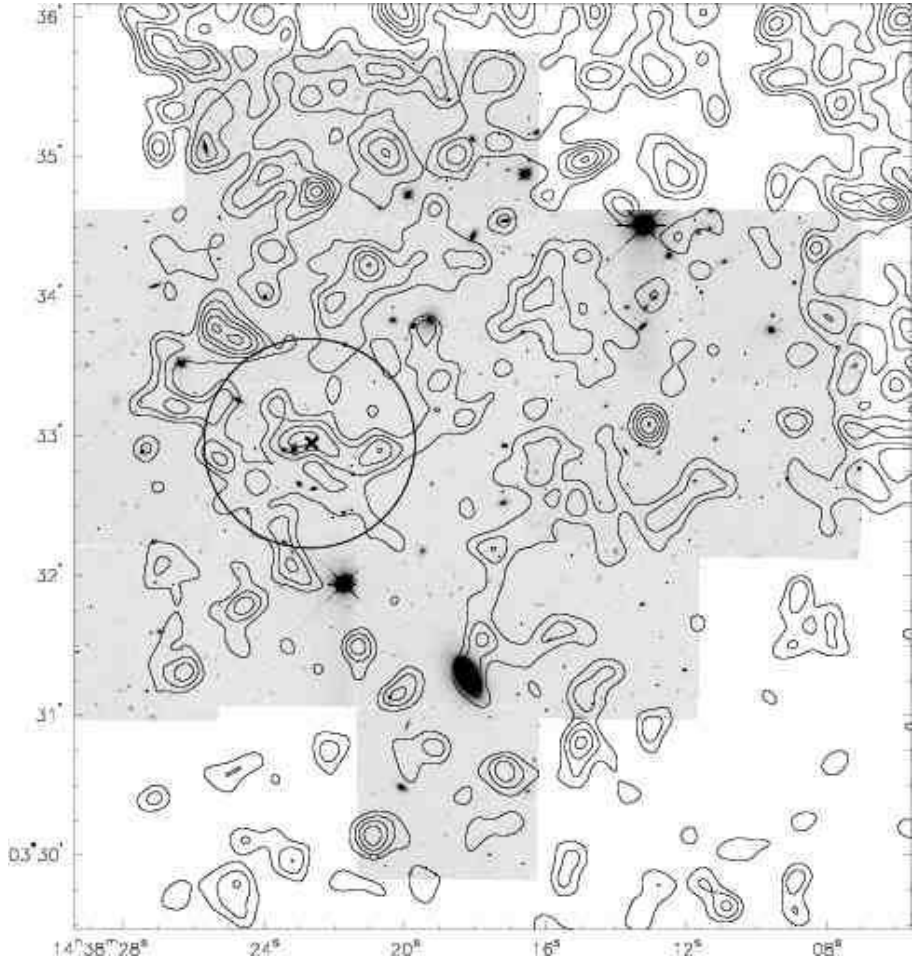}
%}
\caption{ A part of the Chandra image overlaid on the HST image. The point sources
were removed from the image, and it was smoothed with a Gaussian of
$5\arcsec$. The contour levels start at $3.6 \times
10^{-14}$~erg~cm$^{-2}$~s$^{-1}$ and increase in steps of  $0.8 \times
10^{-14}$~erg~cm$^{-2}$~s$^{-1}$. An increase in X-ray emission due to
Abell 1942 (north of 
the dark clump) is visible. In the vicinity of the dark clump, there is a
1.5$\sigma$ source at $\alpha = 14^{\rm h}\,38^{\rm m}\, 22.7^{\rm s},\:
\delta = 3^{\circ}\,32\arcmin\, 57\arcsec$ (marked with ${\mathbf \times}$) in a circular
aperture of radius $45\arcsec$ (indicated).
}
\label{fig:chandra}
\end{center}
\end{figure*}

If the dark clump is a virialized massive dark matter concentration,
one would expect that the baryons trapped by the potential well and
not in galaxies would have been shock-heated in the gravitational
collapse and therefore would emit in X-rays. \citet{ewm00} present the
analysis of the ROSAT HRI image of the A1942 field. They found a hardly
significant X-ray detection at a position $60\arcsec$ away from
the weak lensing mass peak position reported in that paper.

Since then, the field of A1942 was imaged by the Chandra X-ray
observatory in March 2002 (PI: Garmire). It was observed for
58 ks with the ACIS-I detector configuration. We retrieved the
observation from the archive and analysed it using standard techniques
with the CIAO v3.1 package.\footnote{Chandra Interactive Analysis of
Observations (CIAO), http://cxc.harvard.edu/ciao/} and recipes from
the ``ACIS Recipes: Clean the Data'' web
page.\footnote{http://www.astro.psu.edu/xray/acis/recipes/clean.html}
We use the standard set of event grades (0,2,3,4,6) and restrict our
analysis to the 0.5-8.0 keV energy band. We detect sources with
CIAO's WAVDETECT routine using wavelet scales 1, 2, 4, 8 and 16. Only
two sources are detected within $100\arcsec$ of the CFHT data lensing
$\map$ centroid. These two sources are found to be point sources with
clear optical counterparts in the optical data.

Visual inspection does not reveal any obvious extended source at the
position of the dark clump. Nevertheless, in order to check if there
really is any extended source, we created a diffuse emission image
from our original image excising the data where WAVDETECT detected
sources and filling the holes according to CIAO's threads. We also
created a ``blank-sky'' background image to better estimate the
background contribution at the position of the dark clump. We smoothed
the point-source-removed image with and without background subtraction
with various algorithms and scales but no obvious extended source was
detected. The most significant source consistent with the broad
position of the dark clump is a 1.5$\sigma$ source (in a circular
aperture of radius $45\arcsec$) whose coordinates are $\alpha = 14^{\rm
h}\,38^{\rm m}\, 22.7^{\rm s},\: \delta = 3^{\circ}\,32\arcmin\,
57\arcsec$. We measure a count-rate of $(5.9 \pm 4.0) \times 10^{-4}$
counts/s in the 0.5-8.0 keV energy range in a circular aperture of
radius $45\arcsec$ which corresponds to a unabsorbed flux of $(5.3 \pm 3.4)
\times 10^{-15}$ erg cm$^{-2}$ s$^{-1}$ assuming an incident spectrum
of $T=3$keV and a local hydrogen column density of $N_H = 2.61 \times
10^{20}$ cm$^{-2}$. We also measure the count rate in various
apertures centered at the positions of the CFHT and HST data lensing
centroids. The measurements are of lower significance compared to the
previous one. We have fitted a standard beta profile \citep{caf78} to the
azimuthally average radial profile 
centered at the position of this potential source. We obtain best
values for the core radius and beta parameter (slope decline at large
radius) of $5\arcsec$ and 0.55, respectively. These values indicate that if
there really is an extended source, it is rather compact with a small
core radius. Although, one has to keep in mind that these best fit
parameters are highly uncertain due to the low number of counts.
The total count rate predicted integrating the best-fit model would be
$9.4 \times 10^{-4}$ counts/s.

\citet{ewm00} reported a hardly significant ($3.2\sigma$) detection
of X-ray emission in the dark clump area in the ROSAT HRI image of
this field. We have measured the flux in the Chandra image at the same
position, which is very close ($15\arcsec$ away) to the previously reported
source. We find an unabsorbed flux of $(6.8 \pm 3.6) \times 10^{-15}$
erg~cm$^{-2}$~s$^{-1}$ in a $45\arcsec$ aperture. This value is slightly
higher than the previous one only due to the inclusion in the aperture
of another faint source situated at the NE and not included in the
mesurement at the previous position. Therefore, we do not confirm the flux
measurement of the ROSAT HRI source.

Given the faintness and measured uncertainties of our possible
detection, there is little point in speculating about the luminosity
and gas mass of our possible detection. Even if this source was real
and its real flux was the highest allowed by the data, it would not
have enough mass to be considered a rich cluster by its X-ray
properties.  Overall, the Chandra image of this source indicates that
if there is a dark matter concentration producing the lensing signal, this
concentration does not contain the expected 
mass of hot gas that would be expected for its lensing signal if it
were similar to the clusters of galaxies we have observed so far.

\section{Conclusion}

Tangential alignment around the dark clump was detected in three datasets,
which differ in the filter, camera, and telescope used. It can therefore be
ruled out that the alignment is caused by instrument-specific systematics. 
However, the significance of the detection is lowest in the space-based
dataset. If the alignment were due to lensing by a matter concentration, the
highest 
alignment signal would be expected in the HST data, as it extends to
more distant galaxies which should be more affected by the distortions due to
lensing.

The significant alignment signal in the ground-based data is caused mainly
by faint galaxies, for which the individual shape measurement are uncertain
due to background noise. We show in App. \ref{sc:compare} that the shape
measurements agree very well on average when comparing ground-based and
space-based data. Thus, also the amplitude of an alignment signal should be
comparable. Considering the high alignment signal in the two ground-based
datasets, it seems unlikely that background noise boosted the 
signal in both cases. Yet, for the $I$-band image, there is some indication
of such a ``conspiracy'', as the average tangential ellipticity toward
the dark clump of the objects compared is significantly larger for the
ground-based measurements (see \ref{sc:compare-gt}).

There are several issues involved which actually make weak lensing analyses
more difficult to apply to HST data (small field-of-view, complicated PSF
structure). But various tests (App. \ref{sc:hst-systematics} and
\ref{sc:compare-gt})  have not revealed any bias these
problems might cause to the ellipticity measurements. The weaker alignment
signal in the HST data is therefore not a result of these systematics.

For the original detection claim, further evidence for a lensing mass was
lent by the detection of X-ray emission in a ROSAT HRI image. However,
follow-up observations with Chandra measure only a tenth of the flux
originally measured in the ROSAT data, making this more likely to be a
spurious detection. Thus, if it is a lensing mass, the dark clump would be
truly ``dark'', not just in the optical.

We cannot give a definite answer to the question of the nature of the Dark
Clump candidate found by \citet{ewm00}. It has neither been possible to
prove that it is a Dark Matter halo, nor that it is a statistical fluke,
caused by noise in either the intrinsic galaxy ellipticities or the measurement
process. Our analyses 
assert that there is indeed tangential alignment around the
dark clump. However, it remains unclear whether this is caused by a lensing
mass, is just a chance alignment, or a combination of the two.

The lack of signal in the medium magnitude bin of the HST data and the lack
of X-ray emission yield the first hypothesis more unlikely (at least
compared to \citet{ewm00}), but not impossible if one is willing to accept
the existence of very dark, massive halos.

\begin{acknowledgements}
We are very grateful to Ludovic van Waerbeke for many helpful discussions
and suggestions, and to J\"org Dietrich and Tim Schrabback for providing help
and new ideas at various points of this work.
We thank Yannick Mellier for his collaboration on this project, and
Richard Ellis and Meghan Gray  for their
support of the HST follow-up observations.

This work was supported
by the German Ministry for Science and Education (BMBF) through DESY
under the project 05AE2PDA/8,
and by the Deutsche Forschungsgemeinschaft under the project
SCHN 342/3--1.
\end{acknowledgements}

\appendix

\section{Testing for systematics in the HST data}
\label{sc:hst-systematics}

Despite the lack of seeing due to the Earth's atmosphere, weak lensing analyses
of HST data are not applied straightforwardly. Due to the small field-of-view,
there are only few stars available for the anisotropy correction; the pixels
of the WFPC2 camera undersample the PSF, and the camera has a notable charge
transfer inefficiency. In the following, we investigate the effects and
possible bias from these problems.

\subsection{Anisotropy correction}
\label{app-aniso}

\begin{figure*}[btp]
%\vspace{0.4cm}
\begin{center}
%\setlength{\fboxsep}{-\fboxrule}
%\fbox{
\includegraphics*[bb=1.5cm 5cm 19.2cm 23.5cm,width=0.44\hsize]
{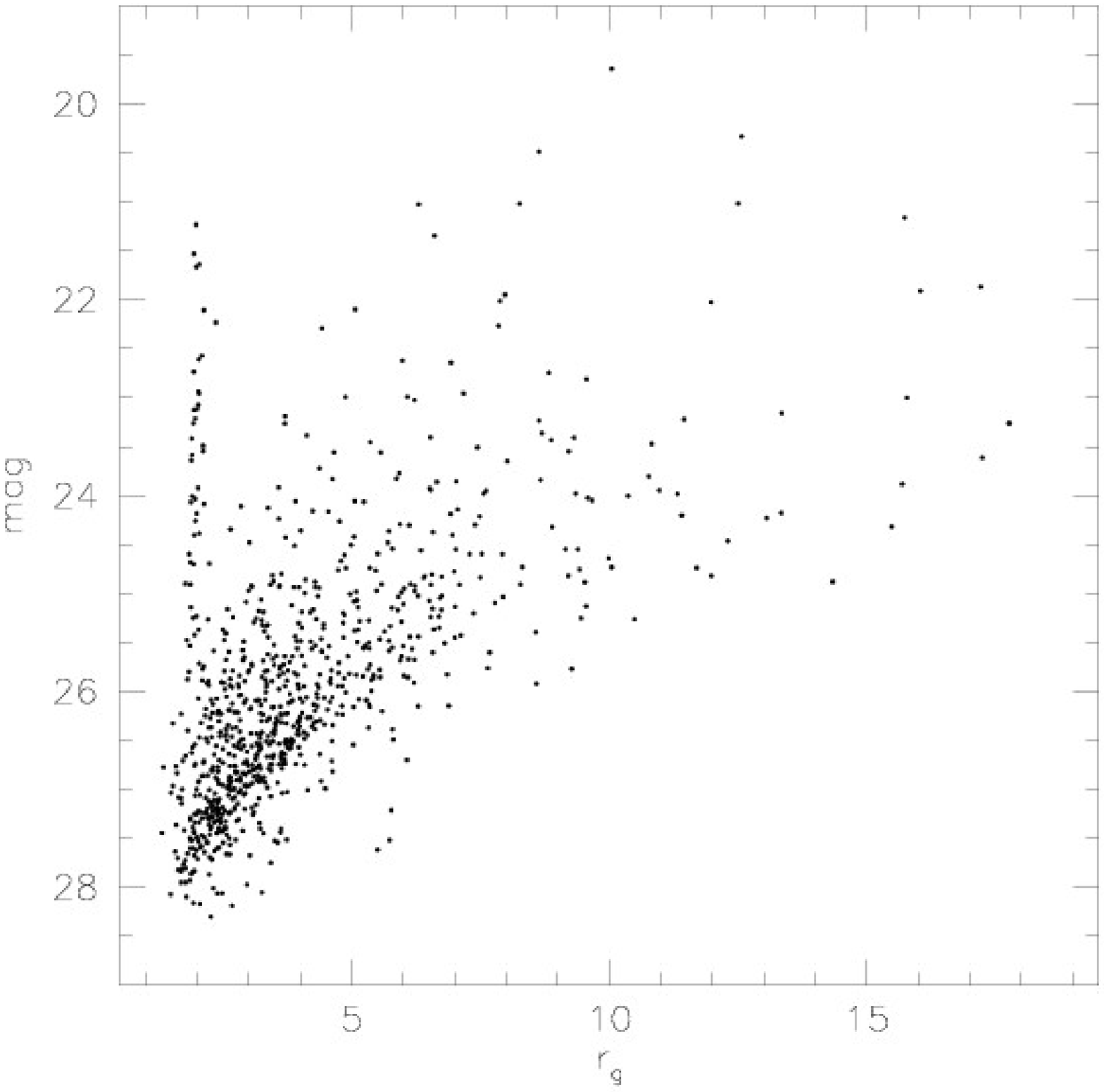}
%}
%\setlength{\fboxsep}{-\fboxrule}
%\fbox{
\includegraphics[bb=0.3cm 6.5cm 20cm 24.5cm,width=0.54\hsize]
{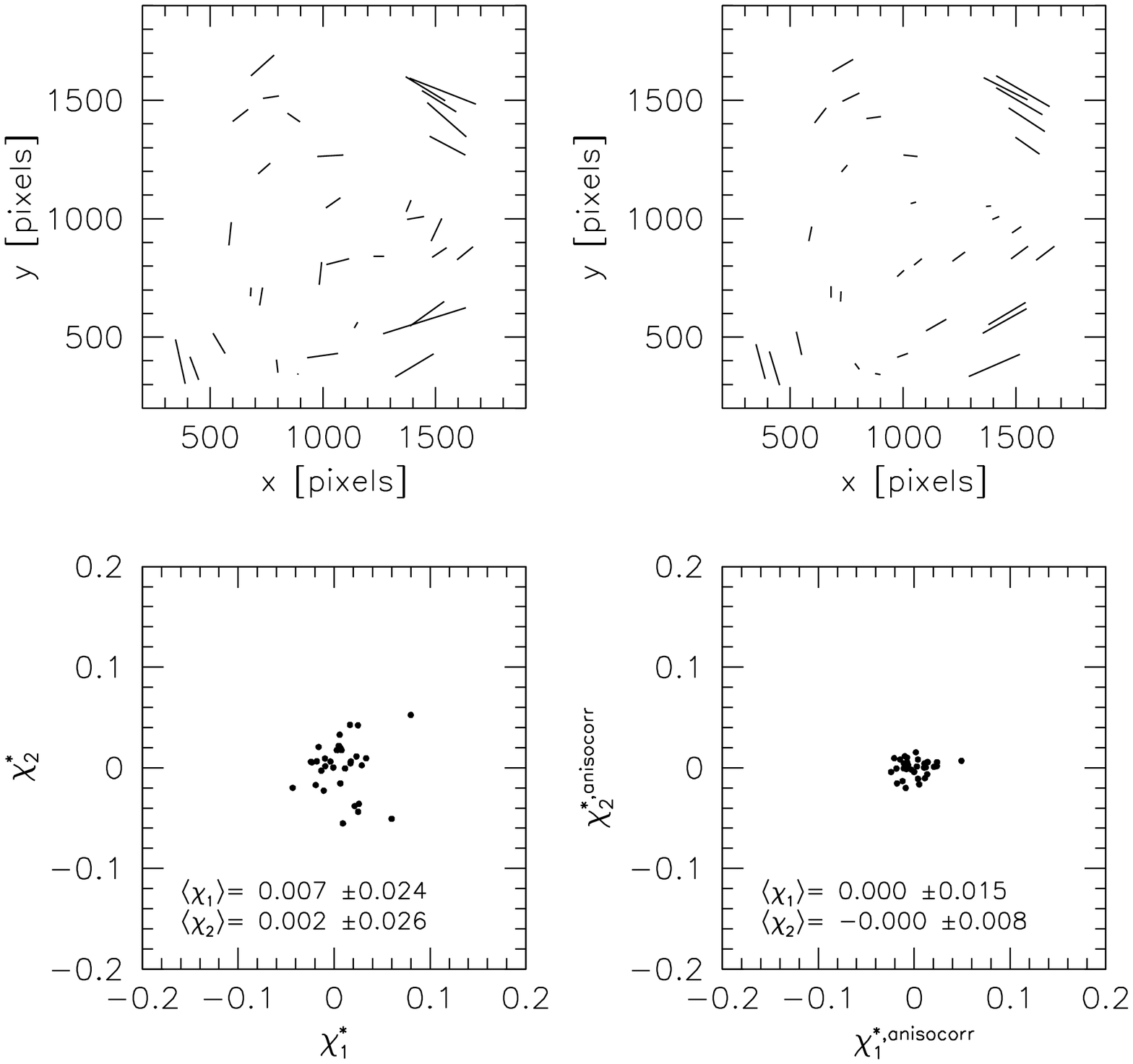}
%}
\end{center}
\caption{The anisotropy correction for Chip 2. On the left, the magnitudes
  of the 850 objects left in the catalog at this stage are plotted vs. their
  half-light radii. 33 objects from the stellar sequence were used to
  determine the stellar anisotropy kernel. The orientation and
  modulus (multiplied with a factor of 2000~[{\tt pixels}]) of $\bfmath{\chi}^*$
  are shown in the middle top box. The fitted
  polynomial, evaluated at the position of these stars, is shown in the top
  right box. In the middle lower box, the ellipticity components of these
  stars before the anisotropy correction are shown; in the lower right box
  they are shown after correction.}
\label{anisocorr-hst-chip2}
\end{figure*}

\begin{figure*}[tbp]
%\vspace{0.4cm}
\begin{center}
%\setlength{\fboxsep}{-\fboxrule}
%\fbox{
\includegraphics*[bb=1.5cm 5cm 19.2cm 23.5cm,width=0.44\hsize]
{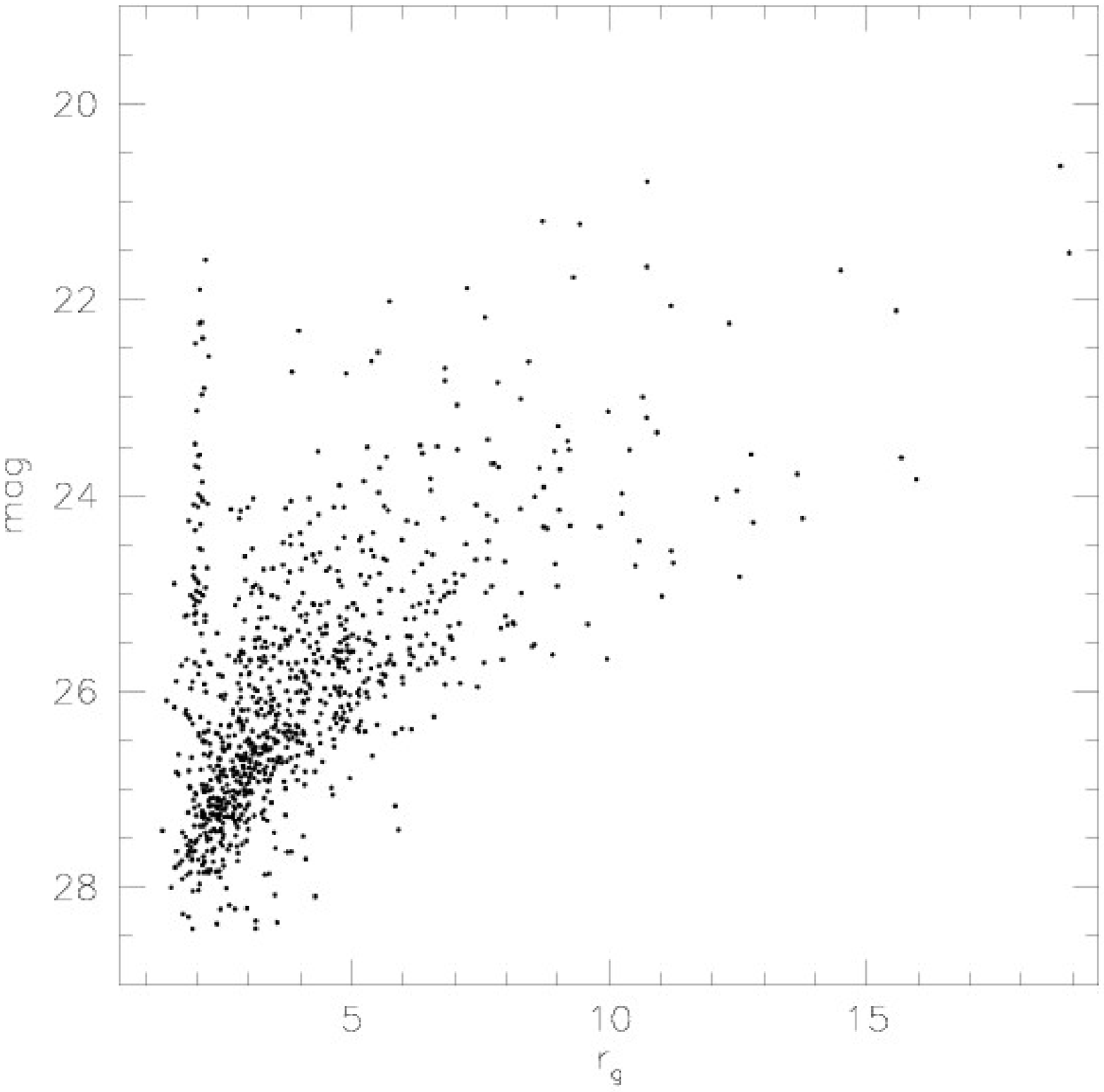}
%}
%\setlength{\fboxsep}{-\fboxrule}
%\fbox{
\includegraphics[bb=0.3cm 6.5cm 20cm 24.5cm,width=0.54\hsize]
{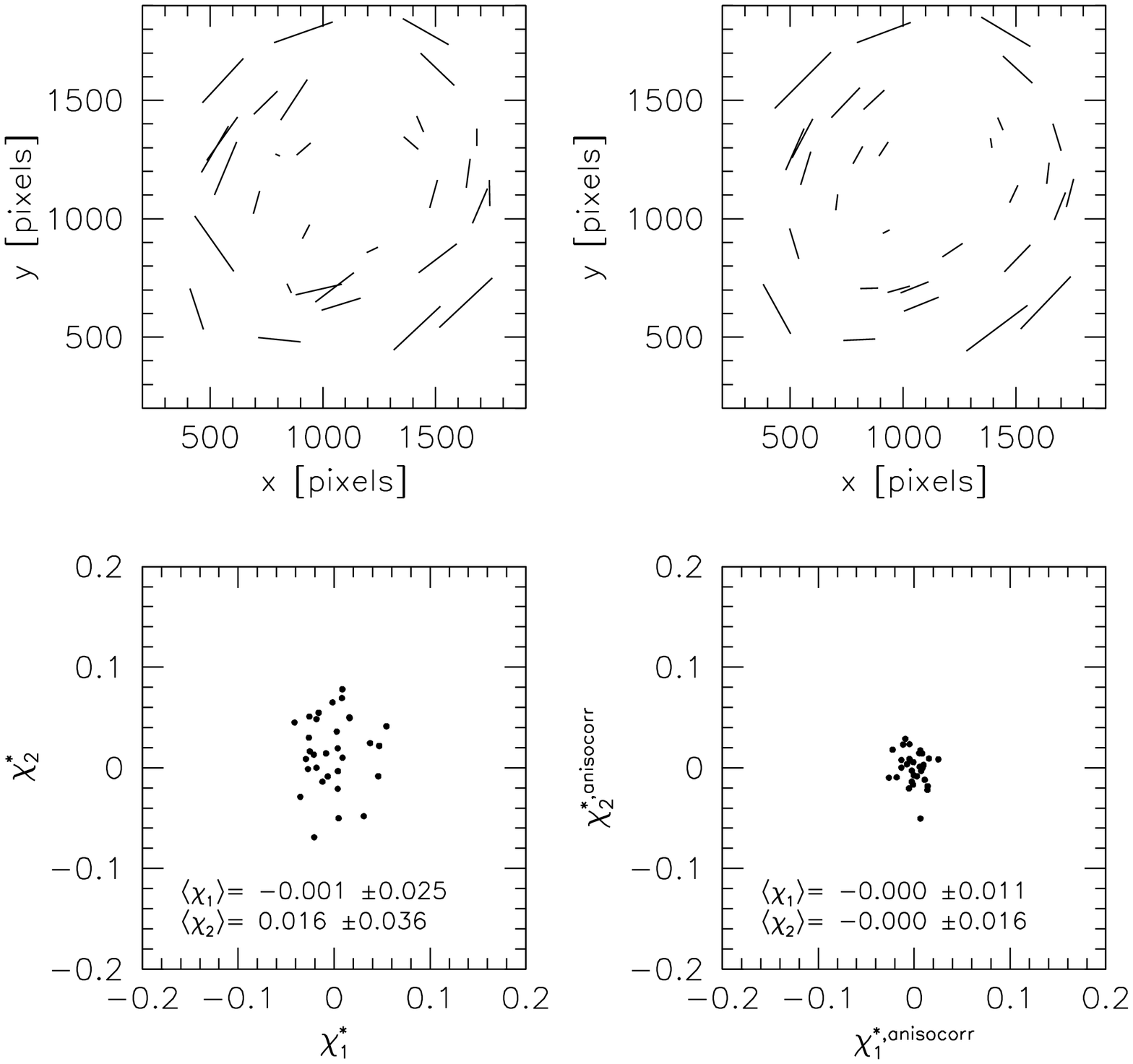}
%}
\end{center}
\caption{Same as Fig. \ref{anisocorr-hst-chip2}, but for Chip 3. 31 objects
  out of 930 were used for the fit.}
\label{anisocorr-hst-chip3}
\end{figure*}

\begin{figure*}[btp]
%\vspace{-0.3cm}
\begin{center}
%\setlength{\fboxsep}{-\fboxrule}
%\fbox{
\includegraphics*[bb=1.5cm 5cm 19.2cm 23.5cm,width=0.44\hsize]
{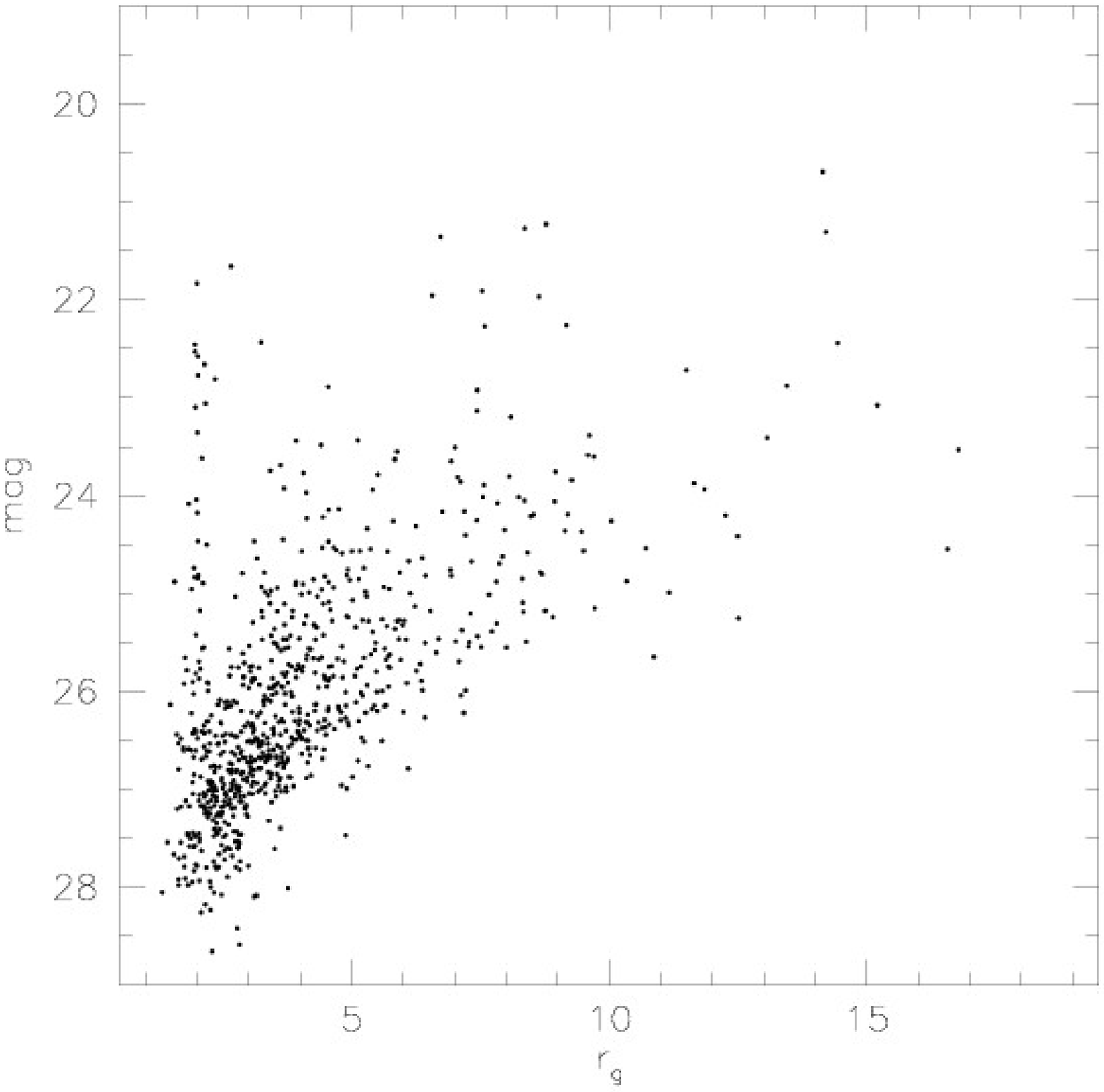}
%}
%\setlength{\fboxsep}{-\fboxrule}
%\fbox{
\includegraphics[bb=0.3cm 6.5cm 20cm 24.5cm,width=0.54\hsize]
{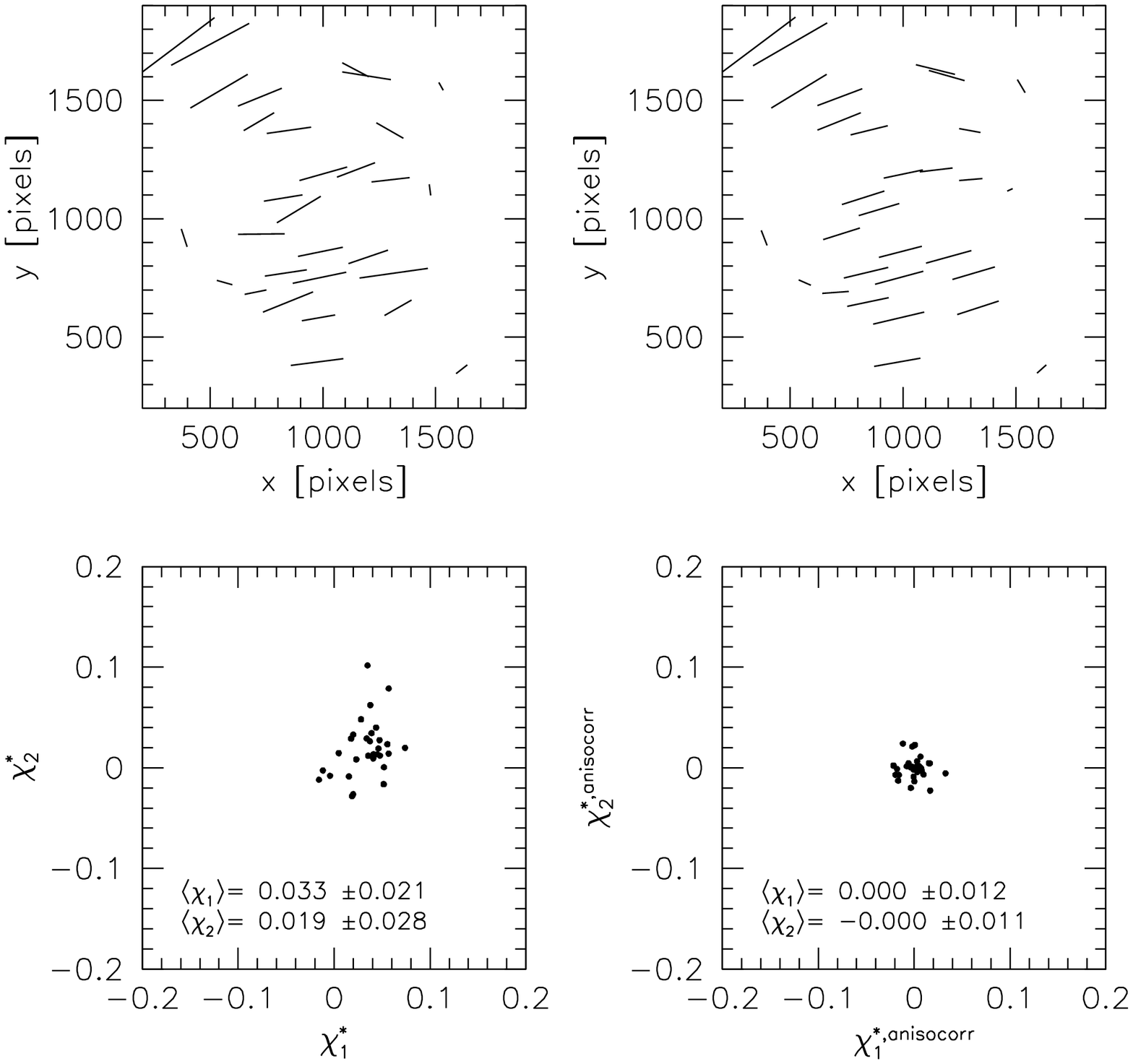}
%}
\end{center}
\caption{Same as Fig. \ref{anisocorr-hst-chip2}, but for Chip 4. 30 objects
  out of 770 were used for the fit.}
\label{anisocorr-hst-chip4}
\end{figure*}

\citet{hfk98} showed that a better anisotropy correction 
(eq. (\ref{eq:aniso})) of WFPC2 images can
be achieved if the weight function applied to the stellar images is adapted
to the size of the object to be corrected.

However, in practice, the ellipticity measurements become increasingly noisy
for larger radii of the weight function. Along with the low number of stars,
this causes fits to the anisotropy kernel to be both noisy and ill-described
by a simple polynomial.

We tested three possible methods of correcting for the anisotropy as
measured from the stars:
\begin{enumerate}
\item the anisotropy kernel $\bfmath{q}^{\star}$ fit to measurements with a
  weight function of stellar size;
\item the anisotropy kernel $\bfmath{q}^{\star}(r)$ fit to measurements with a
  weight function of the objects' size $r$;
\item a factor $a$ is applied to $\bfmath{q}^{\star}$ to minimize the
  difference to the stellar ellipticities measured at the objects's size.
\end{enumerate}
The third method acknowledges that the anisotropy kernel is a function of
object size, but avoids the uncertainties from fitting a polynomial to the
noisy measurements with a larger weight function. However, it can only
compensate for variations in the PSF that are uniform across the chip and of
the same magnitude in both ellipticity components.

The anisotropy patterns produced by these methods are indeed notably
different. But the resulting shear estimates $\varepsilon$ differ only on a
percent level. Accordingly, the $\map$ measurements change only minutely
with a different anisotropy correction. We used the third method as a
compromise solution.

The fits to $\bfmath{q}^{\star}$ are illustrated in
Figs. \ref{anisocorr-hst-chip2} - \ref{anisocorr-hst-chip4}.
For each chip, about 30 stars are available for the anisotropy
correction. This is a very small number, and as can be seen from the
Figures, they are not evenly distributed over the chip. It is therefore
difficult to judge the quality of the anisotropy correction.\\

\textbf{Alternatives?} It is common to use globular cluster fields for
the anisotropy correction rather than the few stars contained in the
field itself. With the large number of stars, the anisotropy can be
very well determined. However, such fields have to be chosen
carefully: they should have been taken temporally close to our field,
have a similar dither pattern, and should have been taken in the same
filter. As it turned out, no images of star clusters, M31, M33, the
Magellanic Clouds, or
galactic fields
were taken with the F702W filter in 2000 - 2002. 

\begin{figure*}[btph]
%\vspace{0.4cm}
\begin{center}
%\sidecaption
%\setlength{\fboxsep}{-\fboxrule}
%\fbox{
\includegraphics[bb=1cm 5.5cm 19.7cm 20.2cm,width=12cm]
{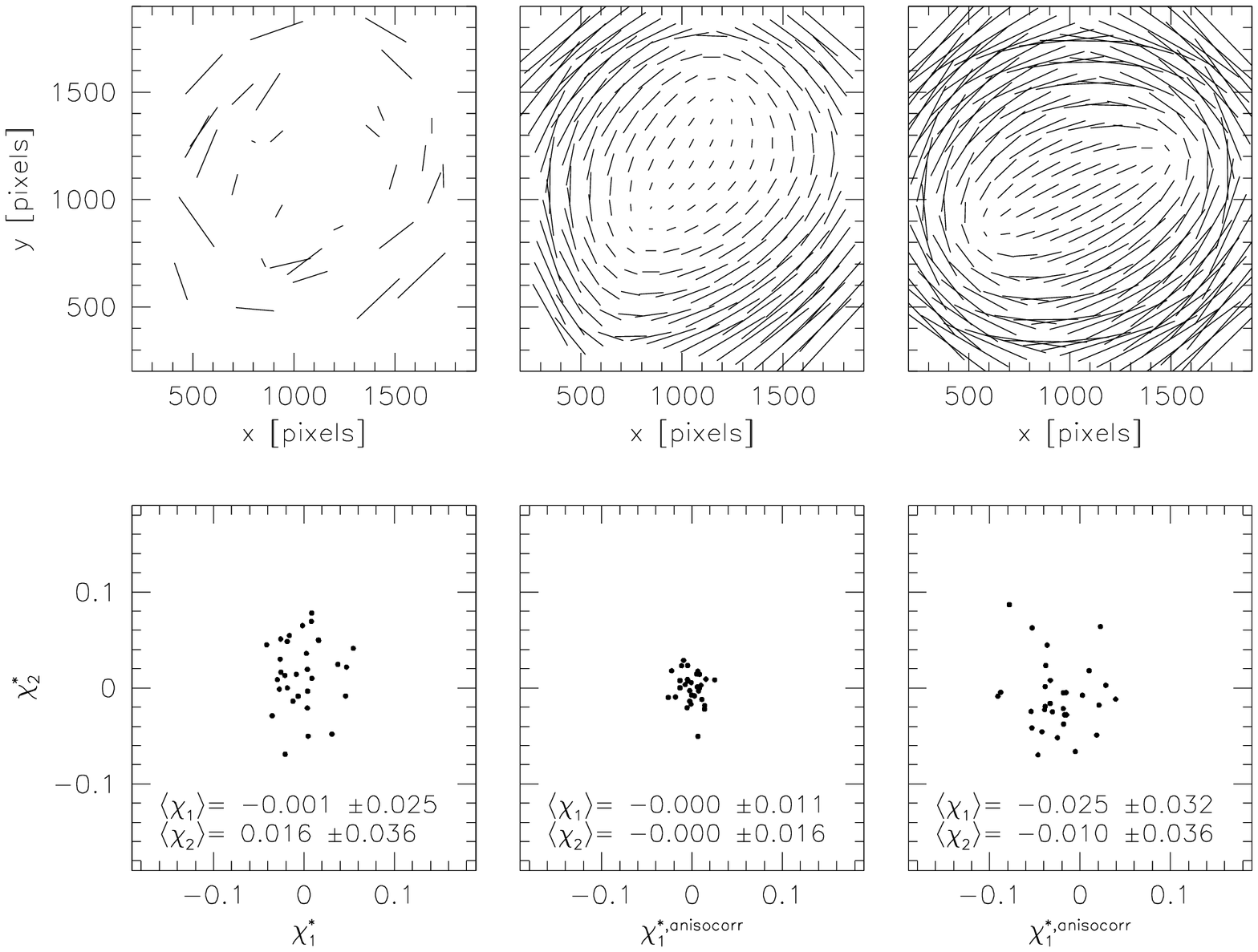}
%}
\caption{Anisotropy correction of Chip 3. In the top left field, we plot the
ellipticities of the 33 stars from the stellar sequence (as in
Fig. \ref{anisocorr-hst-chip3}). Their ellipticity
components are shown in the bottom left panel. In the top middle panel, we
evaluate the anisotropy kernel as found from these 33 stars on a grid. The
residual ellipticities of the stars after correction are shown in the bottom
middle panel. In the top right panel, the correction polynomial found from 372
stars in a star cluster field is shown. Applying this polynomial to our data
yields the residual ellipticities shown in the lower right panel. 
%As the
%dispersion of these is not less than the original, this anisotropy
%correction is inferior to the previous one.
}
\label{anisocorr-chip3-glob}
\end{center}
\end{figure*}

Instead, we retrieved images of a field within the globular cluster 47 Tuc
(NGC 104) taken
in the F606W filter on Oct. 19th, 2002. We found the anisotropy patterns  of
these 
images to be similar to the ones in our images, but with some notable
differences. Using these patterns for the anisotropy correction did not
reduce the ellipticity dispersion of the stars (see
Fig.~\ref{anisocorr-chip3-glob}). At least for the centers of the
images, the previous method is therefore superior. However, due to a
lack of stars, the anisotropy correction is largely extrapolated
toward the edges of each chip. These are also the areas with the
largest anisotropy (as apparent from the figures), so that we are
uncertain of the quality of the anisotropy correction applied to them.

\subsection{Charge Transfer Efficiency}
\label{CTE}

\begin{figure}[btp]
%\vspace{0.4cm}
\begin{center}
%\setlength{\fboxsep}{-\fboxrule}
%\fbox{
\includegraphics[bb=1cm 5.8cm 20cm 24.4cm,width=1\hsize]
{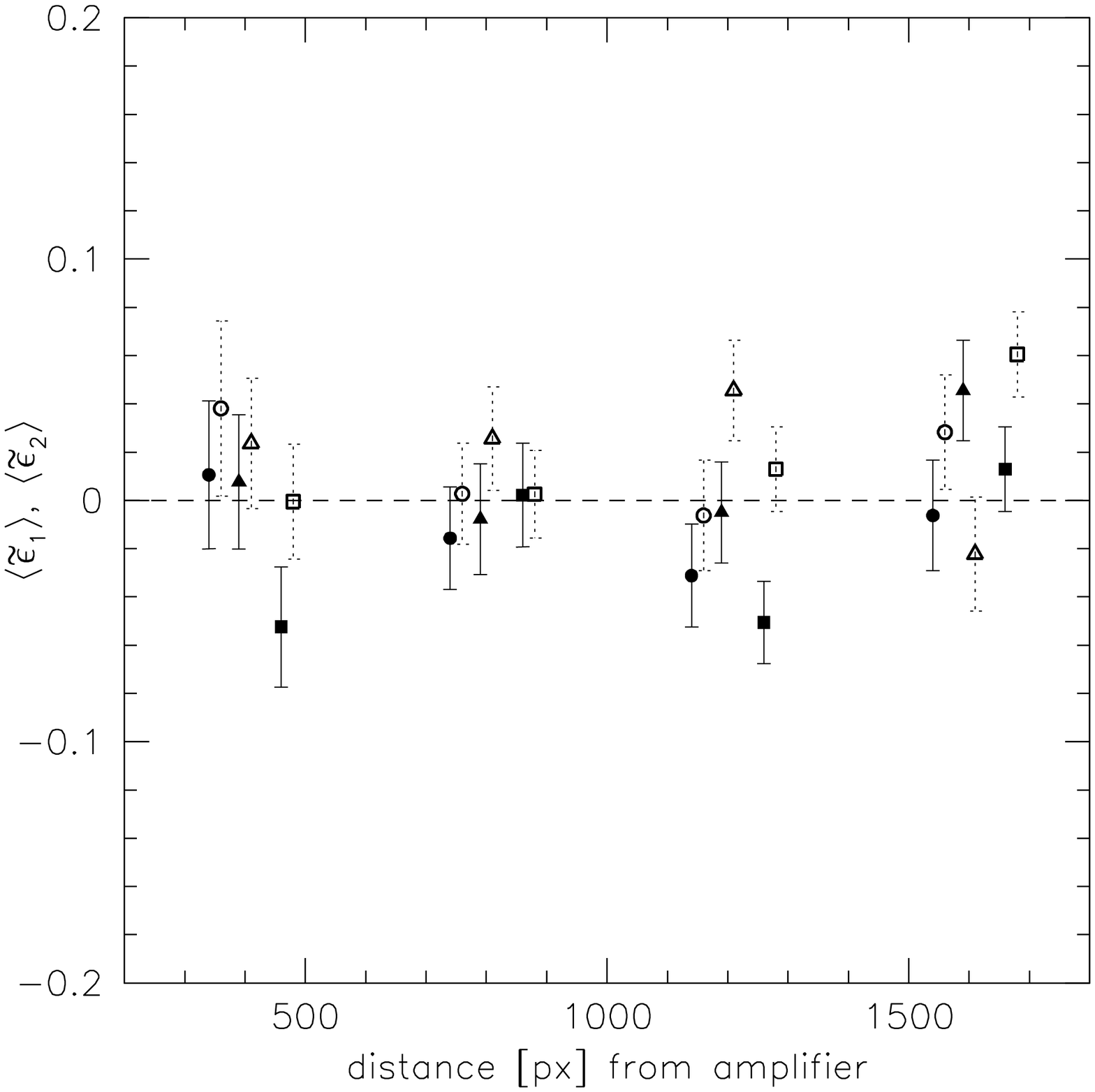}
%}
\caption{The average ellipticity components as a function of distance
from the read-out amplifier. As before, the galaxies are split into
magnitude bins, where circles denotes the faintest, triangles the medium
bright ones, and squares the brightest. $\ave{\tilde{\varepsilon}_1}$ is shown as
solid symbols
with solid error bars, $\ave{\tilde{\varepsilon}_2}$ as open symbols with
dashed error bars. The error bars denote the standard deviation
divided by the square root of the number of galaxies to give the
significance of the mean value.
}
\label{fig:cte-all-ave}
\end{center}
\end{figure}

WFPC2 has a considerable Charge Transfer Efficiency (CTE) problem. Stars bright
enough to show diffraction spikes also show typical ``CTE tails'' in
the direction opposite to the read-out direction. In fainter objects,
these tails are not distinguishable by eye. Unfortunately, this
effect is difficult to compensate for and has been studied only little
for extended objects \citep{rie00}.

The imperfect CTE of the WFPC2 could potentially
bias the measured ellipticities, since it results in charge being
depleted in some pixels and added into others. Unlike for the photometry
of point sources, there is not yet a correction procedure to account
for this effect for shape measurements. According to \citet{rie00},
for extended objects, the 
A deficient CTE causes charge loss in the pixels
closer to the read-out amplifier but adds this charge to the pixels
farther from the amplifier. To first order, this just causes a slight
displacement of the galaxy's centroid. But since depletion and
addition cannot be expected to be symmetric, it may also distort the
galaxy's shape. If so, it would bias the $\varepsilon_1$ components, as this is
measured along rows and columns.

From studies on how CTE affects photometry it is known that it
increases with distance from the read-out amplifier. This is also
visible in our images: stars with large $y$-coordinates have more
pronounced CTE trails.
Also, the charge loss due to CTE depends on the background level of
the image. A high background effectively suppresses CTE losses. The
images of our dataset have a background corresponding to about 35
electrons per pixel, which reduces CTE loss significantly, at least
for photometry. Lastly, the relative CTE losses are largest for faint
objects.

To investigate any possible bias due to CTE, we divide the galaxy
catalog into four bins according to the original $y$ position, which
gives its distance from the amplifier. Unlike for the lensing analysis, we
are not 
interested in the ellipticity with respect to the rectascension axis,
but to the original $x$-axis of the image. We denote these with
$\tilde{\varepsilon}_i$. For
each bin, we calculate the weighted mean of $\tilde{\varepsilon}_1$ and
$\tilde{\varepsilon}_2$.
The results are shown in Fig. \ref{fig:cte-all-ave}.

If the CTE would affect galaxies similar to stars, i.e. it causes them
to trail and thus elongates them in the $\tilde{\varepsilon}_1$ direction, we
would expect that $\ave{\tilde{\varepsilon}_1}$ is consistent with zero for the
first bin and then decreases with increasing distance from the amplifier.
This is clearly not the case for any of the brightness bins. 
The scatter in $\ave{\tilde{\varepsilon}_1}$ about zero is comparable to that
in $\ave{\tilde{\varepsilon}_2}$, which should not be affected by CTE. The
deviations from zero in some bins may well be due to the fact that
there is some degree of tangential alignment in the field, so that
$\ave{\tilde{\varepsilon}_1}$ and $\ave{\tilde{\varepsilon}_2}$ are not necessarily
zero. Yet, Fig. \ref{fig:cte-all-ave} excludes a notable bias due to
CTE. 

%\begin{figure}[bt]
%%\vspace{0.4cm}
%\begin{center}
%%\setlength{\fboxsep}{-\fboxrule}
%%\fbox{
%\includegraphics[bb=1cm 7cm 21.5cm 24.4cm,width=0.32\hsize]
%{images/cte-chip2.ps}
%%}
%\includegraphics[bb=0.8cm 7cm 21.3cm 24.4cm,width=0.32\hsize]
%{images/cte-chip3.ps}
%\includegraphics[bb=0.5cm 7cm 21cm 24.4cm,width=0.32\hsize]
%{images/cte-chip4.ps}
%\end{center}
%\caption{As Fig. \ref{fig:cte-all-ave}, but for each HST chip
%separately: Chip 2 (left), Chip 3 (middle), Chip 4(right).
%}
%\label{fig:cte-chip}
%\end{figure}

%%%%%%%%%%%%%%%%%%%%%%%%%%%%%%%%%%%%%%%%%%%%%%%%%%%%

\section{Comparison of ground-based and space-based measurements}
\label{sc:compare}

With the two datasets - the $I$-band image and the HST image -
we have the opportunity to directly compare shape measurements from
the ground to those from space. Ground-based shape determinations rely
on an accurate compensation of the smearing due to the Earth's atmosphere,
while the HST observations are hampered by the small image size, and the
undersampling of the PSF. In the previous section, we 
confirmed with the HST data the presence of tangential alignment around the
dark clump candidate, which implies that 
ellipticity measurements are to some degree comparable. 
However, we failed to confirm with the HST
data the amplitude of the alignment
signal.

A direct comparison of the objects common to both catalogs tests the
reliability of ground-based shape measurements and may 
also help to identify any systematics in either dataset. The ultimate
goal of this comparison is to find the cause of the discrepancy in the
significance of the alignment signal between the two datasets.

%%%%%%%%%%%%%%%%%%%%%%%%%%%%%%%%%%%%%%%
%                                     %
%         Catalog Correlation         %
%                                     %
%%%%%%%%%%%%%%%%%%%%%%%%%%%%%%%%%%%%%%%

\subsection{Catalog correlation}

As the HST data was astrometrically calibrated by using a reference catalog
extracted from the $I$-band image, objects present in both images can be
identified by their sky coordinates.

To correlate the catalogs, we searched for objects within $1\arcsec$ of
objects detected in the respective other catalog. This radius was found to
be the optimal balance between a high number of matched objects and a
low rate of double detections. 

The catalogs used are the same as those for the lensing analysis, except
that objects with $P^g_s<0.3$ in the ground-based catalog are also
considered. 

%To some degree, it also reflects the
%accuracy of the (external \textit{(?)}) astrometric solution. It turns out that this is
%limited by the distortions present in the ground-based image (see
%Sect. \ref{sect:ground_astrom}). For many of the objects in the
%ground-based image, the position is errant by up to $1\arcsec$.

Within the field covered by the HST mosaic, there are 507 objects in the
$I$-band catalog. 
We find one HST counterpart for 350 of these objects, and
two or three 
counterparts for 17 objects, as the
HST is able to resolve very close objects which were identified as
single objects in the ground-based data. For 140 objects, no
counterpart was found. For most of these, this results from the large
areas that were masked in the HST images. 

\begin{figure}[btp]
%\vspace{0.4cm}
\begin{center}
%\setlength{\fboxsep}{-\fboxrule}
%\fbox{
\includegraphics[bb=1cm 5.8cm 20cm 24.4cm,width=\hsize]
{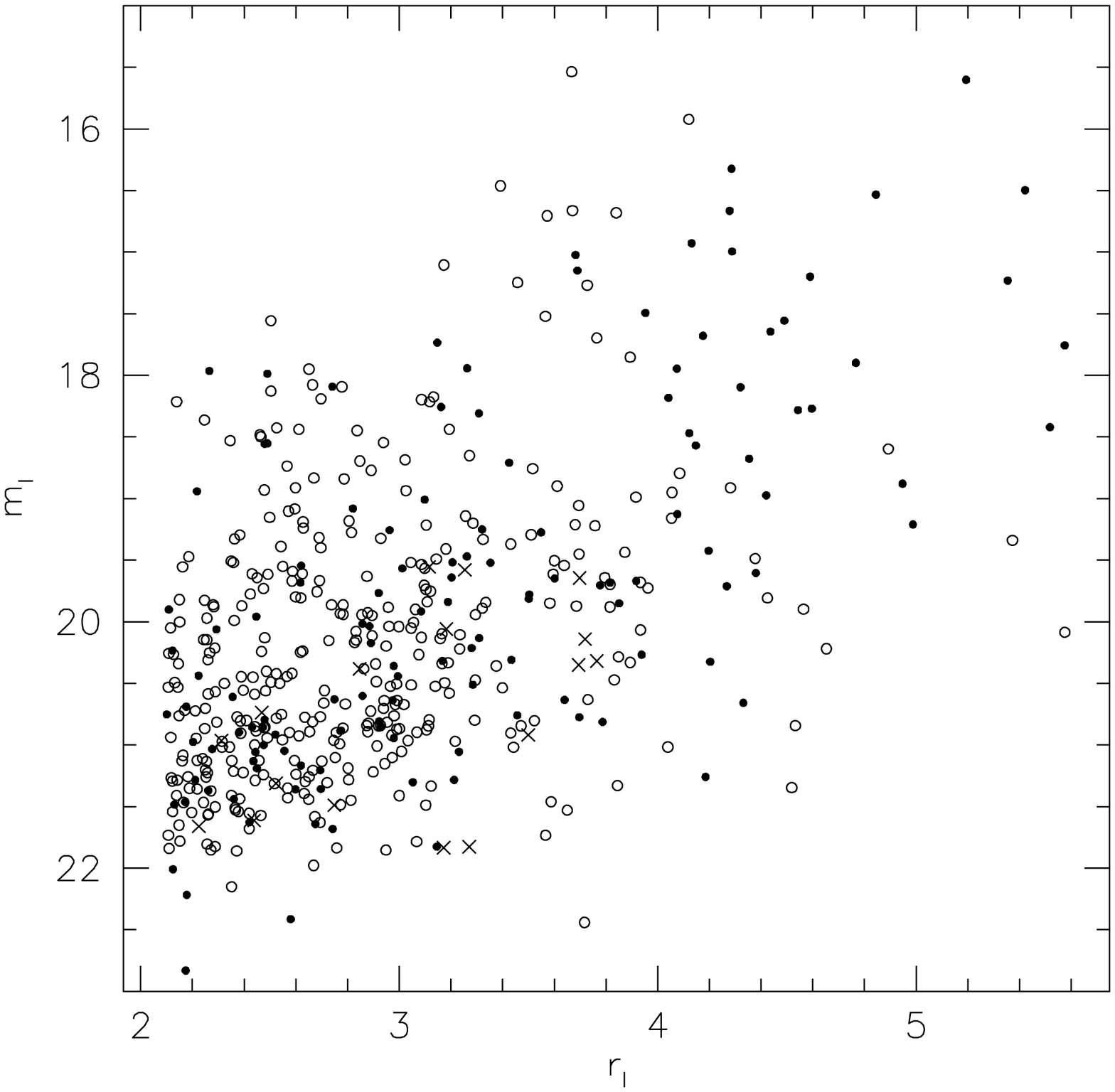}
%}
\caption{The magnitudes and half-light radii of the $I$-band objects
  positioned in the field observed by the HST. Open circles denote objects
  for which one HST counterpart was found, solid ones no counterpart, and
  crosses more than one counterpart. For some of the bright, large objects
  no counterpart was found due to the large masked areas in the HST mosaic.}
\label{fig:m-rg-mag}
\end{center}
\end{figure}

Fig. \ref{fig:m-rg-mag} illustrates the matched objects in an $r_g - m$
diagram of the ground-based data. There is no apparent trend as to for
which objects we are
more likely to find a counterpart in the space-based data. In particular, it
is not less likely to do so for faint objects. This shows that the catalog
was only very little contaminated by noise detections. Since close
objects are often unresolved in the ground-based image, objects with
two HST counterparts have on average a larger radius than those with
one counterpart.

\subsection{Statistical properties}
\label{sect:m-properties}

%%%%%%%%%%%%%%%%%%%%%%%%%%%%%%%%%%%%%%%
%                                     %
%     Ellipticity Measurements        %
%                                     %
%%%%%%%%%%%%%%%%%%%%%%%%%%%%%%%%%%%%%%%

\subsubsection{Ellipticity measurements}
\label{sect:m-ellis}

The basis of weak lensing analyses are the shape measurements of faint galaxy
images. But the fainter an object is, the more difficult the shape
determination is. Our dataset provides an ideal opportunity to test
the reliability of 
shape measurements of ground-based data compared to space-based
measurements. 

\begin{figure*}[bthp]
%\vspace{0.4cm}
\begin{center}
%%\sidecaption
%\setlength{\fboxsep}{-\fboxrule}
%\fbox{
\includegraphics[bb=0cm 6cm 20cm 24.4cm,width=0.48\hsize]
{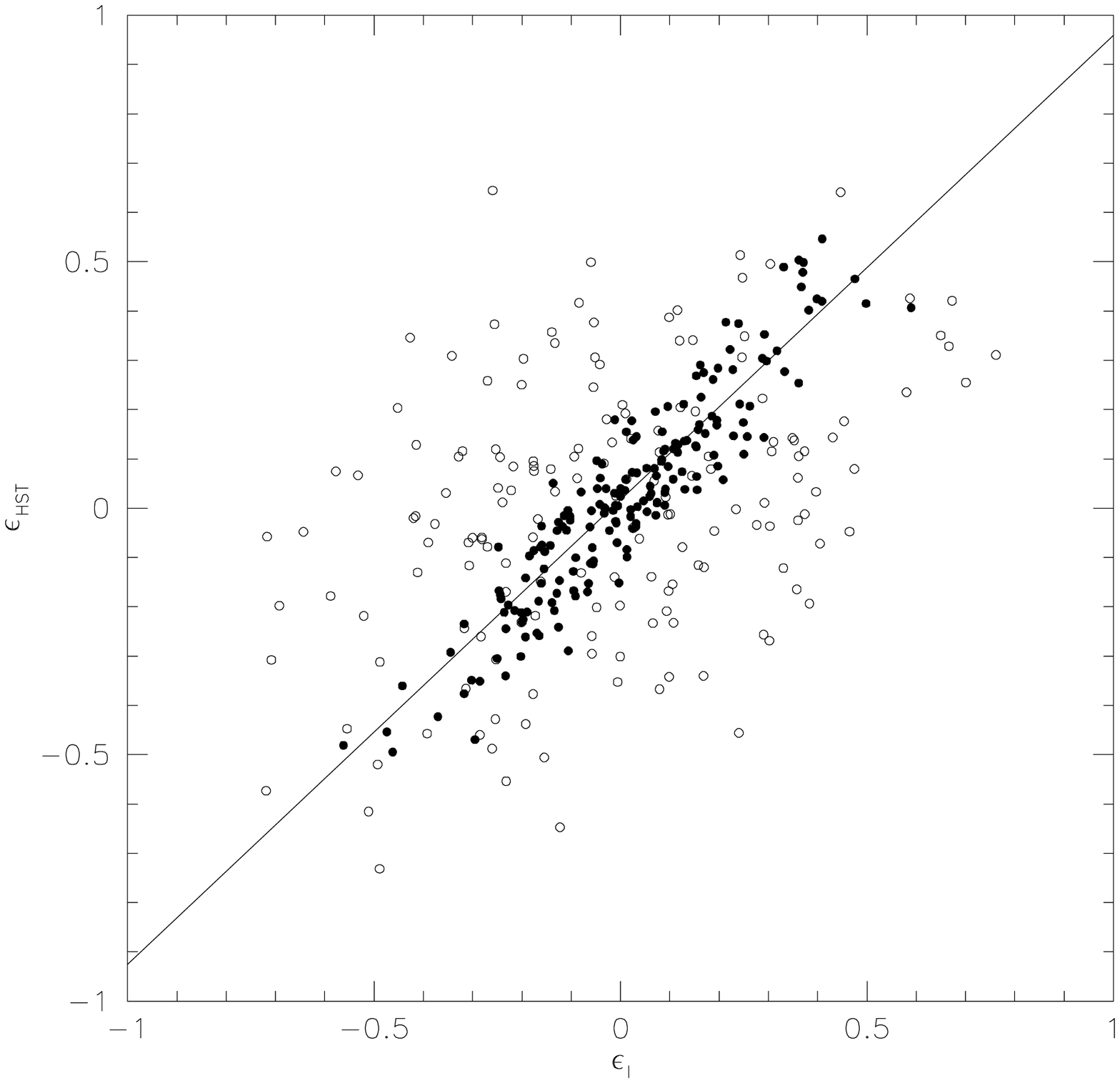}
%}
%\setlength{\fboxsep}{-\fboxrule}
%\fbox{
\includegraphics[bb=0cm 6cm 20cm 24.4cm,width=0.48\hsize]
{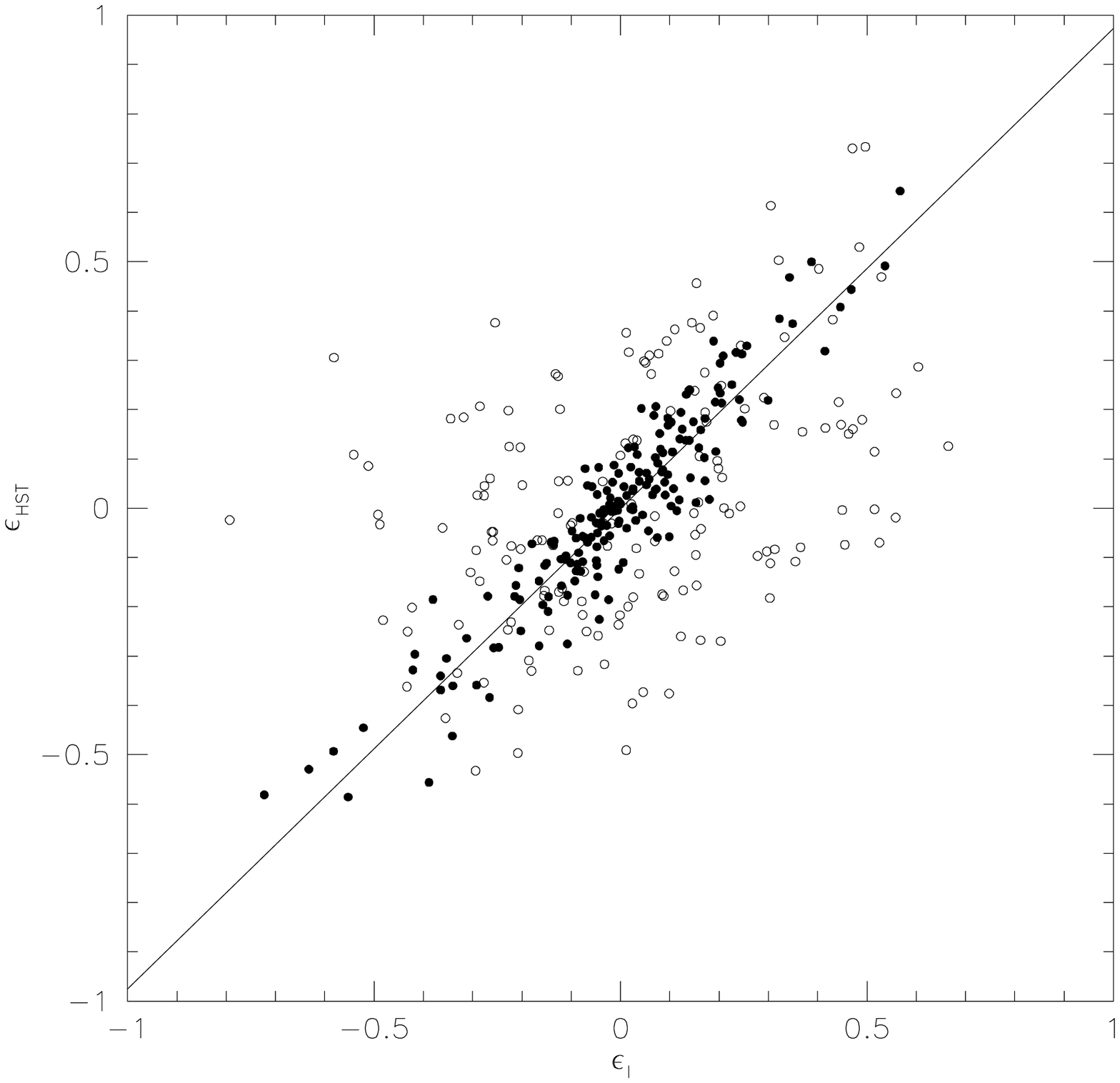}
%}
\caption{Comparison of the ellipticity measurements from the HST images to
  the $I$-band image. Objects for which $\Delta \varepsilon \le 0.2$ [see
  eq. (\ref{eq:delta-g})]  are shown as solid circles, objects with $\Delta
  \varepsilon  > 0.2$ are shown  
  as open circles. The 
  line gives the best linear fit. The error
  bars of individual points are omitted to avoid crowding.}
\label{fig:m-ellis}
\end{center}
\end{figure*}

In Fig. \ref{fig:m-ellis}, we compare the ellipticity measurements in the
space-based data to those of the ground-based data. To these points, we fit a
linear function  
\citep[algorithm {\tt fitexy} from ][ to account for errors in both
coordinates]{numrec}, where we employ the 
same weighting scheme as applied to the lensing analysis. We find:

$$
\renewcommand{\arraystretch}{1.1}
\begin{array}{r@{\;=\;}l@{\qquad\qquad}r@{\;=\;}l}
\varepsilon_{1, \rm hst} & m_1 \cdot \varepsilon_{1, \rm I} + b_1 &
\varepsilon_{2, \rm hst} & m_2 \cdot \varepsilon_{2, \rm I} + b_2 \\[1ex]
m_1 & 0.94 \pm 0.14 & m_2 & 0.97 \pm 0.15\\
b_1 & 0.017 \pm 0.027 & b_2 & -0.001 \pm 0.027
\end{array}
$$
Although the scatter is rather large and a comparison on an
object-to-object basis not feasible, the general agreement for the
is very good, particular for the $\varepsilon_2$ component. 
For the $\varepsilon_1$ component, both slope and offset $b_1$ differ from
unity by about $0.5\sigma$. If the HST ellipticity measurements are
influenced by the CTE problem or by the resampling of the {\tt
drizzle} algorithm, we would indeed expect some bias in the $\varepsilon_1$
component, which is measured along the rows and columns of the CCD chip.\\
\\

The scatter seen in the correlation of both components is of a similar
order of magnitude, so that the deviations in the ellipticity
measurements cannot be attributed to a single component.
For simplicity, we therefore reduce the difference of the
ellipticity measurements to a one-dimensional quantity:
\be
\Delta \varepsilon = 
\sqrt{(\varepsilon_{1, \rm hst}-\varepsilon_{1, \rm I})^2+
(\varepsilon_{2, \rm hst}-\varepsilon_{2, \rm I})^2}
\label{eq:delta-g}
\ee
Ellipticity measurements are considered to be equivalent if $\Delta \varepsilon
\le 0.2$ and 
inconsistent if $\Delta \varepsilon > 0.2$. In the first sample, there are
185 galaxies, whereas in the second there are 165.

%%%%%%%%%%%%%%%%%%%%%%%%%%%%%%%%%%%%%%%
%                                     %
%             Magnitudes              %
%                                     %
%%%%%%%%%%%%%%%%%%%%%%%%%%%%%%%%%%%%%%%

\subsubsection{Magnitudes}

\begin{figure}[btp]
%\vspace{0.4cm}
\begin{center}
%\setlength{\fboxsep}{-\fboxrule}
%\fbox{
\includegraphics[bb=1cm 5.8cm 20cm 24.4cm,width=1\hsize]
{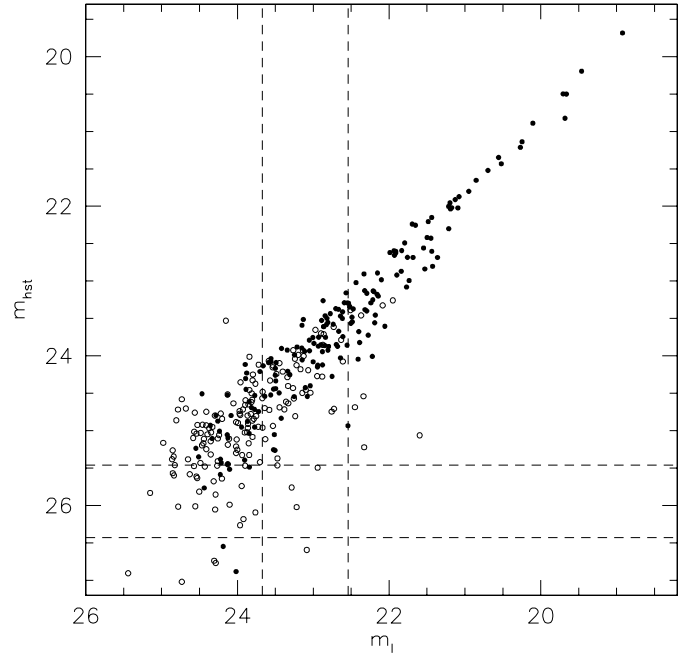}
%}
\caption{Magnitude measured in the HST data vs. magnitude measured in the
  $I$-band image for the matched objects. The symbol coding is identical to
  that of Fig. \ref{fig:m-ellis}. The dashed lines indicate the cuts used to
  divide the respective catalogs into three samples.} 
\label{fig:m-mags}
\end{center}
\end{figure}

For the lensing analyses, the galaxy catalogs were split into three parts
according to brightness to investigate their contribution to the lensing
signal. The division was chosen such that each sample contained an equal
number of galaxies. With the merged catalog, we can now examine how
these samples compare in the two datasets.

Fig. \ref{fig:m-mags} illustrates the magnitudes measured of the matched
objects. Most of the objects found in the $I$-band image are considered
``bright'' 
objects in the HST data. The fact that we also detected
tangential alignment in the bright HST bin confirms the lensing signal
seen in the ground-based image.

As would be expected, the shape measurements agree best for the
brighter objects, while they are inconsistent for most of the faint
objects.

\subsubsection{$P^{\rm g}_s$ correction}

\begin{figure}[bthp]
%\vspace{0.4cm}
\begin{center}
%\setlength{\fboxsep}{-\fboxrule}
%\fbox{
\includegraphics[bb=1cm 5.8cm 20cm 24.4cm,width=1\hsize]
{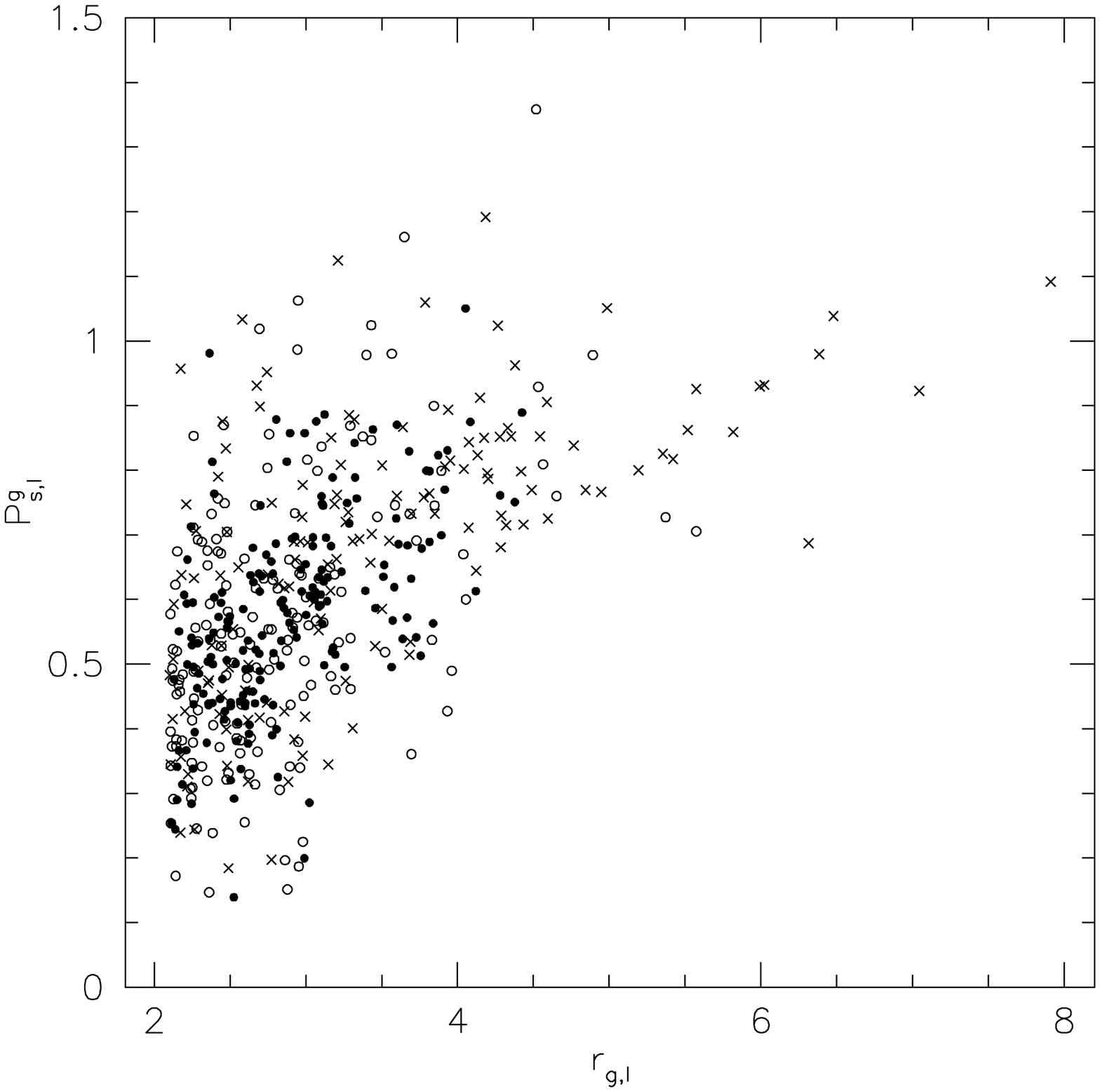}
%}
\caption{The ellipticity correction factor $P^{\rm g}_s$ and half-light
  radii of the $I$-band objects within the HST field. Crosses denote
  objects for which no counterpart was found; the other colors are the same
  as in Fig. \ref{fig:m-ellis}.}
\label{fig:m-pgs}
\end{center}
\end{figure}

One quality attribute we used previously to classify objects was the
factor $P^{\rm g}_s$ by which the smearing due to the PSF was
corrected. In Fig. \ref{fig:m-pgs} 
we plot an $r_g - P^{\rm g}_s$ diagram of the objects found in the $I$-band
image within the field covered by the HST mosaic. There is only a slight
indication that ellipticity measurements with $P^{\rm g}_s < 0.4$ are less
reliable than others. 

\subsubsection{Signal-to-noise ratio and weighting}

\begin{figure}[bthp]
%\vspace{0.4cm}
\begin{center}
%\setlength{\fboxsep}{-\fboxrule}
%\fbox{
\includegraphics[bb=1cm 5.8cm 20cm 24.4cm,width=1\hsize]
{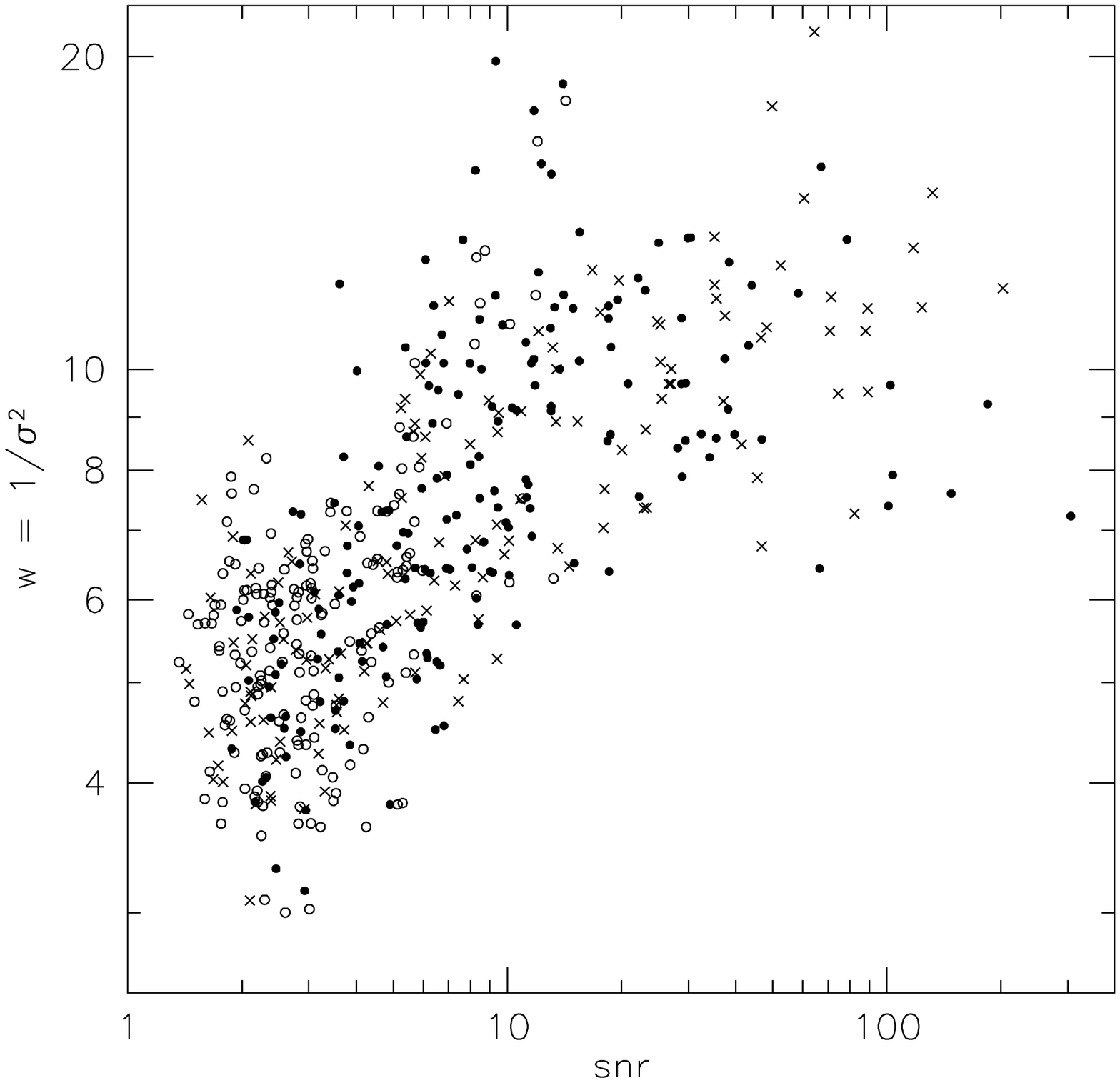}
%}
\caption{The signal-to-noise ratios {\tt snr} and weights of the $I$-band
  objects in the HST field. The color-coding is identical to that of
  Fig. \ref{fig:m-pgs}.} 
\label{fig:m-snratio}
\end{center}
\end{figure}

Two other criteria for the reliability of a shape measurement are the
signal-to-noise ratio of a detection and the
weight it was assigned. The latter one was taken to be inversely
proportional to the variance $\sigma^2$ of the ellipticities of a galaxy
ensemble with 
similar noise properties [see
Sect. \ref{sc:cat-7}]. Fig. \ref{fig:m-snratio} shows the
signal-to-noise ratios and weights of the $I$-band objects in the
HST field.
As expected, the ellipticity measurements deviate for those objects with a
low signal-to-noise ratio. These are in general down-weighted, although the
weight itself is not a clear indicator of the reliability. The weighting
scheme could therefore be improved.

The lensing signal in the ground-based data was carried by faint
galaxies with a signal-to-noise ratio of less than about 4. These are
precisely those galaxies for which the space-based measurements give
different ellipticities. It is therefore not surprising that the amplitude
of the 
$\map$ signal is different in the two datasets.

\subsubsection{Dependence on chip position}

\begin{figure*}[bthp]
%\vspace{0.4cm}
\begin{center}
%%\sidecaption
%\setlength{\fboxsep}{-\fboxrule}
%\fbox{
%\includegraphics[bb=0.3cm 7cm 20.7cm 23.9cm,width=5.9cm]
\includegraphics[bb=0.3cm 5.6cm 20.7cm 23.9cm,width=0.48\hsize]
{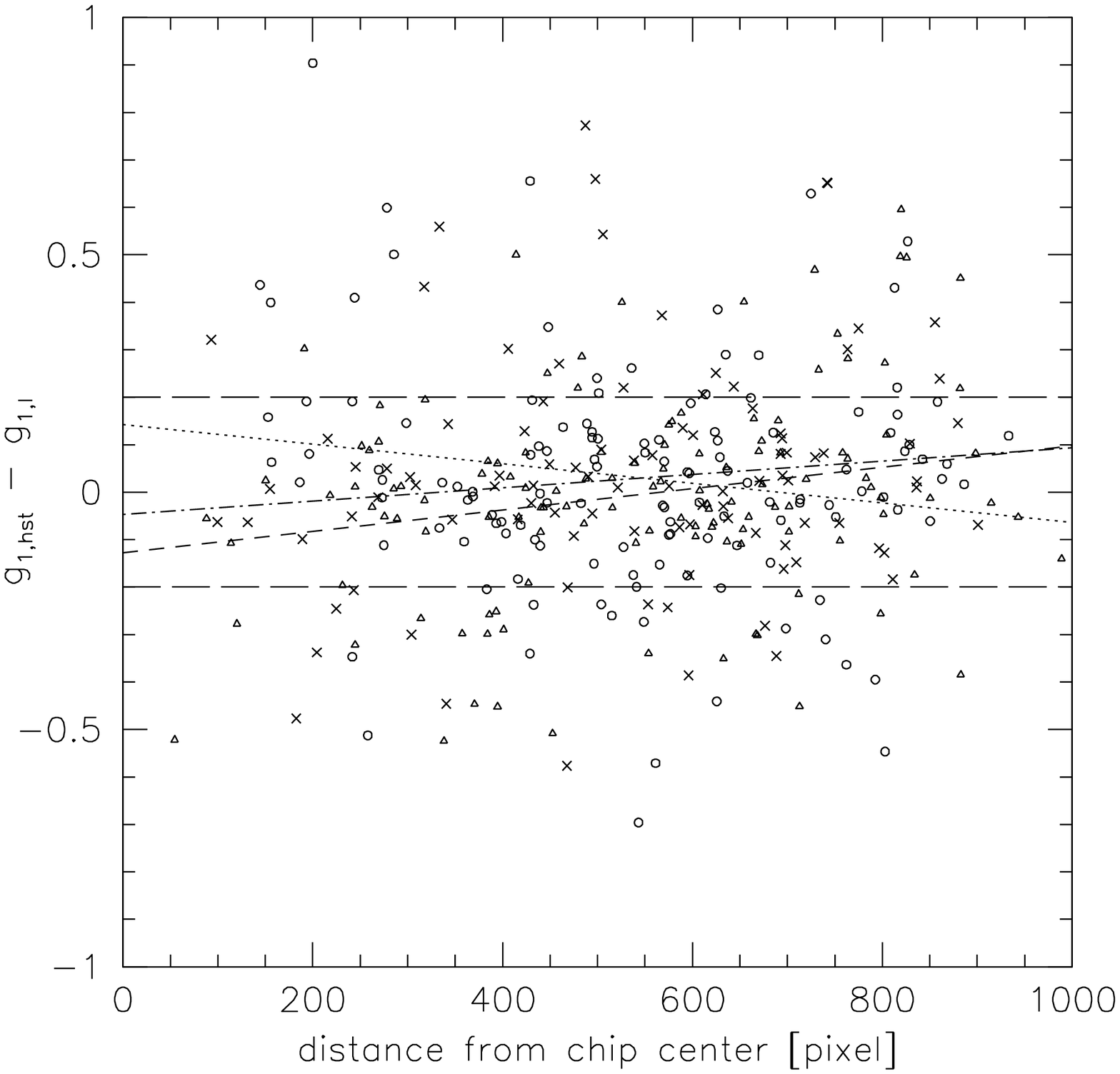}
\includegraphics[bb=0.3cm 5.6cm 20.7cm 23.9cm,width=0.48\hsize]
{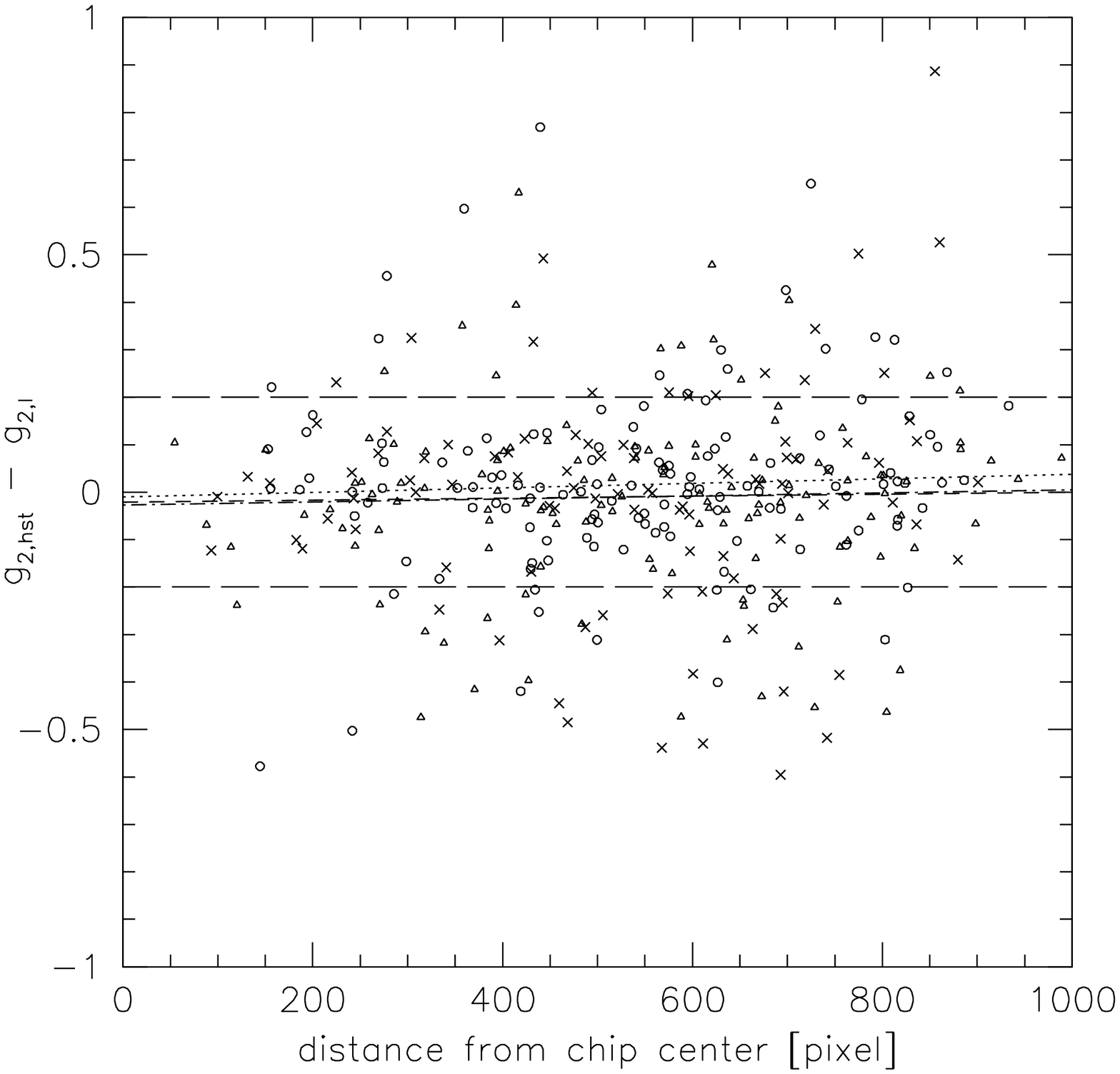}
%}
\caption{Ellipticity differences $\Delta \varepsilon_1$ and $\Delta \varepsilon_2$ as a
  function of position in the 
  HST mosaic. Plotted on the $x$-axis is the distance from the chip
  center (measured in pixel units after the drizzle process);
  objects detected on Chip2 are shown as circles, Chip3 as triangles, and
  Chip4 as crosses. The
  distinction we make between equivalent and inconsistent shape measurements
  is illustrated by the black, long-dashed lines (if for the
  respective other component $\Delta \varepsilon_i = 0$).
  The dotted, short-dashed, and dot-dashed lines give linear
  fits to the respective data.} 
\label{fig:m-dist-center}
\end{center}
\end{figure*}

As mentioned before, we cannot judge the quality of the anisotropy
correction of the HST images due to the small number of stars in the
images. With the 
correlated catalogs, we have another test of the correction: if it is faulty,
then we expect to see a systematic variation of the ellipticity
differences $\Delta \varepsilon_1 = (\varepsilon_{1,\rm hst}-\varepsilon_{1,\rm I})$ and 
$\Delta \varepsilon_2 = (\varepsilon_{2,\rm hst}-\varepsilon_{2,\rm I})$ 
between the three chips and/or with position on the chip. We reduce
the latter one to a one-dimensional quantity by considering the distance $r$
from the respective chip center; this is motivated by the observation that
the anisotropy seen in the HST chips is largest at the edge (see
Fig. \ref{anisocorr-chip3-glob}).

In Fig. \ref{fig:m-dist-center} we plot 
$\Delta \varepsilon_1$ and $\Delta \varepsilon_2$ 
as a function of $r$, where we distinguish between the three chips. 
Clearly, an object's position in the
HST mosaic is not the main cause of discordance in the ellipticity
measurement. 
However, there are trends visible such as a slight overestimation of
$\varepsilon_1$ for small $r$ in Chip 4 and an underestimation of $\varepsilon_1$ for
small $r$ in Chip 3. 
To quantify these, we fit a linear function
$\Delta \varepsilon_{i} = m_i\cdot r / 1000 + b_i$, for each chip. 
We weight the $\Delta \varepsilon_{i}$ values by the same weighting scheme as
before, but do not assign error bars to the distance. The results are:
%$$
%\renewcommand{\arraystretch}{1.1}
%\begin{array}{r@{\;=\;}l@{\qquad\qquad}r@{\;=\;}l@{\qquad\qquad}r@{\;=\;}l}
%m_2 & -0.123 \pm 0.159 & m_3 & 0.016 \pm 0.155  & m_4 & 0.168 \pm 0.176\\
%b_2 & 0.279 \pm 0.092  & b_3 & 0.1807 \pm 0.093 & b_4 & 0.134 \pm 0.099
%\end{array}
%$$
%The results are:
%For the $\varepsilon_1$ component, the results are:
$$
\renewcommand{\arraystretch}{1.1}
\begin{array}{l@{\qquad}c@{\qquad}c@{\qquad}c}
&
\mbox{Chip 2} & 
\mbox{Chip 3} &
\mbox{Chip 4}\\[1.2ex]
m_1 & -0.21\pm0.17 &  0.23\pm0.16 &  0.14\pm0.19 \\
b_1 &  0.14\pm0.10 & -0.13\pm0.10 & -0.05\pm0.11 \\[1ex]
m_2 &  0.05\pm0.17 &  0.02\pm0.16 &  0.03\pm0.19 \\
b_2 & -0.01\pm0.10 & -0.02\pm0.10 & -0.03\pm0.11
\end{array}
$$

For the $\varepsilon_2$ component, the ellipticity difference is remarkably
constant at zero for all chips. However, for the $\varepsilon_1$ component,
there is a significant deviation from such a behavior in Chips 2 and
4. The anisotropy pattern of the HST chips is approximately circular
in appearance (see Fig. \ref{anisocorr-chip3-glob}), so that an
insufficient anisotropy correction should affect both components
equally. The deviation in $\varepsilon_1$ may therefore be a hint that the CTE
and / or the drizzle process affect the ellipticities
systematically. On the other hand, the linear fits are certainly noisy
due to the small number of objects, particularly at low $r$.

Considering that $\Delta \varepsilon_2$ is on average constant at zero, but also
that $\Delta \varepsilon_1$ is consistent with zero for a large number of objects
($400 < r< 800$), strengthens our assumption that the anisotropy
correction worked properly.

%%%%%%%%%%%%%%%%%%%%%%%%%%%%%%%%%%%%%%%
%                                     %
%          LENSING ANALYSIS           %
%                                     %
%%%%%%%%%%%%%%%%%%%%%%%%%%%%%%%%%%%%%%%

\subsection{Lensing analysis}
\label{sect:m-lensing}

\begin{figure*}[bthp]
%\vspace{0.4cm}
\begin{center}
%\sidecaption
%\setlength{\fboxsep}{-\fboxrule}
%\fbox{
\includegraphics[width=12cm]
{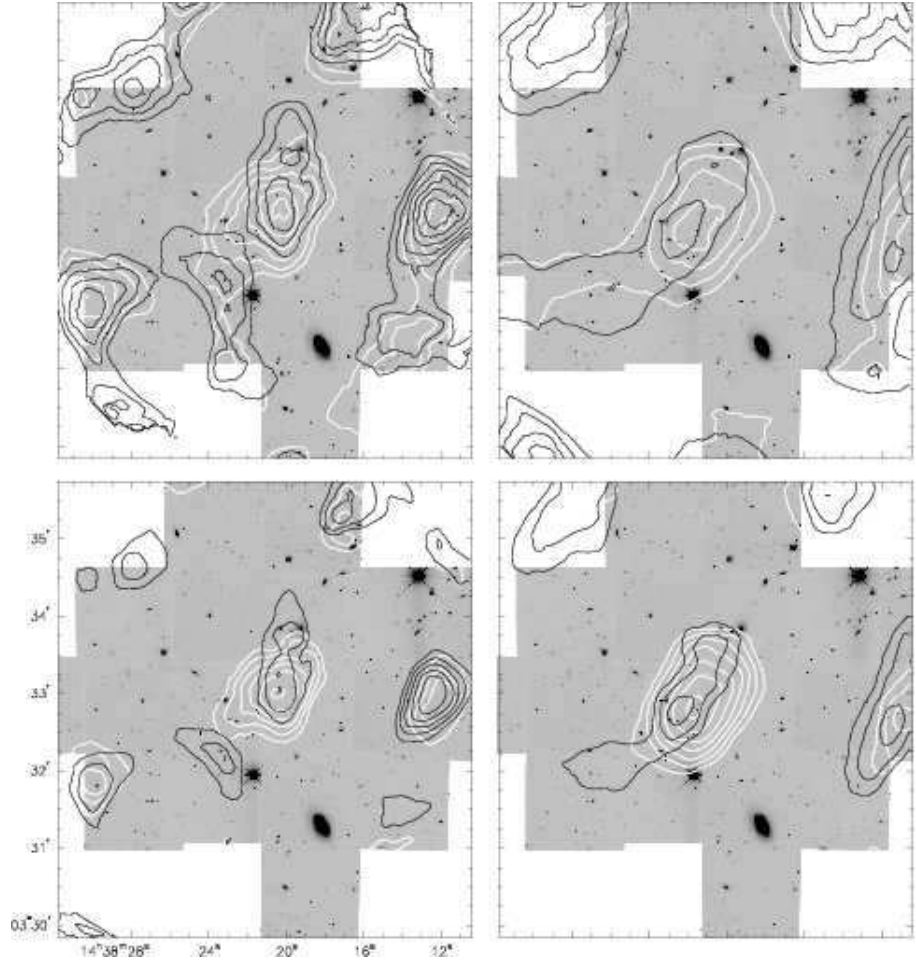}
%}
\caption{$\map$ analysis of the 350 objects common to both the ground- and
  the space-based catalog. The upper two panels show $\map$ contours,
  starting at 0.02 and increasing in 0.02 increments, where a filter scale
  of $80\arcsec$ was used for the left panel, and a filter scale of
  $120\arcsec$ for the right one. The lower two panels give signal-to-noise
  contours, starting at 1.5$\sigma$ and increasing in 0.5$\sigma$
  increments. The white contours correspond to the analysis using ground-based
  ellipticities, and black ones to the one using space-based
  ellipticities. 
}
\label{fig:m-compare}
\end{center}
\end{figure*}

Since the ellipticity measurements agree on average, one can assume that the
lensing analysis of our set of matched objects should yield similar
results. However, the individual analyses indicate otherwise: the faint
galaxies in the ground-based data, which caused most of the lensing signal,
correspond to the bright (and medium bright) galaxies in the HST data, which
gave only a weak signal.

In order to directly evaluate the correlation (or discrepancy) between the
ellipticity measurements and the lensing analysis, we perform several $\map$
analyses of the matched galaxies (see Table \ref{tab:m-map-results} for a
quantitative summary).\\

As a reference, we perform a $\map$ analysis of the 350 matched objects using once
the ground-based ellipticity measurements and once the space-based
measurements (Fig. \ref{fig:m-compare}). It is quite remarkable how well
the $\map$ 
contours agree for these; however, for the dark clump peak, the space-based
values reach only half the height of the ground-based.

\begin{figure*}[bthp]
%\vspace{0.4cm}
\begin{center}
%\sidecaption
%\setlength{\fboxsep}{-\fboxrule}
%\fbox{
\includegraphics[width=12cm]
{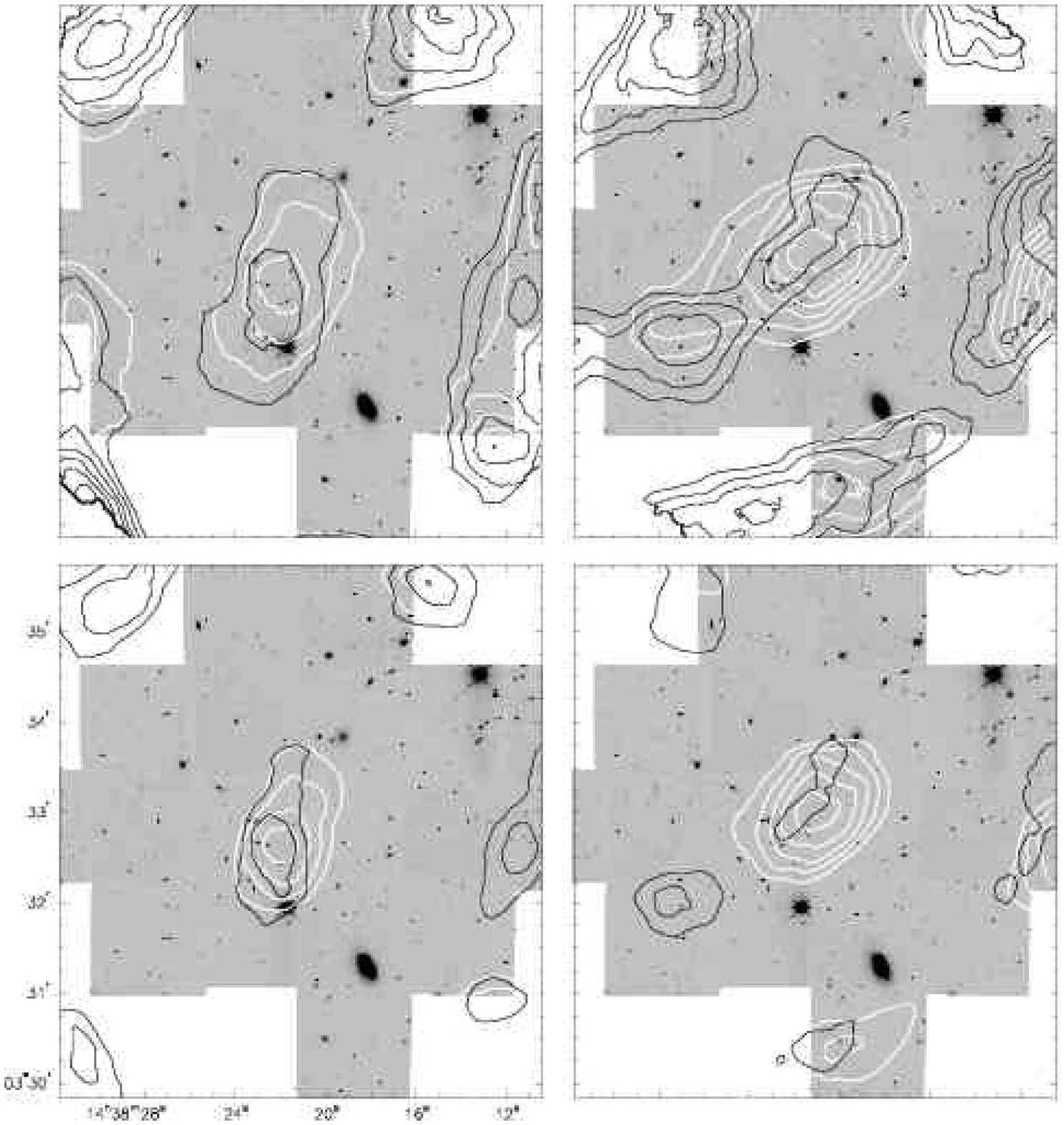}
%}
\caption{$\map$ analysis of objects with $\Delta \varepsilon \le 0.2$ (left side) and
  $\Delta \varepsilon > 0.2$ (right side). The upper panels show the $\map$ values
  measured in the $120 \arcsec$ filter scale, the lower ones the
  corresponding signal-to-noise ratios. The color coding and the contours
  are identical to Fig. \ref{fig:m-compare}.
}
\label{fig:m-diff-same}
\end{center}
\end{figure*}

Restricting the galaxy sample to those for which $\Delta \varepsilon \le 0.2$ further
highlights the agreement, as is expected (left hand side of
Fig. \ref{fig:m-diff-same}). These are mainly the galaxies considered
``bright'' also in the ground-based image (see Fig. \ref{fig:m-mags}), which
contribute to the lensing signal found there.
Even for these, the $\map$ 
values of the dark clump peak are a factor of 1.4 higher for the
ground-based ellipticity measurements. 

For the $120 \arcsec$ scale, the 
discrepancy in the $\map$ measurements at the dark clump position is even larger for those
objects with $\Delta \varepsilon > 0.2$ (right hand side of
Fig. \ref{fig:m-diff-same}). In the ground-based data, there is a $>
3.7\sigma$ peak slightly to the right of the dark clump position, while the
$\map$ values in the space-based data are comparable to the $\Delta \varepsilon \le
0.2$ sample, with a significance of $\sim 2\sigma$. At the dark clump position, the
amplitudes in the ground-based values are on average a factor 3 larger than
the space-based ones for all filter scales. Interestingly, for the smallest
two filter scales, the highest significance found in the vicinity of
the dark clump is equivalent in the two datasets. However, this is
based only on a  very small number of galaxies. \\

\begin{table*}[ptbh]
\begin{center}
%\vspace{-0.2cm}
\caption{Overview of the $\map$ analyses discussed in this appendix. Shown
  are the results at the position of the dark clump as measured in the
  ground-based data, since we want to determine the cause of the high $\map$
  amplitude in the ground-based data at this position. 
We list the ellipticity dispersion, the $\map$ value
  itself, and the signal-to-noise ratio. For those analyses of the 350
  matched objects, we give both the results of using the ground-based and
  of using the space-based data. For the second block from the bottom of the table, the
  ground-based analysis uses the space-based weights and vice
  versa. For the last block, the $\varepsilon_1$ component of the HST measurements is
  modified by applying the inverse of the linear fit between
  space-based and ground-based ellipticities.
Evaluating $\map$ only
  at the ground-based dark clump position biases the result in favor of larger
  ground-based values. For a better comparison of the general values, we quote the highest
  significance found in the vicinity of this position in parentheses.}
\vspace{0.4cm}
\renewcommand{\arraystretch}{1.2}
\renewcommand{\tabcolsep}{1.9mm}
\begin{tabular}{|l|r|c||c|c||c|c||c|c|}
\hline
& $\theta_{\rm out}$ & N &
$\sigma_{\varepsilon}$(I) & $\sigma_{\varepsilon}$(HST) & 
$\map$(I) & $\!\map$(HST)$\!$ & SNR(I) & $\!$SNR(HST)$\!$ \\
\hline
\hline
all objects of &  80$\arcsec$ & 119 & 0.34 & - & 0.056 & - & 2.6 (3.8) & - \\
ground-based   & 100$\arcsec$ & 178 & 0.34 & - & 0.070 & - & 3.5 (4.4) & - \\
catalog in     & 120$\arcsec$ & 267 & 0.35 & - & 0.076 & - & 4.8 (4.8) & - \\
HST field      & 140$\arcsec$ & 364 & 0.34 & - & 0.061 & - & 4.5 (4.6) & - \\
\hline
objects with &  80$\arcsec$ &  76 & 0.33 & 0.32 & 0.064 & 0.017 & 2.4 (4.0) & 0.6 (2.5) \\
1 counterpart& 100$\arcsec$ & 123 & 0.34 & 0.32 & 0.088 & 0.040 & 3.8 (4.6) & 1.8 (2.7) \\
             & 120$\arcsec$ & 181 & 0.34 & 0.32 & 0.087 & 0.043 & 4.6 (4.6) & 2.3 (2.8) \\
             & 140$\arcsec$ & 148 & 0.34 & 0.32 & 0.065 & 0.025 & 4.0 (4.0) & 1.6 (2.2) \\
\hline
$\Delta \varepsilon \le 0.2$
             &  80$\arcsec$ &  43 & 0.27 & 0.29 & 0.041 & 0.014 & 1.4 (3.0) & 0.4 (2.6) \\ 
             & 100$\arcsec$ &  68 & 0.27 & 0.28 & 0.065 & 0.041 & 2.4 (3.1) & 1.4 (2.5) \\
             & 120$\arcsec$ &  95 & 0.28 & 0.28 & 0.074 & 0.050 & 3.2 (3.3) & 2.2 (2.4) \\
             & 140$\arcsec$ & 135 & 0.28 & 0.29 & 0.046 & 0.035 & 2.5 (2.9) & 1.9 (2.4) \\
\hline
$\Delta \varepsilon > 0.2$ 
             &  80$\arcsec$ &  33 & 0.42 & 0.36 & 0.105 & 0.022 & 2.0 (2.8) & 0.5 (3.2)\\ 
             & 100$\arcsec$ &  55 & 0.42 & 0.37 & 0.127 & 0.038 & 3.0 (3.5) & 1.0 (2.2)\\
             & 120$\arcsec$ &  86 & 0.42 & 0.36 & 0.106 & 0.034 & 3.2 (3.8) & 1.1 (2.1)\\
             & 140$\arcsec$ & 113 & 0.42 & 0.36 & 0.095 & 0.012 & 3.1 (3.4) & 0.5 (2.2)\\
\hline
\hline
weights
             &  80$\arcsec$ &  76 & 0.35 & 0.32 & 0.067 & 0.019 & 2.3 (3.6) & 0.7 (2.6) \\ 
swapped      & 100$\arcsec$ & 123 & 0.35 & 0.32 & 0.093 & 0.040 & 3.7 (4.4) & 1.7 (3.0) \\
             & 120$\arcsec$ & 182 & 0.36 & 0.31 & 0.095 & 0.048 & 4.6 (4.7) & 2.6 (3.2) \\
             & 140$\arcsec$ & 249 & 0.35 & 0.32 & 0.073 & 0.030 & 4.2 (4.3) & 1.9 (2.6) \\
\hline
\hline 
HST $\varepsilon_1$
          &  80$\arcsec$ &  76 & - & 0.32 & - & 0.017 & - & 0.6 (2.5) \\
component & 100$\arcsec$ & 123 & - & 0.32 & - & 0.040 & - & 1.8 (2.7) \\
adjusted  & 120$\arcsec$ & 182 & - & 0.32 & - & 0.043 & - & 2.3 (2.8) \\
          & 140$\arcsec$ & 249 & - & 0.32 & - & 0.025 & - & 1.6 (2.2) \\
\hline
\end{tabular}
%\vspace{-0.2cm}
\label{tab:m-map-results}
\end{center}
\end{table*}

As there is a $\map$ peak in all these samples, we can be sure that there is
some tangential alignment around the dark clump candidate. But it is still
unclear why it is measured to be so much larger in the ground-based image.

\subsubsection{Interchanging the weights}

For the lensing analysis, the entries in the ground-based and space-based
catalogs differ mainly in the shape measurements and in the weights assigned
to each object. We have shown that the ellipticity measurements agree on
average, but we have not yet considered the different weights assigned.

\begin{figure}[bthp]
%\vspace{0.4cm}
\begin{center}
%\setlength{\fboxsep}{-\fboxrule}
%\fbox{
\includegraphics[bb=1cm 5.8cm 20cm 24.4cm,width=1\hsize]
{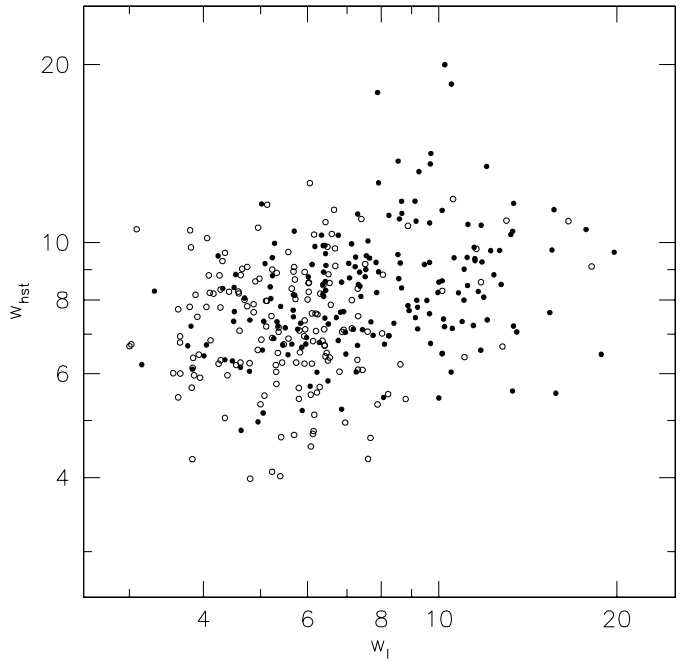}
%}
\caption{A comparison of the weights assigned to the matched objects
in the HST catalog and the ground-based catalog. Again, solid points
denote objects with $\Delta \varepsilon \le 0.2$, open circles $\Delta \varepsilon > 0.2$.
} 
\label{fig:m-weights}
\end{center}
\end{figure}

In Fig. \ref{fig:m-weights}, we directly compare the weights assigned
to the matched 
objects. Certainly, objects which were down-weighted in the ground-based
data have higher weights in the HST catalog. But else, the scatter is fairly
large. 

To test the influence of the weights on the lensing analyses, we perform an
analysis where we interchange
the weights, i.e. we assign to each object in the ground-based image
the weight of its counterpart in the HST image and vice versa. The
results are listed in Table 
\ref{tab:m-map-results}. It is interesting to note that at most filter
scales, this causes the ground-based value to decrease and the space-based
value to increase.
But the $\map$ values are within the error bars of the ones
with the original weights and are therefore only slightly
dependent on the weights.

\subsubsection[Recalibrating the HST $\varepsilon_1$ Component]{Recalibrating the HST $\bfmath{\varepsilon}_1$ component}

We noted earlier that the linear fit applied to the $\varepsilon_1$ components of the
matched objects yields a $y$-intercept with almost $1 \sigma$ significance
(Sect. \ref{sect:m-ellis}). This might point to a problem of the HST data
related to its CTE or the drizzle procedure.

Earlier, the ellipticity measurements from all three chips
were considered. The ellipticities compared were those defined with
respect to the rectascension axis. However, any problem regarding the
CTE would bias the ellipticity measurements with respect to the
read-out direction, which is different for all three chips. Also, the effect
of the anisotropy correction 
is likely to be different for each chip.

\begin{figure*}[bthp]
%\vspace{0.4cm}
\begin{center}
%\sidecaption
%\setlength{\fboxsep}{-\fboxrule}
%\fbox{
\includegraphics*[bb=1.8cm 6cm 19cm 19.7cm,width=12cm]
{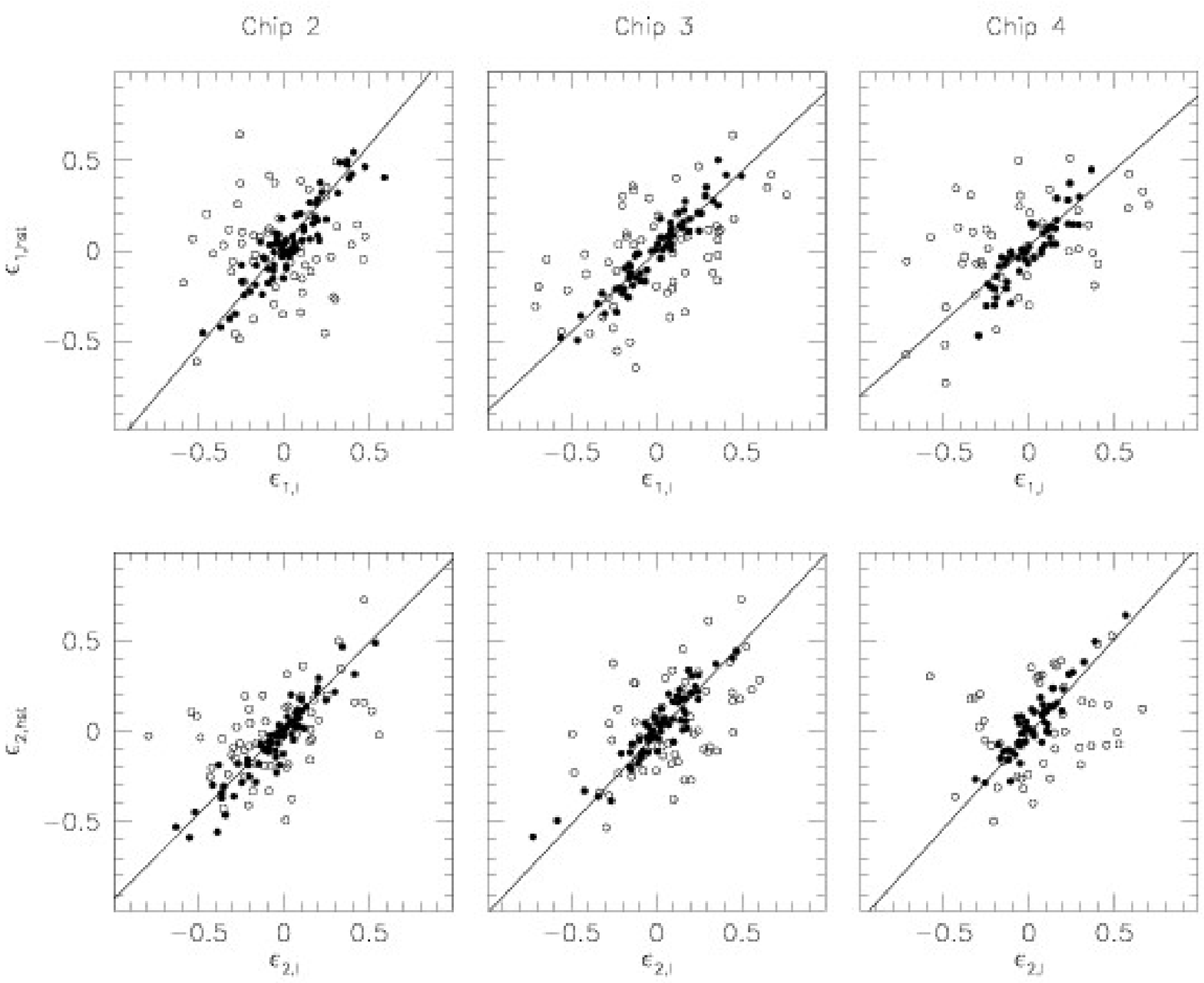}
%}
\caption{Comparison of the ellipticity measurements in the HST image
and the $I$-band image, separated by HST chip. The upper row
shows the $\varepsilon_1$ measurements, the lower the $\varepsilon_2$
measurements. Filled circles denote objects with $\Delta \varepsilon \le 0.2$,
open ones $\Delta \varepsilon > 0.2$. A weighted linear fit is indicated by the solid line.
}
\label{m-chip-fits}
\end{center}
\end{figure*}

Therefore, we repeat the ellipticity comparison for the matched
objects separated according to which HST chips they were measured in. The
linear 
fits yield (with the same definition of the fit-parameters as in
Sect. \ref{sect:m-ellis}): 
$$
\renewcommand{\arraystretch}{1.1}
\begin{array}{l@{\qquad}c@{\qquad}c@{\quad}c}
&
\mbox{Chip 2} & 
\mbox{Chip 3} &
\mbox{Chip 4}\\[1.2ex]
m_1 & 1.09 \pm 0.31 & 0.89 \pm 0.21 & 0.83 \pm 0.22\\
b_1 & 0.035\pm0.047 &-0.006\pm0.043 & 0.032\pm0.047\\[1ex]
m_2 & 0.97 \pm 0.23 & 1.01 \pm 0.25 & 1.07 \pm 0.32\\
b_2 & 0.015\pm0.045 &-0.010\pm0.047 &-0.015\pm0.054
\end{array}
$$
For the $\varepsilon_2$ component, the agreement is still very good for all
three chips. For the $\varepsilon_1$ component, the deviations from a slope of unity
are larger. Particularly, Chip 2 and 4 exhibit a notable non-zero
$y$-intercept. If these deviations are indeed a result of systematics in
the HST data, applying the inverse relation to the HST measurements
should then on average retrieve the ground-based ellipticities.

We modify the HST measurements by the relation
$$
\varepsilon_1 \quad \longrightarrow \quad \frac{\varepsilon_1 - b_i}{m_i}
$$
for each chip $i$. We do not alter the $\varepsilon_2$ components, as we deem their
agreement with the ground-based data satisfactory.

The results of a lensing analysis of the matched objects with these
modified ellipticities are listed in Table
\ref{tab:m-map-results}. Indeed, this transformation yields
$\map$ values 25\% larger than the original ones for the smallest two
filter scales. For the $100 \arcsec$ filter scale, the effect is
smaller, and for the $120 \arcsec$ filter scale it is negligible.

If the height of the $\map$ peak were uncorrelated with the systematic
deviation in the ellipticity measurement, the modification would only
add noise to the $\map$ statistic, as most ellipticities are amplified
due to $m_i < 1$. The $\map$ value of the peak therefore would not be
significantly altered, as is the case at least for the largest
filter scale. For the other filter scales, there is a slight increase
in both $\map$ and SNR value. But the change is at most $0.3 \sigma$,
so the variations are still within the standard deviation of the
original measurement. 

The results of a lensing analysis of the complete HST
catalog with this modification of the ellipticity are very similar
to the original one. But as the linear fit which was the basis of this 
modification  
applies only to bright objects in the HST catalog, it is very
speculative to extrapolate this to fainter objects.

As this modification alone is not able to reproduce the large lensing
signal seen in the ground-based data, particularly at the $120
\arcsec$ filter scale, the offset of the space-based ellipticities is
not the cause of the discrepancy in the $\map$ measurements.

\subsubsection{Rebinning the galaxies in the HST catalog}

For the lensing analysis of the HST data, we had divided the galaxy
sample into three magnitude bins of equal numbers of object. About half of the
objects in the brightest bin were also detected in the $I$-band image
and are used in the comparisons presented in this chapter. For these,
we have confirmed the presence of tangential alignment, even though
the $\map$ amplitude differs in the two datasets. But
we fail to detect alignment in the medium bright HST bin.

The completion limit of the ground-based image falls within the
brightest HST bin (cf. Fig. \ref{fig:m-mags}), and so a number of
objects in the latter, though of similar brightness, were not detected
in the ground-based image. To test how much the additional objects in
the HST image contribute to the lensing signal, we rearrange the
HST brightness bin: instead of the magnitude cut between the brightest and
medium bin, we split the
galaxies according to whether or not they are a counterpart of an object
detected in the $I$-band image. From the latter sample the faintest bin is
split with the same magnitude cut as before.
This effectively moves several objects
from the bright into the medium bin and only a few the other way. The
faint bin remains almost unaltered.

\begin{figure*}[bthp]
%\vspace{0.4cm}
\begin{center}
%\sidecaption
%\setlength{\fboxsep}{-\fboxrule}
%\fbox{
\includegraphics[width=12cm]
{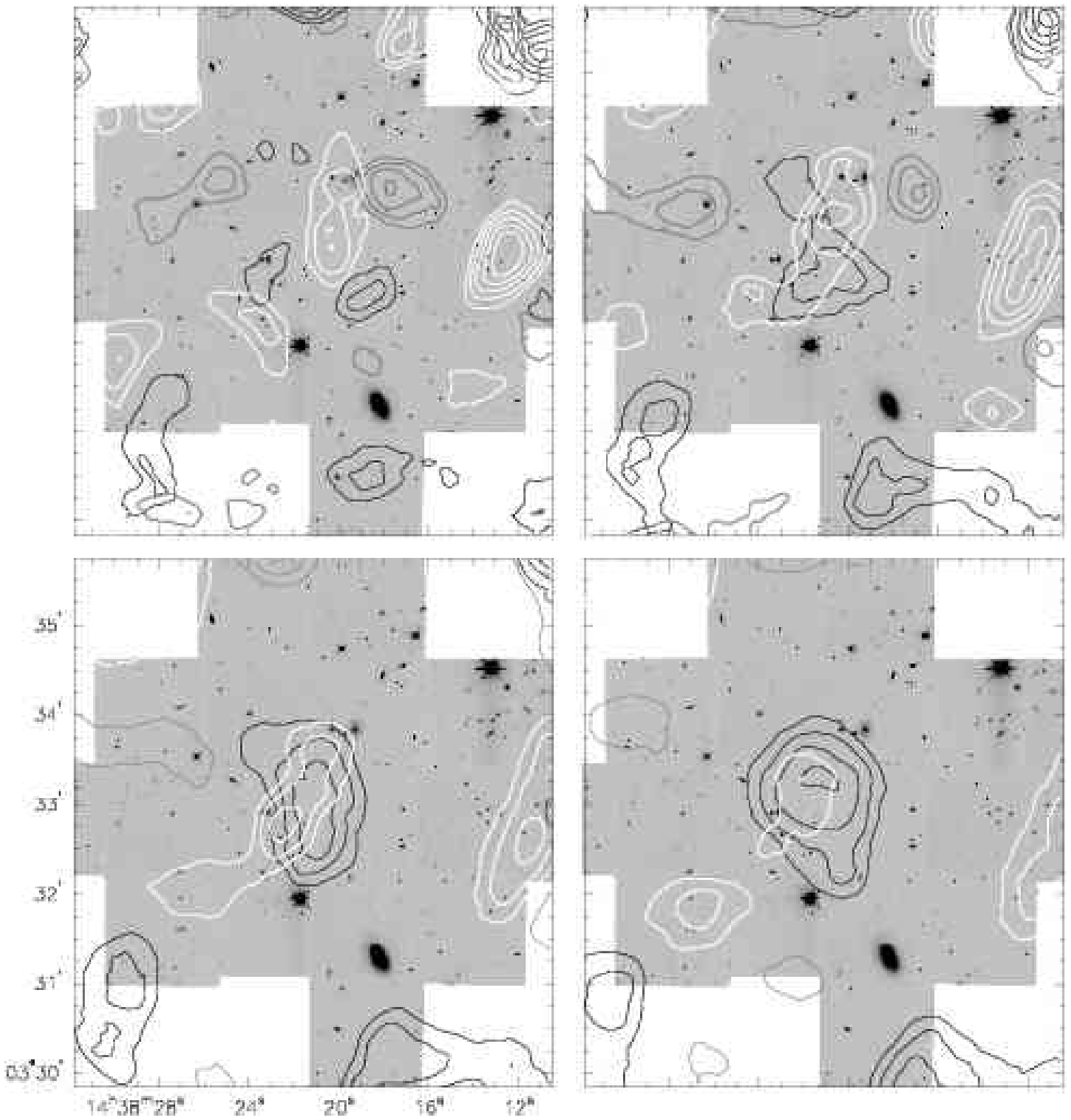}
%}
\caption{$\map$ analysis of the HST data, with the galaxies split into
three samples according to whether they are a counterpart of an object
detected in the $I$-band image (white contours). Those objects not
detected in the $I$-band image are split by a magnitude cut at
$m=26.5$, where gray contours corresponds to galaxies brighter than this
limit, black contours to those fainter. The contours start at $1.5
\sigma$ and increase in $0.5 \sigma$ increments. The filter scales are 
80$\arcsec$, 100$\arcsec$, 120$\arcsec$, and 140$\arcsec$ (from upper left to
lower right).}
\label{fig:m-hst-nomatch}
\end{center}
\end{figure*}

In Fig. \ref{fig:m-hst-nomatch}, we present the signal-to-noise contours of a $\map$
analysis of these three bins. The medium bin, i.e. galaxies brighter
than $m=26.5$ that were not detected in the ground-based image
contains the most galaxies (835 compared to 350 in the bright bin and
594 in the faint bin), so that the noise is smallest. Yet,
the $\map$ values in the vicinity of the dark clump are compatible
with zero within $1 \sigma$.

Apparently, there is no tangential alignment present in the objects
that were moved from the bright bin to the medium bin. This is
compliant with the observation that for those objects also detected in
the $I$-band image, the signal-to-noise ratio of the $\map$ peak is
larger than for the bright magnitude bin (cf.
Fig. \ref{hst-tomo}). 

Intriguingly, this implies that of the faint galaxies at the
completion limit of the $I$-band image, those with a tangential
alignment to the dark clump were preferentially detected. However, the
fact that there is no tangential alignment for about half of the
objects in the HST image, which are neither at the bright nor the
faint end of the magnitude distribution, clearly disfavors the
hypothesis that the tangential alignment in the brighter galaxies is
caused by a lensing mass.

\subsection{Comparison of the tangential ellipticities}
\label{sc:compare-gt}

So far, we have not found any systematic that could be the single
cause for the discrepancy in the lensing analyses. However, it is
clear that for the faint galaxies which mainly caused the lensing
signal in the ground-based image, the ellipticities agree with the HST
measurements only on average, but not on an object-to-object basis.
This is not surprising as the noise in the ground-based image is
likely to influence the shape measurements. Yet, it is unlikely that
noise can cause (or amplify) tangential alignment around a certain
point. 

\begin{figure}[bthp]
%\vspace{0.4cm}
\begin{center}
%\setlength{\fboxsep}{-\fboxrule}
%\fbox{
\includegraphics[bb=1cm 5.8cm 20cm 24.4cm,width=1\hsize]
{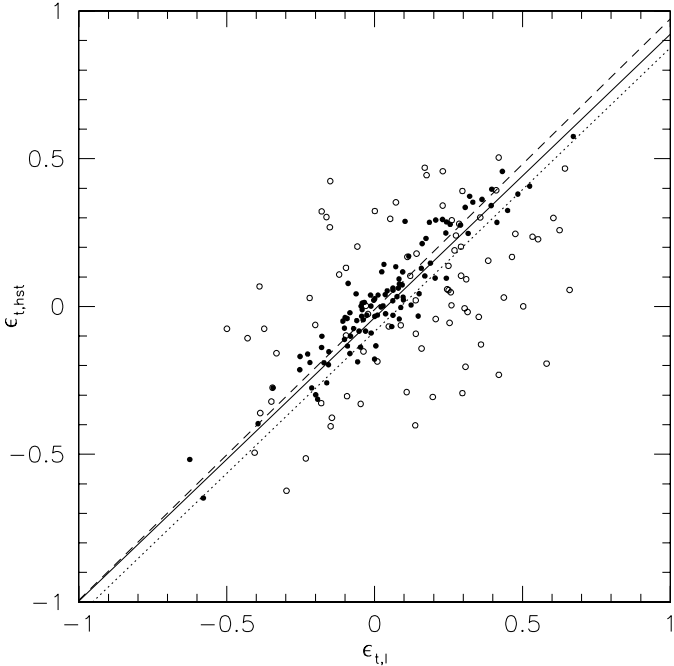}
%}
\caption{Comparison of the tangential ellipticities measured toward
the dark clump (as measured in the ground-based data) of 
space-based and ground-based objects. Objects with $\Delta \varepsilon \le 0.2$
are shown as solid, objects with $\Delta \varepsilon > 0.2$ as open circles. The dashed
line indicates a linear fit to the first sample, the dotted
line one to the latter sample. The solid line represents the fit to
the combined set.} 
\label{fig:m-gt-compare}
\end{center}
\end{figure}

In Fig. \ref{fig:m-gt-compare} we compare the tangential ellipticity
$\varepsilon_{\rm t}$
of the matched objects with respect to the dark clump centroid found
in the ground-based data. Only objects within $120 \arcsec$ of that
position are shown. Evidently, most objects in the ground-based data have a
positive $\varepsilon_{\rm t}$. 
The scatter is comparable to the comparison of the individual
ellipticity components (Sect. \ref{sect:m-ellis}). Likewise, we fit
a linear function, where we distinguish between the
complete sample of matched galaxies and how well the ellipticity
measurements agree (as before, the fit is expressed such that the
ground-based data are the independent variable):
$$
\renewcommand{\arraystretch}{1.1}
\begin{array}{l@{\qquad}c@{\qquad}c@{\quad}c}
& \mbox{all} & \Delta \varepsilon \le 0.2 & \Delta \varepsilon > 0.2 \\
m & 0.98 \pm 0.21 & 1.03 \pm 0.28 & 0.93 \pm 0.32 \\
b &-0.047 \pm 0.043 &-0.018 \pm 0.055 & -0.086\pm 0.071
\end{array}
$$
For each of these fits, the $y$-intercept is negative, which indicates
larger ground-based values. While it lies well within the error bars
of the fit for those galaxies with $\Delta \varepsilon \le 0.2$, it is
significantly non-zero for those with $\Delta \varepsilon > 0.2$. For those,
also the slope indicates a preferentially larger $\varepsilon_{\rm t}$ value in
the ground-based data.\\

On the basis of the correlation of the individual measurements of the
tangential ellipticity, the agreement between the two datasets is
comparable to that of the ellipticity components, as would be
expected. Yet, the mean tangential ellipticity is different: for the
ground-based data, it is $\ave{\varepsilon_{\rm t,I}} = 0.086 \pm 0.017$, for
the space-based 
data, $\ave{\varepsilon_{\rm t,hst}} = 0.034 \pm 0.017$. These are very similar to the
results of the $\map$ analysis at the same position at the
$120\arcsec$ filter scale, $\map^I(120\arcsec) = 0.087 \pm 0.019$ and
$\map^{\rm hst}(120\arcsec) = 0.043 \pm 0.019$ (compare Table
\ref{tab:m-map-results}). This again demonstrates that the coherent
alignment is present over a range of distance from the dark clump, as
the $\map$ statistics is a weighted average tangential ellipticity,
where the weight is dependent on the distance. 

\begin{figure}[bthp]
%\vspace{0.4cm}
\begin{center}
%\setlength{\fboxsep}{-\fboxrule}
%\fbox{
%\includegraphics[bb=1cm 6cm 20.3cm 24.4cm,width=1\hsize]
\includegraphics[width=1\hsize]
{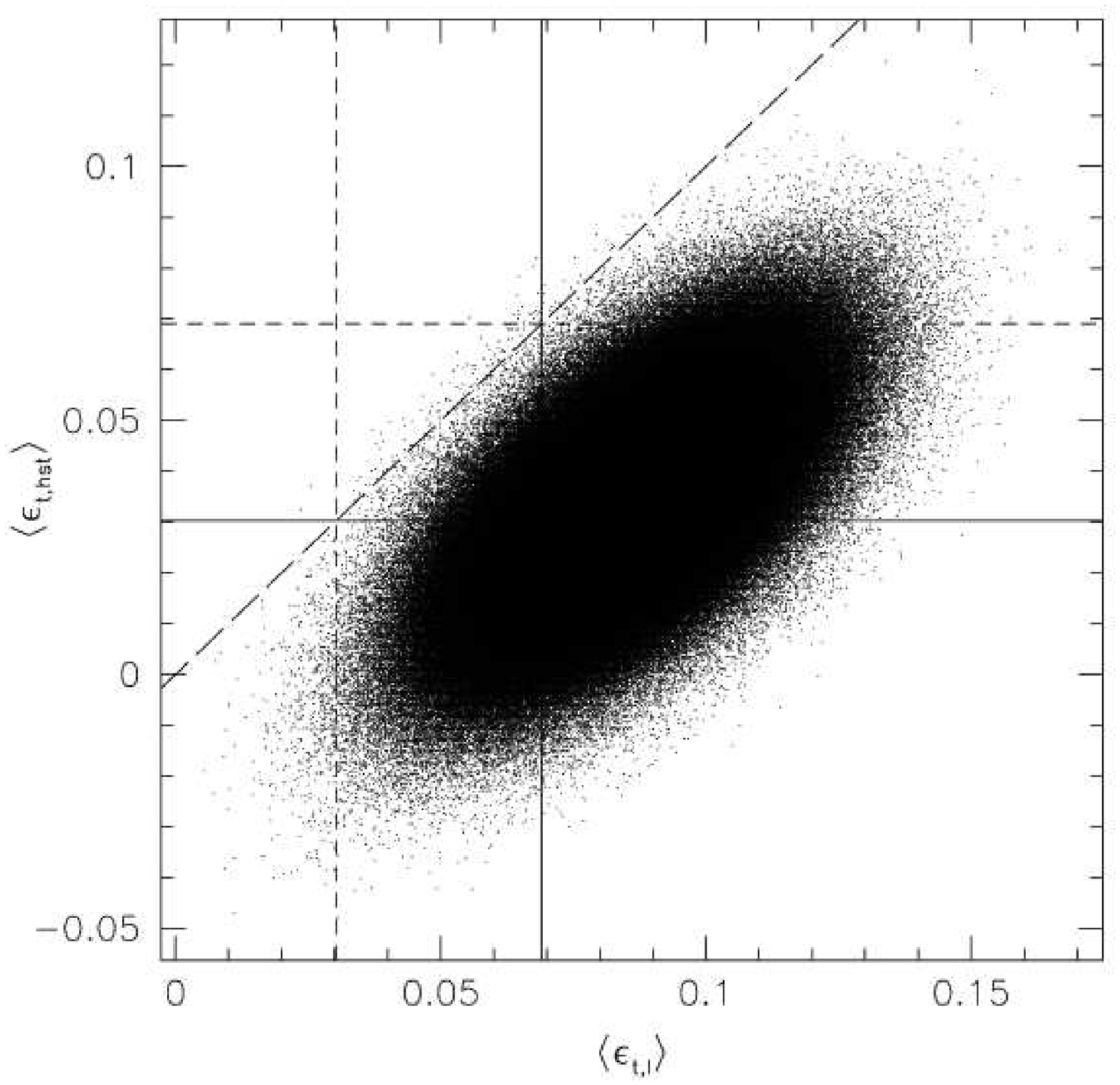}
%}
\caption{Results of a bootstrap analysis of the mean tangential
ellipticity of the matched objects within $120\arcsec$ of and with
respect to the dark clump position in the ground-based data. For
objects below the long-dashed line, $\ave{\varepsilon_{\rm t,I}} > \ave{\varepsilon_{\rm
t,hst}}$, above it, $\ave{\varepsilon_{\rm t,I}} < \ave{\varepsilon_{\rm t,hst}}$. The
solid black (red) line and dashed black (red) line indicate $\ave{\varepsilon_{\rm
t,I}}$ ($\ave{\varepsilon_{\rm t,hst}}$) without bootstrapping.} 
\label{fig:gt-bootstrap}
\end{center}
\end{figure}

Within the standard
deviations of the cited values, the two analyses are compatible with
each other. However, these errors are not independent, as they are based
on the ellipticity dispersion of the same galaxies. To estimate the
significance with which the ground-based mean is larger than the
space-based mean, we perform a bootstrap analysis, i.e. from the 195
galaxies in the sample, we draw at random 195 objects, with replacement. From these
we calculate $\ave{\varepsilon_{\rm t,I}}$ and $\ave{\varepsilon_{\rm t,hst}}$ and repeat the
procedure 1,000,000 times. The result is shown in Fig. \ref{fig:gt-bootstrap}. We
find that the space-based mean exceeds the ground-based mean in only
176 cases, i.e. the latter one is larger with $3.8 \sigma$ confidence.

This analysis is biased in the way that the reference position is the
dark clump position as measured in the ground-based data. Therefore,
we repeat the analysis for the centroid found in the space-based
data. Even for this case, the ground-based mean value exceeds the
space-based mean with $3.1 \sigma$ confidence. 

%We had found before
%that the $\map$ amplitudes in the vicinity of the dark clump are about
%twice as high for ground-based ellipticities than for space-based.

It is quite puzzling that despite the general agreement of the
ellipticity measurements, the mean tangential alignment can differ so
much. As yet, we have no explanation for the cause of this.\\

\begin{figure*}[bh!]
\begin{center}
\includegraphics[width=0.49\hsize,trim=0 0 -0.7cm 0]{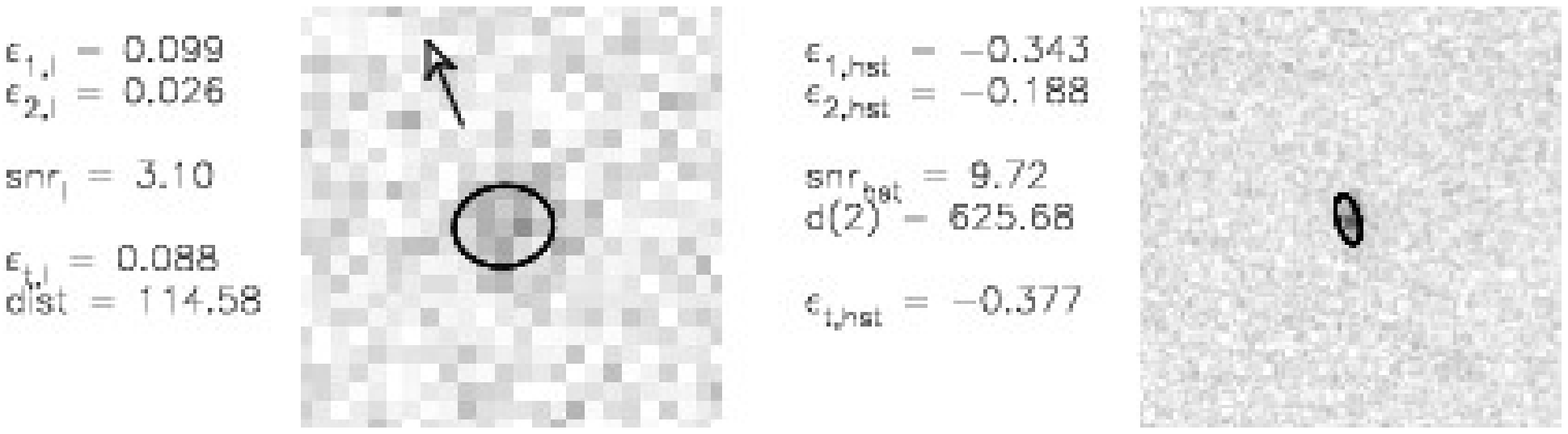}\vspace{0.4cm}
\includegraphics[width=0.49\hsize,trim=-0.7cm 0 0 0]{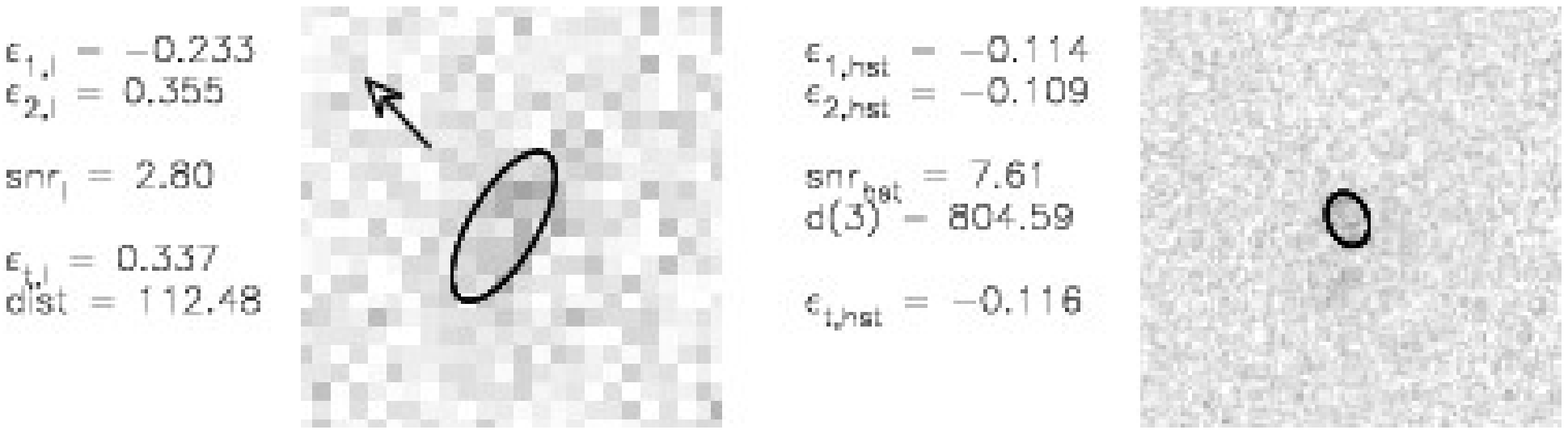}
\includegraphics[width=0.49\hsize,trim=0 0 -0.7cm 0]{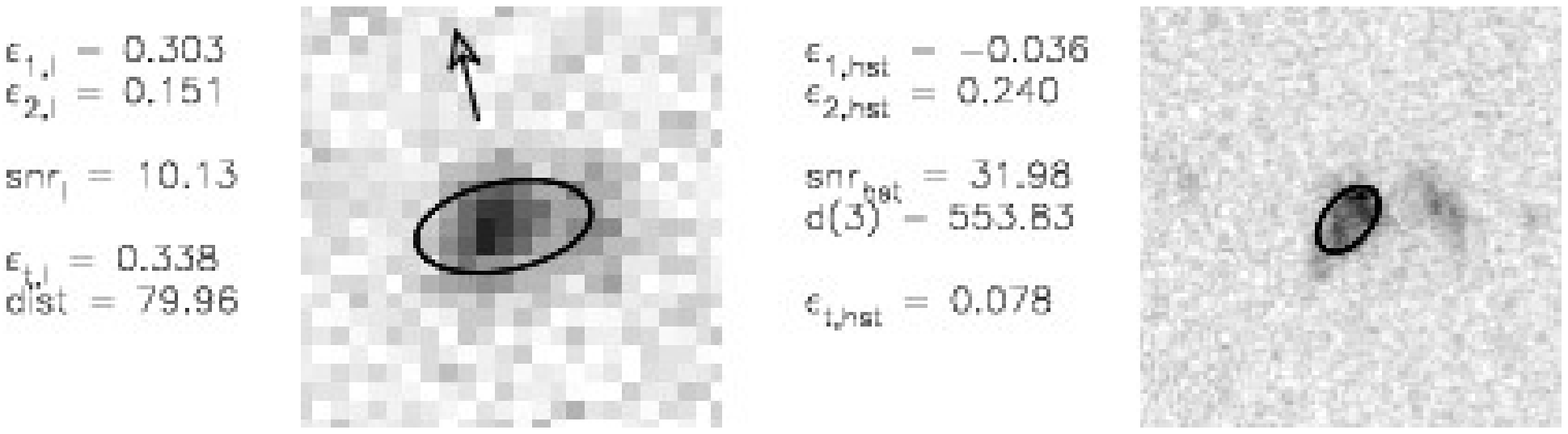}\vspace{0.4cm}
\includegraphics[width=0.49\hsize,trim=-0.7cm 0 0 0]{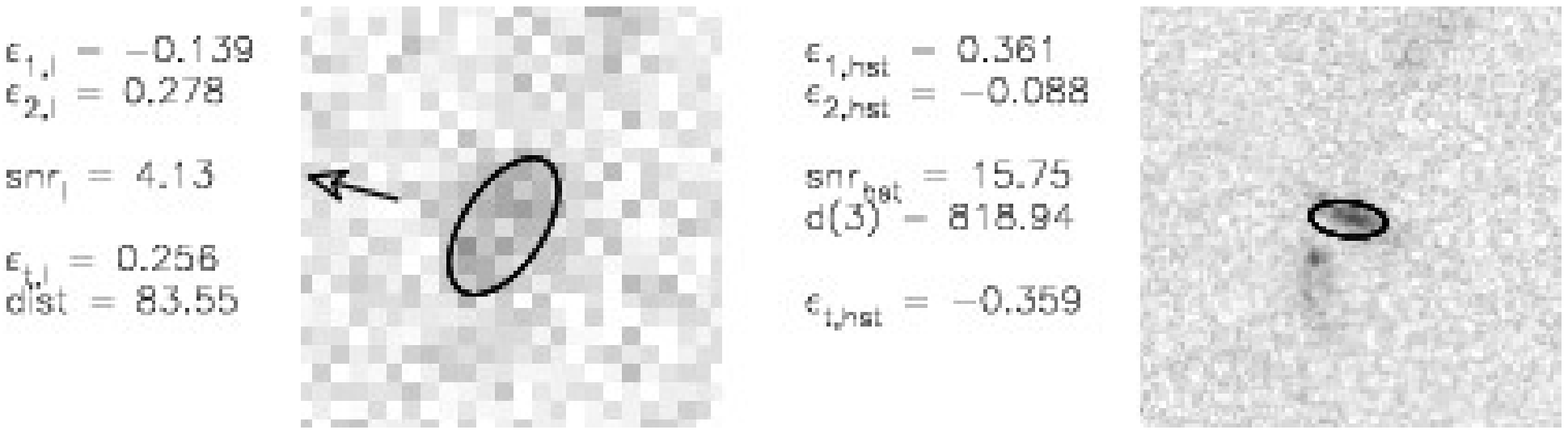}
\includegraphics[width=0.49\hsize,trim=0 0 -0.7cm 0]{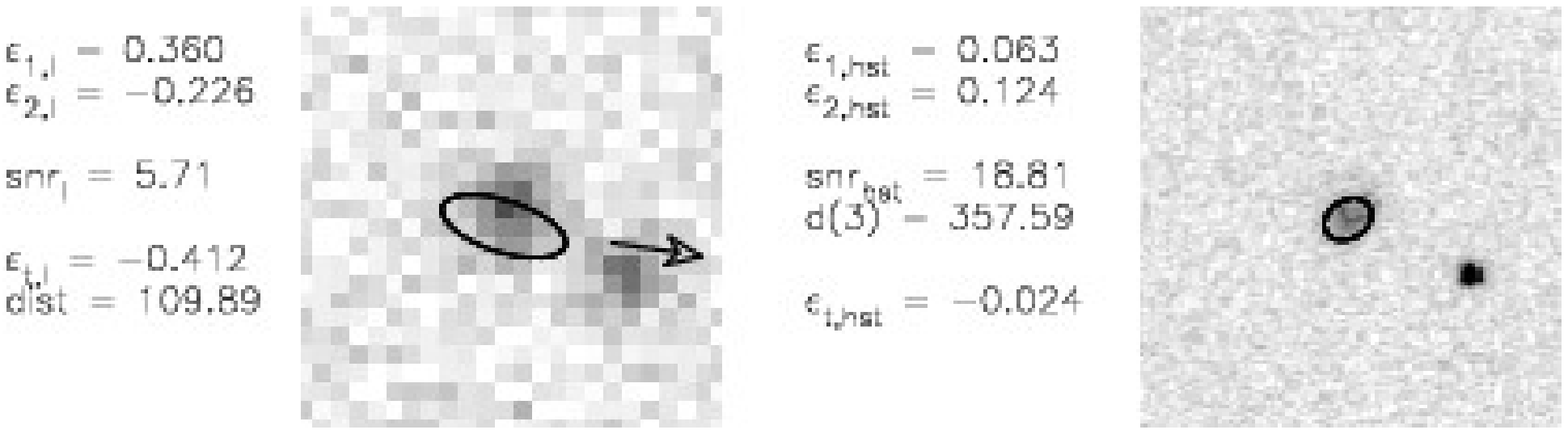}\vspace{0.4cm}
\includegraphics[width=0.49\hsize,trim=-0.7cm 0 0 0]{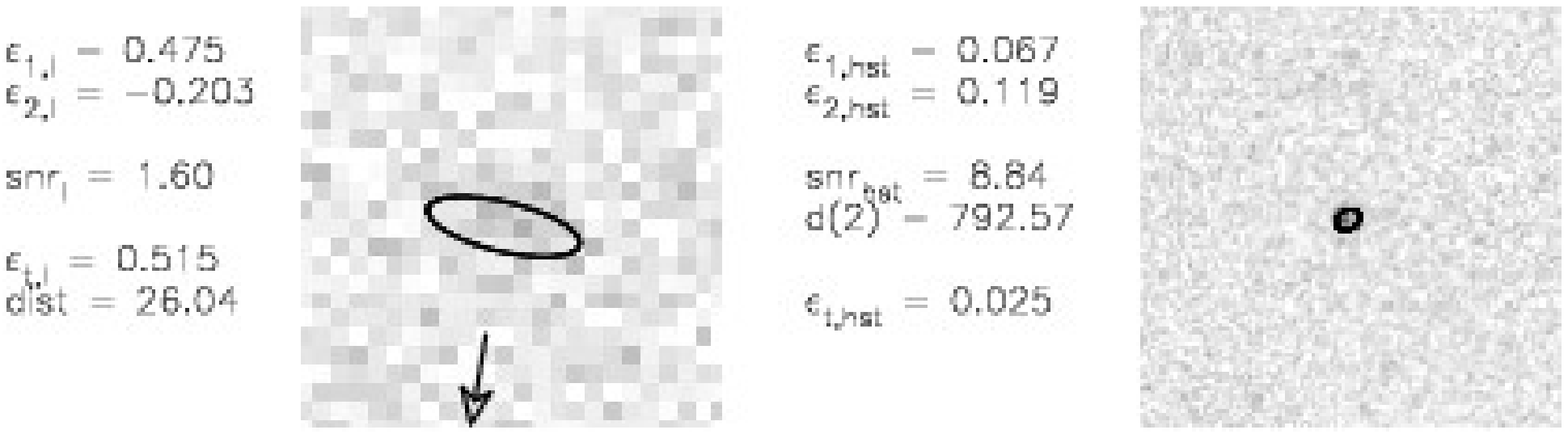}
\includegraphics[width=0.49\hsize,trim=0 0 -0.7cm 0]{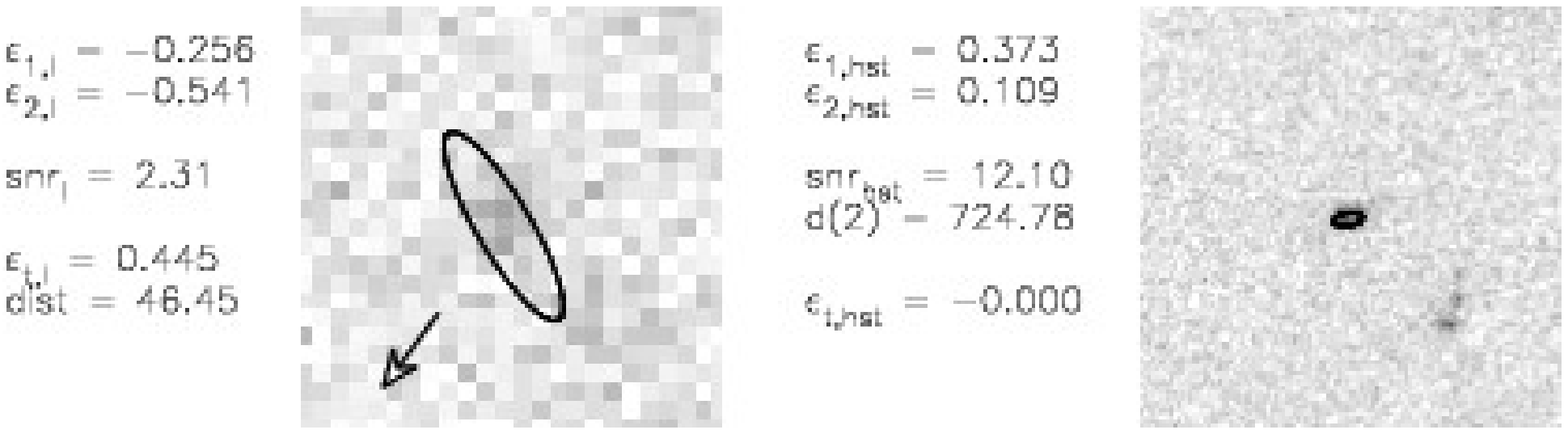}\vspace{0.4cm}
\includegraphics[width=0.49\hsize,trim=-0.7cm 0 0 0]{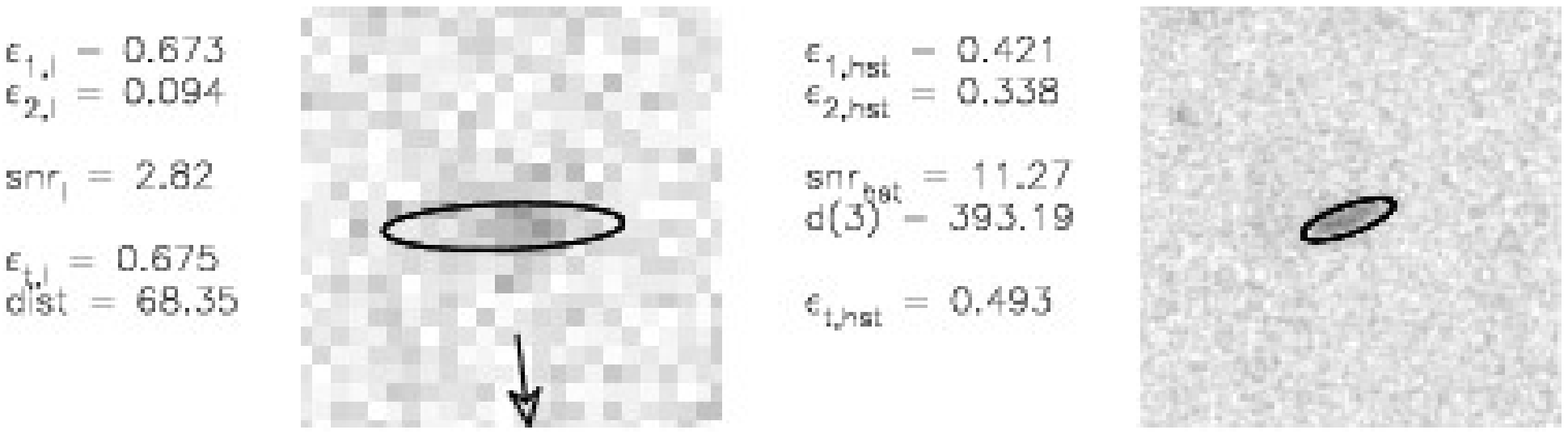}
\begin{picture}(0,0)
\put(-252,-5){\line(0,1){305}}
\end{picture}
\end{center}
\caption{
Direct comparison of objects with $\Delta \varepsilon > 0.2$ within a
$120 \arcsec$ radius around the dark clump (position measured in the
ground-based data) in the $I$-band (left side) and HST image (right
side). For each object, we list its ellipticity components $\varepsilon_1$ and
$\varepsilon_2$ in both images, which are also indicated by ellipses
superposed on the images. The signal-to-noise ratio {\tt snr} of each detection
is listed as well. In the ground-based image, we also indicate
the direction to the dark clump (by an arrow in the image) and its
distance $dist$, and 
list the tangential ellipticity $\varepsilon_{\rm t}$ towards the dark clump for both
images. For the HST images, we also give each object's distance $d(i)$ to the
center of the chip $i$ it was measured on. Only 8 exemplary objects are shown.
}
\label{fig:bildchen}
\end{figure*}

In Fig. \ref{fig:bildchen}, we present clippings of some of these galaxies from
the two images to convey a visual impression on how they compare in the
images and how the shape measurements relate to the image, especially
for the ground-based image. We find that in the HST image, the
measured ellipticity very well represents the shape of the
object. This is to be expected as all these objects are detected with
a high signal-to-noise ratio in the HST image. In the $I$-band image,
these objects have a low signal-to-noise ratio and the ellipticity
measurements only vaguely reflect the shape of the objects. 
About 10\% of the objects are actually two or more objects which were not
resolved in the ground-base image; for about 20\% a nearby (though
resolved) object seems to influence the ellipticity measurement.
It is
especially striking that for many, the ellipticity modulus $|\bfmath{\varepsilon}|$ is
clearly overestimated.

\subsection{Summary}

We have shown that on average, ground-based and space-based
ellipticity measurements agree very well. This justifies the
assumption that weak lensing analyses based on ground-based data yield
reliable results despite the smearing of object shapes by the Earth's
atmosphere. 

For the HST ellipticities, we have found only little evidence that
they might be biased due to the CTE problem, the PSF undersampling, or
the anisotropy correction. This holds at least for bright objects, for
fainter ones we have no possibility of comparison. Yet, previous
analyses (Sect. \ref{CTE}) have shown that these problems have little
influence.

The fainter an object, the less reliable its shape measurement is
(Fig. \ref{fig:m-mags}). The strong lensing signal seen in the
ground-based data is caused mainly by faint objects, for which the
ellipticity measurements in the two datasets deviate. In both datasets,
there is a $\map$ peak at the position of the dark clump, but the
degree of tangential alignment is much stronger in the ground-based
data. For many of
these objects, the ellipticity modulus $|\bfmath{\varepsilon}|$ is overestimated in the
ground-based measurements, and many of these seem to be accidentally
tangentially aligned to the dark clump, thus causing a high $\map$ signal.

\clearpage

\bibliographystyle{aa}
\bibliography{refs.bib}

\end{document}